\documentclass[11pt]{article}
\usepackage{xcolor}
\usepackage{textcomp}
\usepackage{cite}
\usepackage{multirow}
\usepackage{enumerate}
\usepackage{chet}
\usepackage[utf8]{inputenc}

\usepackage{graphicx}
\usepackage{tikz}
\usepackage[subrefformat=parens,labelformat=parens]{subfig}

\usepackage{amsfonts,amsmath}
\usepackage{braids}
\usepackage{mathtools}
\usepackage{colortbl}
\usepackage{multirow,multicol}

\newcommand{\bZ}{\mathbb{Z}}

\newcommand{\bC}{\mathbb{C}}
\newcommand{\bR}{\mathbb{R}}

\newcommand{\cN}{\mathcal{N}}

\newcommand{\minibraidd}[5]{\begin{tikzpicture}[scale=#1]\braid[rotate=90,style strands={1}{#3},style strands={2}{#4},style strands={3}{#5}] #2; \end{tikzpicture}}
\newcommand{\minibraidc}[4]{\begin{tikzpicture}[scale=#1]\braid[rotate=90,style strands={1}{#3},style strands={2}{#4}] #2; \end{tikzpicture}}
\newcommand{\minibraid}[2]{\begin{tikzpicture}[scale=#1]\braid[rotate=90] #2; \end{tikzpicture}}
\newcommand{\ab}[2]{\ensuremath{(a,b)=(#1,#2)}}

\newcommand{\nicemod}{~\text{mod}~}
\newcommand{\im}[1]{\textit{Im\ensuremath{(#1)}}}
\newcommand{\re}[1]{\textit{Re\ensuremath{(#1)}}}
\newcommand{\II}[0]{\textit{II}\xspace}
\newcommand{\III}[0]{\textit{III}\xspace}
\newcommand{\IV}[0]{\textit{IV}\xspace}
\newcommand{\IN}[0]{\textit{I}\xspace}

\newcommand{\rep}[1]{\ensuremath{\boldsymbol{#1}}\xspace}

\begin{document}
\title{Dualities of Deformed $\cN=2$ SCFTs\\ \vspace{.4cm} from Link Monodromy on D3-brane States}
\author{\small Antonella Grassi$^{1,}$\email{grassi@sas.upenn.edu}, James Halverson$^{2,}$\email{j.halverson@neu.edu}, Fabian Ruehle$^{3,}$\email{fabian.ruehle@physics.ox.ac.uk} and Julius L. Shaneson$^{1,}$\email{shaneson@math.upenn.edu}}

\affiliation{%
$^{1}$ Department of Mathematics, University of Pennsylvania, \\ \vspace{-.1cm}
David Rittenhouse Laboratory, 209~S~33rd~Street, Philadelphia, PA 19104, USA\\
\vspace{.2cm}
$^{2}$ Department of Physics, Northeastern University, Boston, MA 02115, USA\\
\vspace{.2cm}
$^{3}$ Rudolf Peierls Centre for Theoretical Physics, Oxford University, 1 Keble Road, Oxford, OX1 3NP, UK \\
}



\abstract{We study  D3-brane theories that are dually described as
  deformations of two different $\cN=2$ superconformal theories with
  massless monopoles and dyons. These arise at the self-intersection
  of a seven-brane in F-theory, which cuts out a link on a small
  three-sphere surrounding the self-intersection. The spectrum is
  studied by taking small loops in the three-sphere, yielding a
  link-induced monodromy action on string junction D3-brane states,
  and subsequently quotienting by the monodromy. This reduces the
  differing flavor algebras of the $\cN=2$ theories to the same flavor
  algebra, as required by duality, and projects out charged states,
  yielding an $\cN=1$ superconformal theory on the D3-brane. In one, a
  deformation of a rank one Argyres-Douglas theory retains its $SU(2)$
  flavor symmetry and exhibits a charge neutral flavor triplet that is
  comprised of electron, dyon, and monopole string junctions.  From
  duality we argue that the monodromy projection should also be
  imposed away from the conformal point, in which case the D3-brane
  field theory appears to exhibit confinement of electrons, dyons, and
  monopoles. We will address the mathematical counterparts in a
  companion paper.}

\maketitle

\toc

\clearpage

\newsec{Introduction}
\label{sec:Introduction}

Historically, the study of D3-branes has led to a rich array of
physical phenomena in supersymmetric quantum and conformal field
theories.  For example, at orbifold singularities D3-branes give rise
to rich quiver gauge theories \cite{Douglas:1996sw}; in F-theory
\cite{Vafa:1996xn} compactifications they realize
\cite{Banks:1996nj,Douglas:1996js} a variety of Seiberg-Witten
theories \cite{Seiberg:1994rs,Seiberg:1994aj} and superconformal field
theories (SCFTs) of Argyres-Douglas
\cite{Argyres:1995jj,Argyres:1995xn} and Minahan-Nemeschansky
\cite{Minahan:1996fg,Minahan:1996cj}, and also other theories \cite{Aharony:1996bi,Fayyazuddin:1997cz,Fayyazuddin:1998fb,Aharony:1998xz,Heckman:2010qv}; finally, most famously $N$ D3-branes
give rise to gravity \cite{Maldacena:1997re,Witten:1998qj} in the large $N$ limit. Many of the most
interesting results exist at strong coupling, but are tractable due to
$SL(2,\bZ)$ invariance.

F-theory itself has a plethora of strongly coupled phenomena and SCFTs
beyond its D3-brane sectors. For example, its seven-brane
configurations may realize exceptional gauge symmetry and seven-brane
structures, which is central to certain phenomenological aspects of
F-theory GUTs \cite{Donagi:2008ca,Beasley:2008dc,Beasley:2008kw};
there is growing evidence that non-trivial seven-brane structures,
so-called non-Higgsable clusters
\cite{Morrison:2012np,Morrison:2012js}, are generic
\cite{Morrison:2012js,Grassi:2014sda,Halverson:2015jua,Taylor:2015xtz,Taylor:2015ppa,Halverson:2016vwx}
in F-theory; and in recent years there has been a resurgence of
interest in $6d$ $(1,0)$ \cite{Heckman:2013pva,Heckman:2015bfa} and
$4d$ $\cN=1$ SCFTs \cite{Heckman:2010qv,Morrison:2016nrt} that arise
from F-theory and in $\mathcal{N}=2$ SCFTs in general \cite{Argyres:2015ffa,Argyres:2015gha,Argyres:2016xmc}.  All of these typically involve strongly coupled
physics.

In this paper we initiate the study of string junctions on D3-brane
theories that probe non-trivial seven-brane configurations in lower
(than eight) dimensional compactifications of F-theory. Specifically,
we will develop a mathematical and physical formalism for studying the
spectrum of D3-brane theories at certain isolated seven-brane
singularities (non-trivial self-intersections of an $I_1$-locus) that
should be extendable to broader classes of singularities.

One physical aspect we will study is how duality arises
geometrically from deforming rather different $\cN=2$ SCFTs. 
Specifically, the D3-brane theory we study in this paper,
which we call Theory $h$ for brevity, is a deformation of two different $\cN=2$
SCFTs realized on D3-branes in simpler F-theory backgrounds. We will
call the latter two Theory $f$ and Theory $g$, and denote their flavor
symmetries as $G_f$ and $G_g$, which can take values $G_f\in \{SU(2),
SO(8), E_7 \}$ and $G_g\in \{\emptyset, SU(3), SO(8), E_6, E_8\}$ with
$G_f\neq G_g$ in general. Schematically, D3-brane positions relative
to the seven-brane configurations in these theories appears as
\begin{center}
\includegraphics[height=5cm]{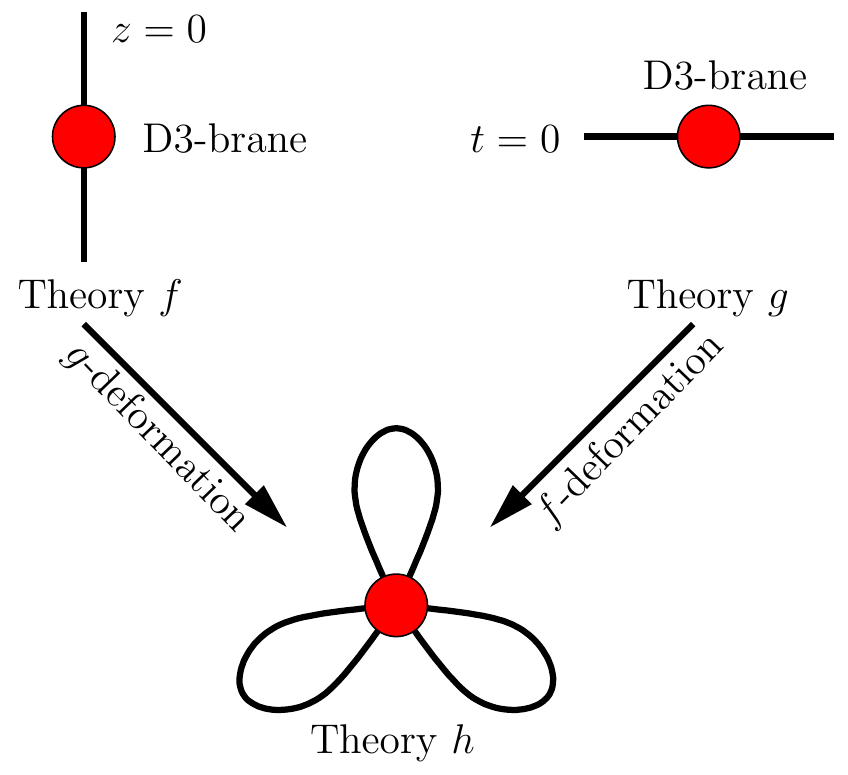}
\end{center}
\noindent where the D3-brane theory $h$ with flavor group $G_h$ in the non-trivial seven-brane
configuration at the bottom may be obtained either from a deformation of the
vertical or horizontal seven-branes of theories $f$ and $g$. The deformed
$\cN = 2$ theories are necessarily dual since the D3-brane theory in the
non-trivial background can be obtained from either deformation. Said
differently, the coordinates that parameterize the Coulomb branches of
the $\cN=2$ theories are on equal footing as spatial coordinates from a ten-dimensional perspective,
and in the deformed theory the seven-branes that the D3-brane probes spread
out in both directions.

The seven-brane backgrounds that we study are easily described in
F-theory. We will study specific backgrounds, but our techniques should
be generalizable to others as well.  They are described by an elliptic fibration $X_h$ over
$\bC^2$ with coordinates $(z,t)$ and with fiber coordinates $(x,y)$, which in Weierstrass form are given
by
\begin{align}
X_h: \ \ y^2 = x^3 - z^a\, x + t^b,
\label{eqn:introWeier}
\end{align}
where $a\in \{1,2,3\}$, $b\in \{1,2,3,4,5\}$; see work
\cite{Grassi:2001xu} of Grassi, Guralnik, and Ovrut for the
$(a,b) = (1,1)$ case. A seven-brane is localized on the locus
$\Delta = 0$ where $\Delta = -4z^{3a}+27t^{2b}$. 
Note  that  for $b >1$ the elliptic threefold defined by  \eqref{eqn:introWeier} has an isolated singularity  at $z=t=x=y=0$; we will address the role of the singularity in this context in  a sequel paper \cite{Grassi:2017xxx}. The worldvolume theory of the D3-brane
at $z=t=0$ is Theory $h$, and the $\cN=2$ SCFTs Theory $f$ and Theory
$g$ are obtained by turning off the terms $-z^a x$ and $t^b$ in
(\ref{eqn:introWeier}):
\begin{align}
X_f: \ \ y^2 = x^3  + t^b\,,\qquad\qquad
X_g: \ \ y^2 = x^3  -z^ax\,.
\end{align}
Theories $f$ and $g$ have different flavor
symmetries, which must be reduced to a common one by the deformation
to Theory $h$.  The ``paradox'' can be seen directly in the
background (\ref{eqn:introWeier}), since Theory $h$ may be obtained by
taking a D3-brane to $z=t=0$ via coming in along the locus $z=0$ or
the locus $t=0$. These processes naively look like turning off mass deformations
of Theory $f$ and Theory $g$, respectively, but this cannot be the full story
since then the flavor symmetries would disagree. This (incorrect) conclusion is obtained
by looking too locally in the geometry, and by looking more globally the issue 
is resolved. Specifically, torus knots or links on which seven-branes are localized
arise naturally in the geometry, and we will use them to reconcile the
naive flavor symmetry discrepancy between theories $f$, $g$, and $h$.

Throughout this work, our focus will be on the implications of the geometry for
the D3-brane spectrum, but there are many interesting questions for future work.

The sketch of our results are as follows. As is
well-known, D3-brane probes of seven-brane backgrounds in
eight-dimensional F-theory compactifications have $(p,q)$ string
junctions stretching between the D3-brane
and the seven-brane. These describe a rich spectrum of states in
non-trivial flavor representations that are generally charged both
electrically and magnetically under the $U(1)$ of the D3-brane.
Mathematically, these string junctions are topologically described by
elements of relative homology; they are two-cycles in an elliptic
fibration $X$ over a disc $D$ relative a chosen fiber $E_p$ above a
point $p$, which means they are two-chains that may have boundary in $E_p$. Thus, topologically
a junction $J$ is $J \in H_2(X,E_p)$. Here the elliptic
fiber $E_p$ is the elliptic fiber over the D3-brane, so the
``asymptotic charge'' $a(J):=\partial J\in H_1(E,\bZ)$ gives the
electromagnetic charge of the junction ending on the D3-brane. There
is a pairing $\langle\cdot,\cdot \rangle$ to the integers on
$H_2(X,E_p)$ that is the intersection pairing $(\cdot,\cdot)$ on
closed classes, i.e. those with $a(J)=0$. Finally, following
\cite{DeWolfe:1998zf,Grassi:2013kha,Grassi:2014ffa}, the set
\begin{align} 
  R:=\{J\in H_2(X,E_p)\,\,\, | \,\,\, (J,J)=-2,~a(J)=0\}
\end{align}
has the structure
of an ADE root lattice. In particular, we can use the intersection pairing to compute the Cartan matrix
\begin{align}
 \label{eq:CartanMatrix}
  A_{ij}:=2\frac{(\alpha_i,\alpha_j)}{(\alpha_i,\alpha_i)}\,,\qquad i,j=1,\ldots, rk(G)\,,
\end{align}
where the $\alpha_i\in R$ are those junctions that form simple roots
of an underlying ADE algebra.  We will label the sets $R$ with
subscripts $f$, $g$, $h$ to denote the relevant objects in theories
$f$, $g$, $h$, and in particular $R_f$, $R_g$, $R_h$ define the flavor
algebras $G_f$, $G_g$, $G_h$.  Non-trivial flavor representations and
BPS states of $G_f$ ($G_g$) can be constructed
\cite{DeWolfe:1998bi,Halverson:2016vwx} from string junctions\footnote{$X_{f,g}$ and $E_{f,g}$ are particular elliptic surfaces and elliptic fibers in those elliptic surfaces; they will be defined in Sections \ref{sec:GTube} and \ref{sec:FTube}.} $J\in H_2(X_f,E_f)$
($J\in H_2(X_g,E_g)$)  with $a(J)\ne 0$, i.e.\ they are charged under the $U(1)$ of the D3-brane. 

\begin{figure}[t]
 \centering
 \includegraphics[width=.55\textwidth]{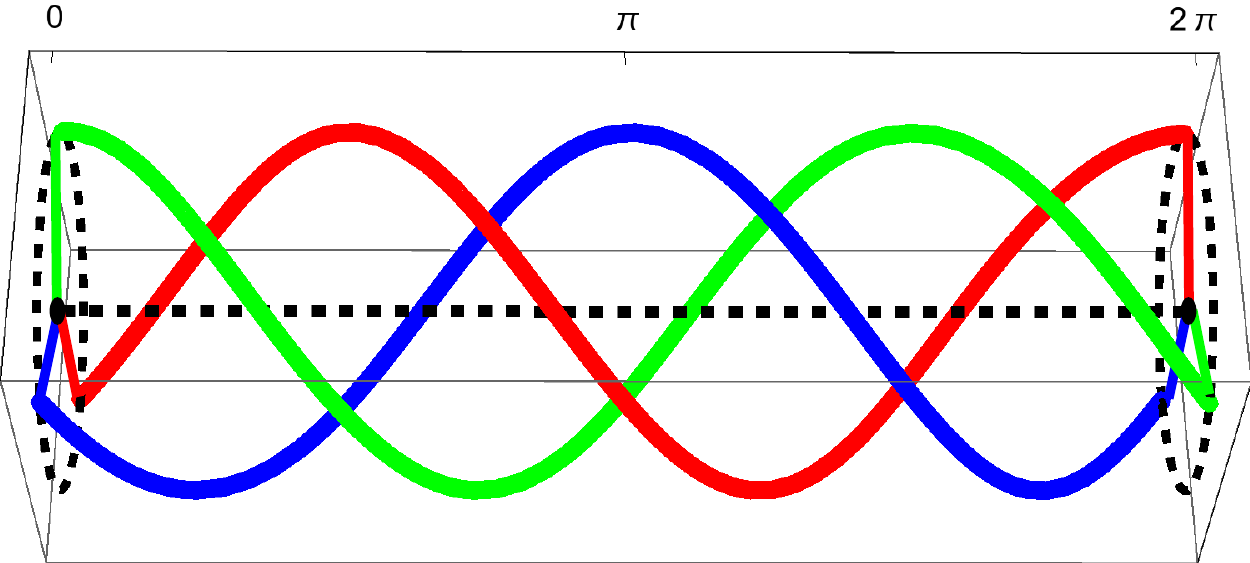}\qquad\qquad
 \includegraphics[width=.35\textwidth]{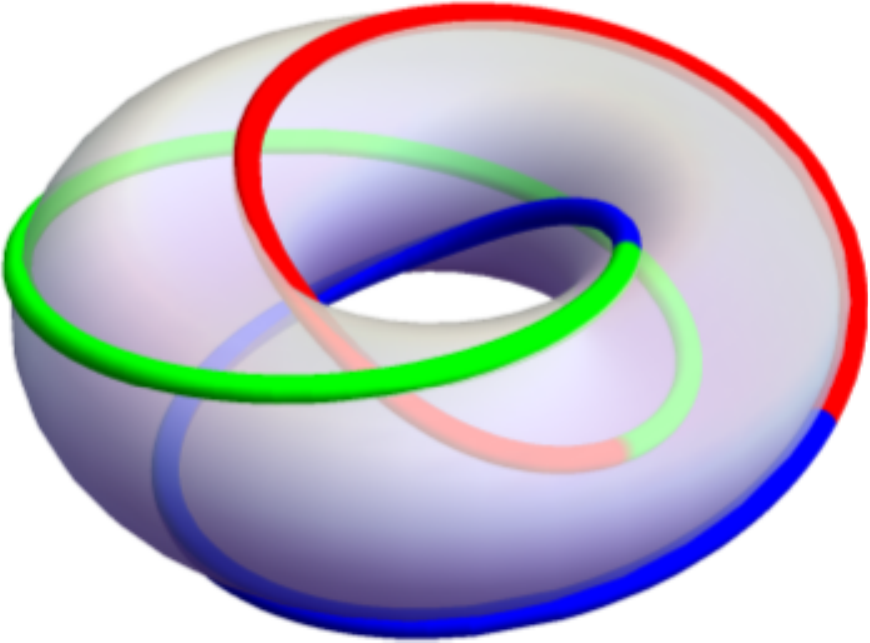}
 \caption{We show the (trefoil) torus knot with \ab11 along the $f$-tube. The braid cyclically permutes the three points (green$\;\rightarrow\;$blue$\;\rightarrow\;$red$\;\rightarrow\;$green) corresponding to the vanishing cycles. Hence they have to be identified, since the torus knot is actually closed, i.e.\ the circles at 0 and $2\pi$ on the left hand side are identified, cf.\ the right hand side. This identification gives rise to a monodromy action.}
 \label{fig:equivalentJunctions}
 \end{figure}

What changes geometrically for the D3-brane in this paper is that the
lower-dimensional F-theory background that it probes has seven-branes
extending in multiple directions. The seven-brane wraps the divisor defined by
\begin{align} 
\label{LINKT}
 -4z^{3a} + 27t^{2b}=0\,,
 \end{align} 
and locally cuts out a knot or a link on a three-sphere near the singularity $t=z=0$.
String junctions with one end on the D3-brane then have
their other end on the link, and as the D3-brane traverses the link and
eventually comes back to its initial position
there is an associated monodromy action on the string junction
states. 
The knot, or link, associated to equation \eqref{LINKT} has  two  canonical braids representations, the $a$-braid  with $3a$ strands and and the $b$-braid with $2b$ strands.
These braids define two solid tubes, which we call respectively the $f$-tube and the $g$-tube.  A transverse section of the $g$-tube, for example, is a disc, which we call $D_t (\theta_z)$,  parameterized by the angle $\theta_z$  and centered at $z=0$. A transverse section gives a natural string junction interpretation of the singularity of $X_g$, we then study the associated
action on states. 
 Mathematically, these are monodromies
\begin{align}
\label{eq:FAndGMonodromies}
M_f: H_2(X_f,E_f)\to H_2(X_f,E_f) \qquad \qquad M_g: H_2(X_g, E_g)\to H_2(X_g,E_g)\,,
\end{align}
obtained from studying two one-parameter families of elliptic
fibrations, and we will compute them explicitly. See Figure~\ref{fig:equivalentJunctions} for a pictorial representation of the $f$-tube with \ab11, its relation to the $a$-braid with $3a=3$ strands, and the monodromy induced by identifying the various strands of the braid upon traversing the torus.

Though the
flavor symmetries of the $\cN=2$ theories generally differ, as captured
by the fact that generally $H_2(X_f,E_f)\neq H_2(X_g,E_g)$,
one of our main results is that
\begin{align}
\frac{H_2(X_f,E_f)}{M_f}\cong \frac{H_2(X_g,E_g)}{M_g}\,.
\end{align}
That is, the string junctions that are invariant under the
link-monodromy, and thus may exist as massless states on the D3-brane
theory at the isolated singularity, generate the same lattice
regardless of whether one takes the $f$ or $g$ perspective. Specifically, the link-monodromy
associated with the deformations\footnote{For the sake of brevity, we will from now on implicitly talk about these deformed fibers/theories without mentioning it explicitly every time.} reduces the flavor algebras
$G_f$ and $G_g$ to a common algebra $G_h\in \{\emptyset, SU(2), SU(3)\}$.
This leaves us mathematically with two Lie algebras at each point, which share a common reduction.
Interestingly, though $G_h$ is sometimes non-trivial, no
$U(1)$ charged string junctions are monodromy-invariant.

In summary, the theories we study are dual deformations of two
different $\cN=2$ SCFTs and the geometry shows that the deformations
sometimes break the flavor symmetry of the $\cN=2$ theories, but
always break the $U(1)$ gauge symmetry as deduced by the absence of
charged string junctions. This deformation yields an $\cN=1$ SCFT for
the D3-brane at $z=t=0$. One such theory, which is a deformation of
the rank one Argyres-Douglas theory $H_1$, exhibits a charge neutral
$SU(2)$ flavor triplet that is comprised of electron, dyon, and
monopole string junctions, even though none of those charged junctions
survive the monodromy projection themselves.  

We argue that duality
also requires imposing the monodromy projection for theories away from
$z=t=0$, in which case the $\cN=1$ D3-brane theories are related to
deformations of massive $\cN=2$ field theories, or deformation of one
massive $\cN=2$ field theory and one $\cN=2$ SCFT. Then the
geometry implies that the D3-brane theory can exhibit massive charge-neutral monodromy-invariant string junctions in non-trivial flavor representations that
are comprised of electron,
monopole, and dyon string junctions. The presence of this massive state, 
together with the absence of charged states, suggests an
interpretation as  confinement of an electron, monopole, and dyon.

\newsec{Review of Seven-branes and String Junctions}
\label{sec:Review}
There is a rich literature on string junctions, and we review some aspects of them here.

String junctions have been introduced \cite{Aharony:1996xr,Schwarz:1996bh,Bergman:1997yw,Gaberdiel:1997ud} as a generalization of ordinary open strings stretching between D-branes in Type II theories. They occur as non-perturbative objects in these theories and are hence closely related to F-theory, as first pointed out by Sen in \cite{Sen:1996vd,Sen:1996sk}. One introduces $(p,q)$-strings that carry $p$ units of NS-charge and $q$ units of Ramond-charge. In this notation a fundamental Type II string corresponds to a $(1,0)$ string.  Alternatively, in the context of Seiberg-Witten theory \cite{Seiberg:1994rs,Seiberg:1994aj} one can think of them as states carrying $p$ units of electric charge and $q$ units of magnetic charge. Via an $SL(2,\bZ)$ action a $(1,0)$ string can be turned into a $(p,q)$ string \cite{Witten:1995im}. The $(p,q)$ seven-branes are then defined as seven-branes on which $(p,q)$ strings can end. Note that, since D3-branes are $SL(2,\bZ)$ invariant, any $(p,q)$ string can end on them and we need not attach a $(p,q)$ label to them. In the worldsheet description of the D3-branes, the $7$-branes act as flavor branes. String junctions arise if several $(p,q)$ strings join at a common vertex. Since the overall charge needs to be conserved at each vertex, this means that the sum of the incoming $(p_i,q_i)$ charges is zero.

The mathematics of string junctions has been worked out in \cite{Gaberdiel:1997ud,DeWolfe:1998bi,DeWolfe:1998zf} and in \cite{Grassi:2013kha,Grassi:2014sda,Grassi:2014ffa}. We will review the latter description since it makes direct contact with
F-theory geometries, as will be useful for describing the seven-brane backgrounds utilized in this paper. This description can
be related to the former if paths from the base point to seven-branes can be chosen so as to reproduced the $(p,q)$
labels of \cite{DeWolfe:1998bi}, for example.

We describe an elliptically fibered Calabi-Yau $n$-fold $X$ via a
Weierstrass model, i.e.\ we start with the anti-canonical hypersurface
$E$ in $\mathbb{P}_{231}$ with homogeneous coordinates $[x,y,w]$,
\begin{align}
 E:\qquad y^2-(x^3+f x w^4+g w^6)=0\,.
\end{align}
To describe a fibration over some $(n-1)$-dimensional base $B$ with
canonical bundle $K_B$ such that the whole space $X$ is CY, 
$f$ and $g$ are sections of $\mathcal{O}(-4K_B)$
and $\mathcal{O}(-6K_B)$, respectively. Models of this type always
have a holomorphic section, the so-called zero section, at
$[x,y,w]=[1,1,0]$. The elliptic fiber becomes singular if $E=dE=0$,
which means that the zero section is non-singular. We thus set $w=1$
from now on when we wish to study the singularities. In this case $E$
becomes singular if the discriminant $\Delta=4f^3+27g^2$ vanishes,
i.e.\  on the locus $\{\Delta = 0\} \subset B$. 

\begin{figure}
\centering
\includegraphics[width=.6\textwidth]{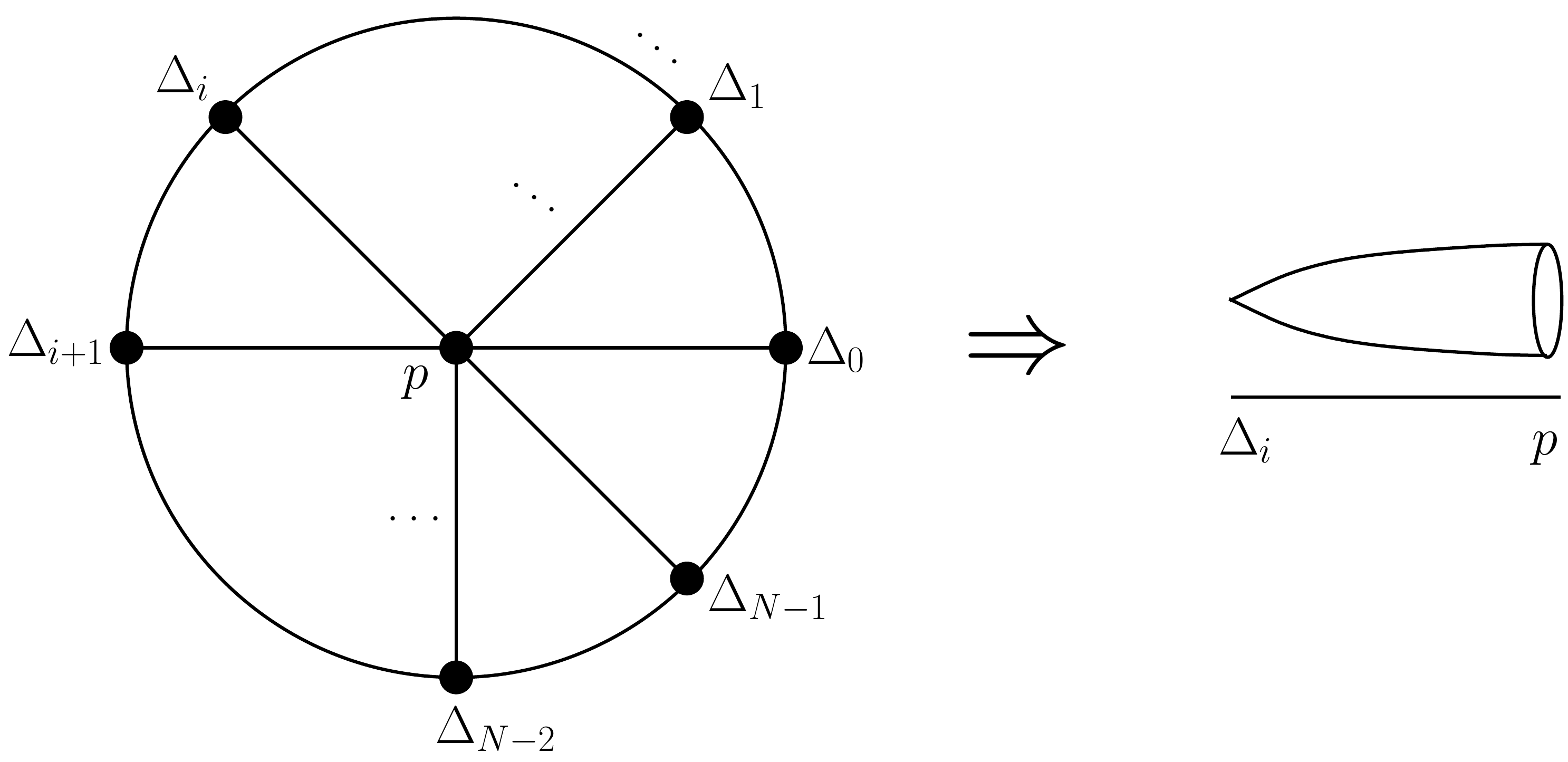}
\caption{We choose a junction basis as straight line paths from $p$ to the vanishing loci $\Delta_i$. As illustrated on the right hand side, a cycle vanishes as one starts from $p$ and approaches $\Delta_i$.}
\label{fig:JunctionBasis}
\end{figure}

For $p\in \Delta$ such that $\pi^{-1}(p)$ is a Kodaira type $I_1$
fiber (as will be the case when string junctions are utilized), the
singular fiber is an elliptic fiber where a one-cycle has vanished. In
this way a vanishing cycle is associated with a zero of $\Delta$. In
more detail, this association works as follows: Consider a
discriminant with $r$ vanishing loci $\Delta_i$ with Kodaira $I_1$
fibers. We fix a base point $p$ of $B\backslash\Delta$ and a basis of
the first homology of the fiber above $p$ and choose a path $\gamma_i$ starting
at $p$ and ending at $\Delta_i$. Upon reaching $\Delta_i$, a cycle
$\pi_i\in H_1(E)$
vanishes, and if a basis on $H_1(E)$ is chosen this can be written as
$\pi_i=(p_i,q_i)^T$. This corresponds to a $(p_i,q_i)$ 7-brane along $\Delta_i=0$
in F-Theory, cf.\ Figure~\ref{fig:JunctionBasis}.

The inverse image of the path $\gamma_i$, $\Gamma_i:=
\pi^{-1}(\gamma)$, is a Lefschetz thimble, which looks like the
surface of a cigar and is commonly referred to as a ``prong'' in the
junction literature.  This prong is a $(p_i,q_i)$ string and has
boundary $\partial\Gamma_i=\pi_i$, which is called the asymptotic
charge in the literature.  A multi-pronged string junction is then
\begin{align}
 J=\sum J_i \Gamma_i,
\end{align}
with $J_i\in \bZ$.  The absolute value $|J_i|$ corresponds to the
number of prongs ending on the $(p_i,q_i)$ 7-brane and the sign
specifies their orientation. The asymptotic charge $a(J)$ of
a general junction $J$ is given as $a(J)=\partial(J)=\sum J_i
\, \partial\Gamma_i$. Note that string junctions with asymptotic
charge $a(J)=(0,0)$ are two-spheres and can be thought of as $7-7$
strings (perhaps passing through a D3-brane at $p$), whereas junctions
with $a(J)\neq 0$ are $3-7$ strings, or perhaps a part of a larger
junction, the remainder of which ends on a seven-brane.

The picture is simplified for elliptic surfaces, which can be useful
for higher dimensional fibrations since, when restricted to a patch, a
local model for an elliptic fibration can be thought of as a family of
elliptic surfaces.  Consider the case of a disc $D$, a neighborhood in
one of the bases of those elliptic surfaces that is centered at $p$,
where $\Delta_i$ intersects $D$ at a point $q_i$ and the paths
$\gamma_i$ are chosen to be straight lines from $p$ to $q_i$, which gives rise to an ordered set of
vanishing cycles~\cite{Grassi:2014ffa}. Let $X\xrightarrow{\pi} D$ be the elliptic surface
and $E_p:=\pi^{-1}(p)$. Then the $\Gamma_i$ form a basis, the
``junction basis'' on the relative homology $H_2(X,E_p)$. There is a
pairing $\langle \cdot, \cdot \rangle$ on $H_2(X,E_p)$ that becomes
the intersection pairing $(\cdot,\cdot)$ on closed classes in
$H_2(X,E_p)$, i.e.\ those elements of $H_2(X,E_p)$ that are also in
$H_2(X,\bZ)$.  In certain cases, such as the $q_i$ being obtained from
the deformation of a Kodaira singular fiber \cite{Kodaira:1963} with
associated ADE group $g$, there is a distinguished set of interesting
junctions
\begin{equation}
R := \{J \in H_2(X,E_p)\,\, | \,\, a(J)=0 \,\,, (J,J)=-2 \}
\end{equation}
that furnish the non-zero weights of the adjoint representation of $g$
from the collection of seven-branes.  This is the gauge symmetry on
the seven-brane of the singular (undeformed limit) in which the $q_i$
collide, or alternatively the flavor symmetry on the D3-brane probing
the seven-brane.

\newsec{D3-branes near Seven-brane Self-intersections: Traversing Links}
\label{sec:LinkTheory}

In this paper we are interested in D3-brane theories located at
certain isolated singularities in non-trivial seven-brane
backgrounds; the isolated singularity is located at the self-intersections of the
seven-brane. To study these theories we will first consider D3-brane
theories \emph{near} these singularities, and the effect on the
spectrum of moving them around loops in the geometry. We will study the implications
for the D3 brane at the singularity in Section~\ref{sec:D3Theories}.

\subsection{The Seven-brane Background, Links, and Braids}
\label{sec:Links}
We will take the F-theory description of the seven-brane background, utilizing a Weierstrass model
as discussed. If the base $B$ of the Weierstrass model is comprised of multiple patches, then the associated
\emph{global} Weierstrass model across the entirety of $B$ may be restricted to a patch, giving a \emph{local}
Weierstrass model, which suffices here since the D3-brane sits at a point in the elliptic fibration and is affected
only by local geometry.

We study a D3-brane in a particular collection of self-intersecting seven-brane backgrounds defined by 
the local Weierstrass model
\begin{align}
\label{eq:WeierstrassModel}
y^2=x^3-z^a x + t^b, \qquad \qquad \Delta = -4z^{3a}+27t^{2b}\,,
\end{align}
and the integers $a\in\{1,2,3\}$, $b\in\{1,2,3,4,5\}$.
The seven-branes are localized on $\Delta = 0$ and the D3-brane will move around near the origin $(z,t)=(0,0)\in \bC^2$,
where the seven-brane self-intersects (technically, where it is singular in the base).
In Section~\ref{sec:D3Theories} we will study the D3-brane theory at $z=t=0$.

We wish to study the local structure of this codimension two
singularity by surrounding it with a three-sphere and moving the
D3-brane around on the three-sphere. The knot, or link, associated to equation (\ref{LINKT})
has two canonical braid representations, the $a$-braid with $3a$ strands
and the $b$-braid with $2b$ strands. These braids define two solid
tubes, which we will call the $f$-tube or $g$-tube. Writing $z=r_ze^{i\theta_z}$ and
$t=r_te^{i\theta_t}$, the three-sphere of radius $R$ is
$|z|^2+|t|^2=r_z^2+r_t^2=R^2$ and the discriminant locus is
$4r_z^{3a}e^{3ia\theta_z}=27r_t^{2b}e^{2ib\theta_t}$. On the
discriminant $3a\theta_z = 2b\theta_t$ modulo $2\pi$. Intersecting the
discriminant locus with the three-sphere gives a link $L_\Delta$
\begin{align}
L_\Delta := \{S^3\}\cap\{\Delta = 0\}\,,
\end{align}
which is a $(3a,2b)$ torus link (torus knot if $3a$ and $2b$ are coprime); that is the seven-branes intersect
the three-sphere at a torus link.
It can be described by either of the equations
\begin{align}
4(R^2-r_t^2)^{3a/2}e^{3ia\theta_z}=27r_t^{2b}e^{2ib\theta_t}, \qquad \qquad 4r_z^{3a}e^{3ia\theta_z}=27(R^2-r_z^2)^{b}e^{2ib\theta_t}\,.
\end{align}
Consider a one-parameter family of discs, $D_{t}(\theta_z)$, centered at $t=0$ with parameter $\theta_z$.  The first equation intersects each member of the family at a collection of points, and as $\theta_z$ is varied in the positive direction from $0$ to $2\pi$ the intersection points encircle the origin, creating a spiral that could be thought of as sitting on a tube. For a pictorial representation see e.g.\ Figure~\ref{fig:IITorusLinkTube}. 
Call this the $g$-tube. Alternatively, there is also a one-parameter family of discs $D_z(\theta_t)$; the second equation intersects a member of this family at some points, and the whole family at a spiral that sits on the $f$-tube, see e.g.\ Figure~\ref{fig:IIITorusLinkTube}.

Formally, associated to the $g$-tube and $f$-tube, respectively, are periodic one real parameter families of elliptic surfaces
\begin{equation}
X_{\theta_z} \xrightarrow{\pi_{\theta_z}} D_t(\theta_z), \qquad \qquad \qquad X_{\theta_t}\xrightarrow{\pi_{\theta_t}} D_z(\theta_t).
\end{equation}
We will be interested in studying the string junctions in the members of these families $X_g:=X_{\theta_z=0}$ and $X_f:=X_{\theta_t=0}$,
and also the monodromy action on string junctions associated with taking a loop in the family. For consistency of notation, we will
will also define $\pi_g:=\pi_{\theta_z=0}$ and $\pi_f:=\pi_{\theta_t=0}$.

\subsection{General analysis of the \texorpdfstring{$g$}{g}-tube}
\label{sec:GTube}

We now study string junctions emanating from the seven-brane link and
ending on a D3-brane sitting on the three-sphere, as well as
the seven-brane action on the D3-spectrum associated with traversing
the $g$-tube.  We must specify the initial location of the
D3-brane. We choose this point $p$ to be $r_z^2=R^2$, $\theta_z=0$,
$t=0$, which sits on the three-sphere and at the origin of the disc in
the $g$-tube at $\theta_z=0$. Mathematically, the selection of this
point selects a distinguished fiber in the elliptic fibration
$E_{g}=\pi_{g}^{-1}(p)$ from which to build the relative homology
associated with string junctions. We will be more precise about
this definition in a moment.

We must study $E_g$, define a basis of cycles there, and determine the action on this basis of cycles as $\theta_z$ varies from $0$ to $2\pi$, i.e.\ as the D3-brane travels down the $g$-tube. At $t=0$, the Weierstrass model simplifies to
\begin{align}
y^2 = x(x+z^{\frac{a}{2}})(x-z^{\frac{a}{2}})=x(x+R^{\frac{a}{2}}e^{i\theta_z \frac{a}{2}})(x-R^{\frac{a}{2}}e^{i\theta_z \frac{a}{2}})
\end{align}
and defines a one-parameter family of elliptic curves depending on $\theta_z$. At $p$, where $\theta_z=0$, $y^2=x(x+R^{\frac{a}{2}})(x-R^{\frac{a}{2}})$ and the elliptic curve is a double cover of the $x$-plane with four branch points a $0$, $\pm R^{\frac{a}{2}}$, and $\infty$. We will study the first three points, which sit on the real axis. If $a=1$, note that as $\theta_z$ passes from $0$ to $2\pi$ the points at $\pm \sqrt{R}$ swap via a counterclockwise rotation. In general we find that the root at
$R^{\frac{a}{2}}$ becomes the root at $(-1)^aR^{\frac{a}{2}}$. This determines some monodromy $M_g\in SL(2,\bZ)$ that can be computed explicitly, and to do so it is convenient to choose a basis of one-cycles. 

\begin{figure}[t]
  \centering
  \subfloat[t][Cycle choice in basis one.]{
    \includegraphics[width=0.47\textwidth]{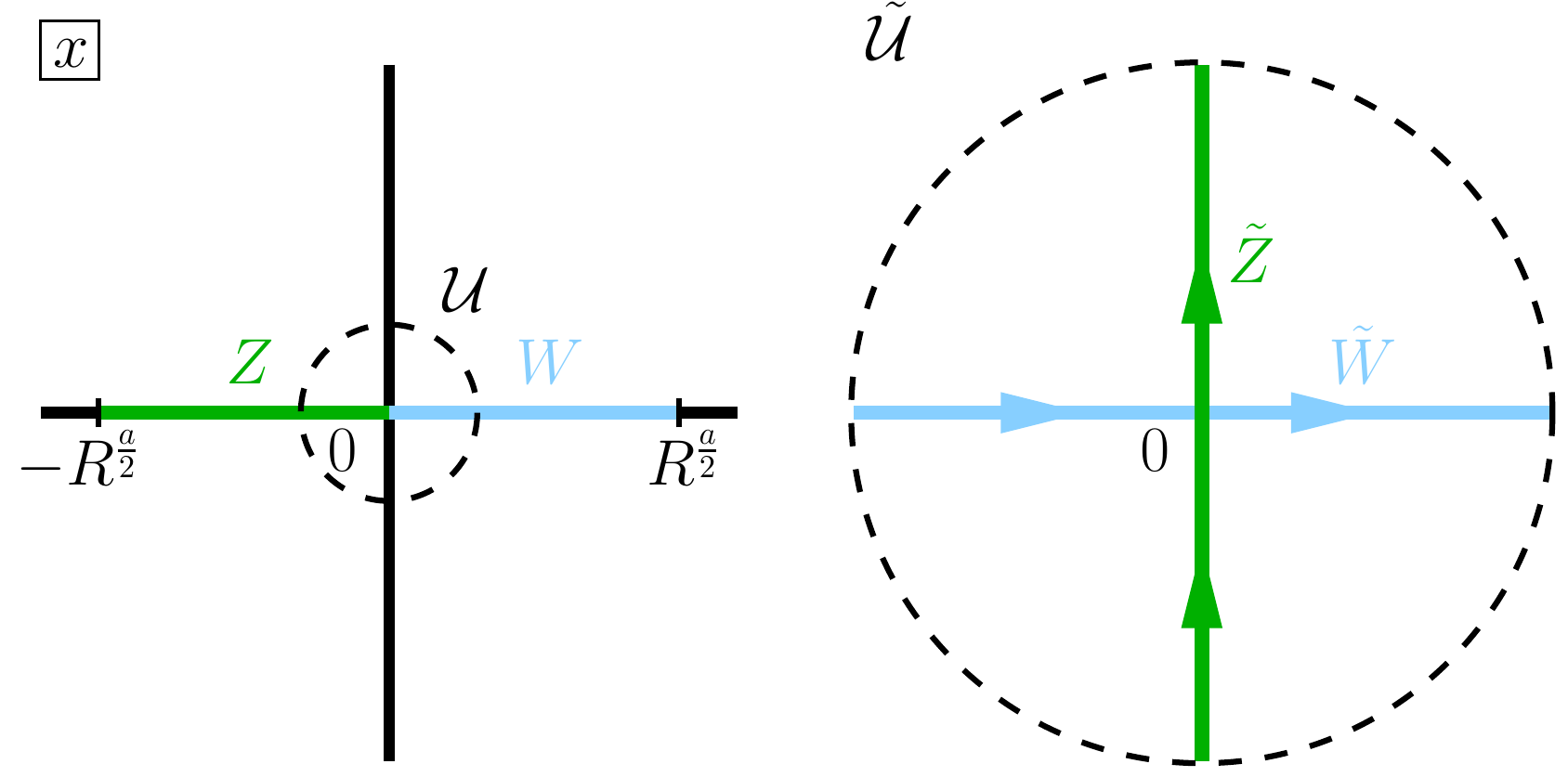}
  }
  ~ 
  \subfloat[t][Cycle choice in basis two.]{
    \includegraphics[width=0.47\textwidth]{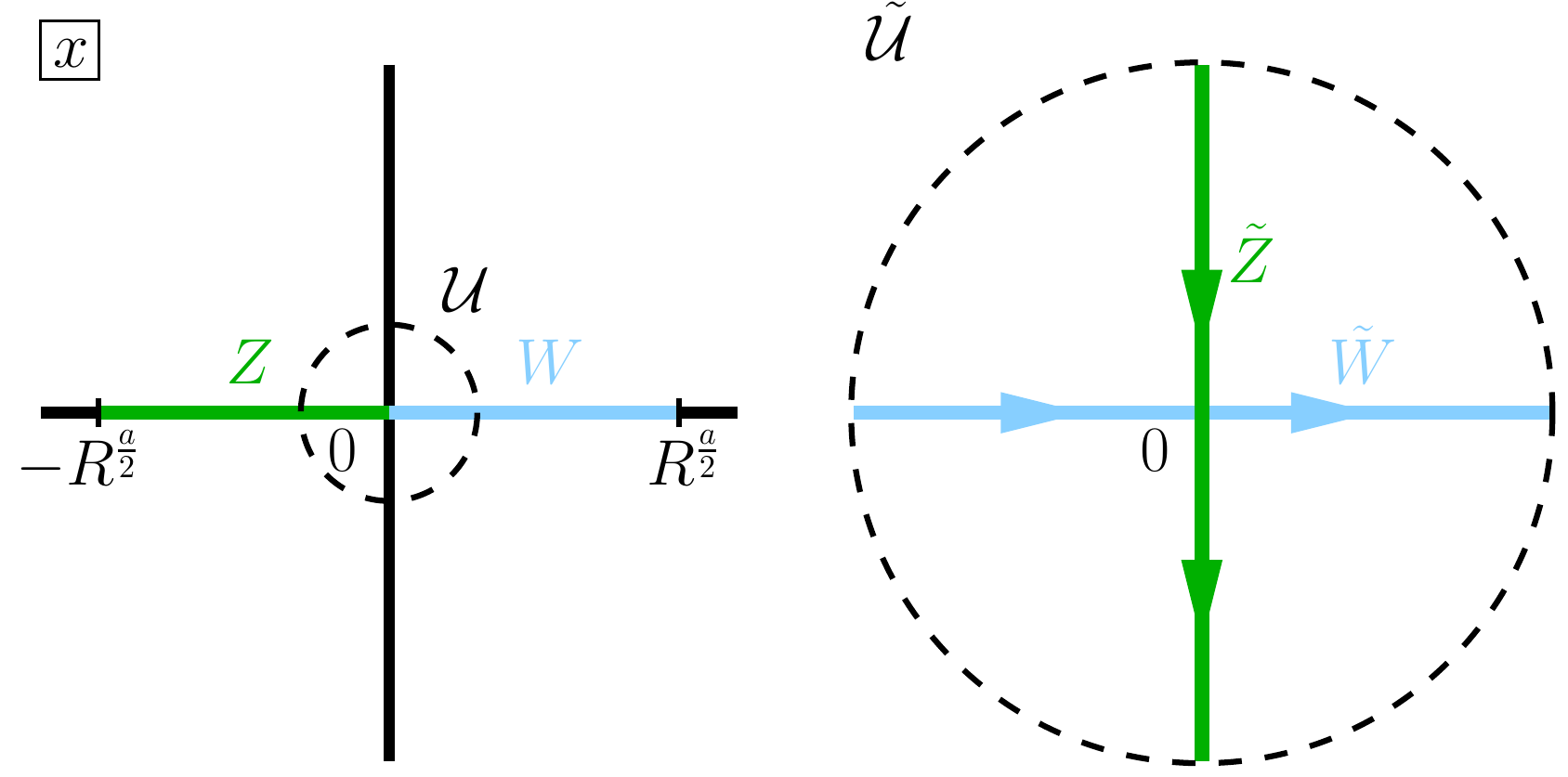}
  }
  \caption{In this Figure we illustrate two different choices of bases for the cycles in the $x$- and $y$-plane for the $g$-tube.}
  \label{fig:GTubeBasisChoice}
\end{figure}

We will consider two different bases and ensure that they give the same theory of string junctions. The elliptic curve is a double cover of the $x$-plane with branch points at $0$, $-R^{\frac{a}{2}}$ and $R^\frac{a}{2}$. Let $Z$ ($W$) be a straight line connecting $0$ to $-R^{\frac{a}{2}}$ $(+R^{\frac{a}{2}})$ on the $x$-plane. Some details of the analysis can be easily understood in a small neighborhood $\mathcal{U}$ of $x=0$ and its inverse image $\tilde{\mathcal{U}}$ in the double cover. Let $\tilde u$ be a local coordinate on $\tilde{\mathcal{U}}$ and $u=\tilde u^2$ a local coordinate on $\mathcal{U}$, and $\tilde Z$ and $\tilde W$ the inverse images of $Z$ and $W$ in the double cover. $Z|_{\mathcal{U}}$ sits along the negative $\re{u}$-axis and $W|_{\mathcal{U}}$ along the positive $\re{u}$ axis; therefore $\tilde Z|_{\tilde{\mathcal{U}}}$ sits along the entire $\im{\tilde u}$-axis and $\tilde W|_{\tilde{\mathcal{U}}}$ sits along the entire $\re{\tilde u}$-axis. $\tilde Z$ $(\tilde W)$ is an $S^1$, and there is an orientation according to whether the circle is traversed coming out from $\tilde u =0$ along the positive or negative $\im{\tilde u}$ $(\re{\tilde u})$ axis. Note that this distinction is lost in $\mathcal{U}$, since coming out from $\tilde u=0$ along the positive or negative axis corresponds to the same path of exit from $u=0$ due to the double cover.

The two bases of $H_1(E_{g},\bZ)$ that we study are defined as follows. In basis one, abusing notation,  define $Z$ to be $\tilde Z$ with the orientation associated with departing $\tilde u=0$ along the positive $\im{\tilde u}$-axis, and $W$ to be $\tilde W$ with the orientation associated with departing $\tilde u=0$ along the positive $\re{\tilde u}$-axis. In basis two, $W$ is the same as in basis one, but $Z$ is instead defined to be $\tilde Z$ with the opposite orientation, i.e.\ departing $\tilde u=0$ along the negative $\im{\tilde u}$-axis. In the usual complex structure on $\tilde{\mathcal{U}}$, defined by the phase of $\tilde u$ going counter-clockwise rather than clockwise\footnote{Equivalently, $i=\sqrt{-1}$ versus $i=-\sqrt{-1}$.}, $\{W,Z\}$ $(\{Z,W\})$ determine a positive basis on the tangent space $T_{0}(\tilde{\mathcal{U}})$ in basis one (two), and therefore $W\cdot Z=1$ $(Z\cdot W=1)$ in basis one (two). The intersection product of arbitrary one-cycles in the usual complex structure is $(p,q)\cdot (r,s)=ps-rq$, where $(p,q)$ and $(r,s)$ are one-cycles in some basis. To use this intersection product, we can choose $W=(1,0)^T$ and $Z=(0,1)^T$ in basis one and $W=(0,1)^T$ and $Z=(1,0)^T$ in basis two. Using the notation of \cite{Grassi:2014sda}, we define
\begin{align}
 \pi_1=\begin{pmatrix}1\\0\end{pmatrix},~~\pi_2=\begin{pmatrix}-1\\-1\end{pmatrix},~~\pi_3=\begin{pmatrix}0\\1\end{pmatrix},\quad \pi_1+\pi_2+\pi_3=0.
 \label{eq:defPis}
\end{align}
with the usual intersection product, and we will map onto this language later. The cycle $\pi_2$ will be used in the next section.

Traveling down the $g$-tube via $\theta_z$ passing from $0$ to $2\pi$ rotates counter-clockwise in $x$ by $a \pi$ and counter-clockwise in $y$ by $a \pi/2$. The latter gives an action on the bases
\begin{align*}
\text{Basis One:}\qquad &(W,Z)\mapsto (Z,-W)~\text{for}~a=1,\quad (W,Z)\mapsto (-W,-Z)~\text{for}~a=2,\\
			&(W,Z)\mapsto (-Z,W)~\text{for}~a=3,\\
\text{Basis Two:}\qquad &(W,Z)\mapsto (-Z,W)~\text{for}~a=1,\quad (W,Z)\mapsto (-W,-Z)~\text{for}~a=2,\\
			&(W,Z)\mapsto (Z,-W)~\text{for}~a=3.  
\end{align*}
All of these can be seen by direct inspection of Figure~\ref{fig:GTubeBasisChoice}. The associated monodromy matrices are $M_1^a$ ($M_2^a$) with
\begin{align}
 M_1=\begin{pmatrix}0&-1\\1&0\end{pmatrix},\qquad M_2=\begin{pmatrix}0&1\\-1&0\end{pmatrix}\,.
\end{align}

Having determined the bases on $H_1(E_{g},\bZ)$, let us determine the vanishing cycles. We do this on the three-sphere at $\theta_z=0$, where we read off the vanishing cycles by following straight line paths from the D3-brane at $p$ to the seven-branes. The seven-branes intersect the three-sphere at a link, and at $\theta_z=0$ this determines a set of points in a disc centered at
$t=0$ that are the solutions to the equation
\begin{align}
4(R^2-r_t^2)^{\frac{3a}{2}}=27r_t^{2b}e^{2ib\theta_t}\,,
\label{eq:g-tubevcdisc}
\end{align}
which requires
\begin{align}
\theta_t = \frac{\pi k}{b}, \qquad k\in \{0,1,\dots,2b-1\}\,.
\end{align}
The $r_t$-dependent part of \eqref{eq:g-tubevcdisc} is satisfied for some $r_t^*\in \bR_+$, and therefore the seven-branes intersect the disc $D_t(0)$ at the $2b$ points $p_k=r_t^*e^{i \frac{\pi k}{b}}$, and each of the vanishing cycles may be read off by following a straight line path from $p$ to $p_k$. Let us determine the vanishing cycles explicitly using a simple analysis from calculus. The Weierstrass model over $D_t(0)$ is 
\begin{align}
y^2 = x^3-(R^2-r_t^2)^{\frac{a}{2}} x + r_t^{b}e^{ib\theta_t}
\end{align}
which on any straight line path from $t=0$ to $p_k$ simplifies to 
\begin{align}
y^2 = x^3-(R^2-r_t^2)^{\frac{a}{2}} x + r_t^{b}e^{i\pi k}=:v_k(x)\,.
\end{align}
At $r_t=0$ (that is, at $p$), the cubic $v_k(x)$ has three real roots, and it is positive for real $x\in [-R,0]$ and negative for real $x\in [0,R]$. Letting $r_t$ vary from $0$ to $r_t^*$, all of the roots remain real, but two of them collide at $r_t=r_t^*$. To determine which two roots collide, note that 
\begin{align}
\frac{\partial v_k}{\partial r_t} = a r_t(R^2-r_t^2)^{\frac{a}{2}-1}x+(-1)^k b\,  r_t^{b-1}
\end{align}
so that $\partial v_k/\partial r_t|_{x=0}$ is positive for $k$ even and negative for $k$ odd. Then the center and right root collapse for $k$ even as $r_t$ goes from $0$ to $r_t^*$, and the center and left root collapse for $k$ odd. That is, if $k$ is even (odd) the vanishing cycle is $W$ ($Z$). Since we choose to index our seven-branes starting from $k=0$, the ordered set of vanishing cycles is
\begin{align}
\{W,Z,W,Z, \ldots\}
\end{align}
where the $W$, $Z$ pair repeats $b$ times, for a total of $2b$ vanishing cycles. Note that this set applies to both bases discussed above since vanishing cycles do not have a sign, but the basis choice must carefully be taken into account when studying monodromy associated with taking closed paths in the
geometry (as we will see).

Finally, before studying examples we briefly discuss the map on seven-branes that is induced by the braid upon traveling down the $g$-tube. At $\theta_z=0$ the $k^{\text{th}}$ seven-brane is at an angle in the $t$-plane given by $\theta_t = \pi k / b$, due to points on the discriminant satisfying $2b\theta_t = 2\pi k$. Upon traveling down the $g$-tube, $\theta_z$ varies from $0$ to $2 \pi$ and the $z$ term in the discriminant picks up a phase $e^{6\pi i a}$ so the associated phase condition on discriminant points becomes
\begin{align}
\label{eq:GTube7BraneMap}
2b\theta_t = 2\pi k + 6\pi a = 2\pi \tilde k\,,
\end{align}
where $\tilde k = (k+3 a)$ mod $2b$. So a seven-brane that starts with index $k$ spirals down the $g$-tube and becomes the seven-brane with index $k+3a$ modulo $2b$. This seven-brane mapping, together with undoing the $SL(2,\bZ)$ action on $E_g$ associated with traveling down the $g$-tube will induce a map on string junctions, allowing for the comparison of closed cycles representing simple roots and the determination of whether or not traveling down the $g$-tube gives an outer automorphism on string junctions.

\subsection{General analysis of the \texorpdfstring{$f$}{f}-tube}
\label{sec:FTube}

In the discussion of the $f$-tube we can proceed similarly to the analysis of the previous Section. This time we choose the point $p$ of the D3-brane to be at $r_t=R^2$, $\theta_t=0$, $z=0$ and study the string junctions with respect to the elliptic fiber $E_{f}=\pi_{f}^{-1}(p)$. The corresponding one-parameter family of Weierstrass models at $z=0$ read
\begin{align}
 y^2=x^3+t^b=\left(x+R^{\frac{b}{3}}e^{\frac{2\pi i}{3} \frac{b \theta_t}{2\pi}}\right)\left(x+R^{\frac{b}{3}}e^{\frac{2\pi i}{3}(1+\frac{b\theta_t}{2\pi})}\right)\left(x+R^{\frac{b}{3}}e^{\frac{2\pi i}{3}(2+ \frac{b\theta_t}{2\pi})}\right)\,.
\end{align}

\begin{figure}[t]
  \centering
 \includegraphics[width=0.48\textwidth]{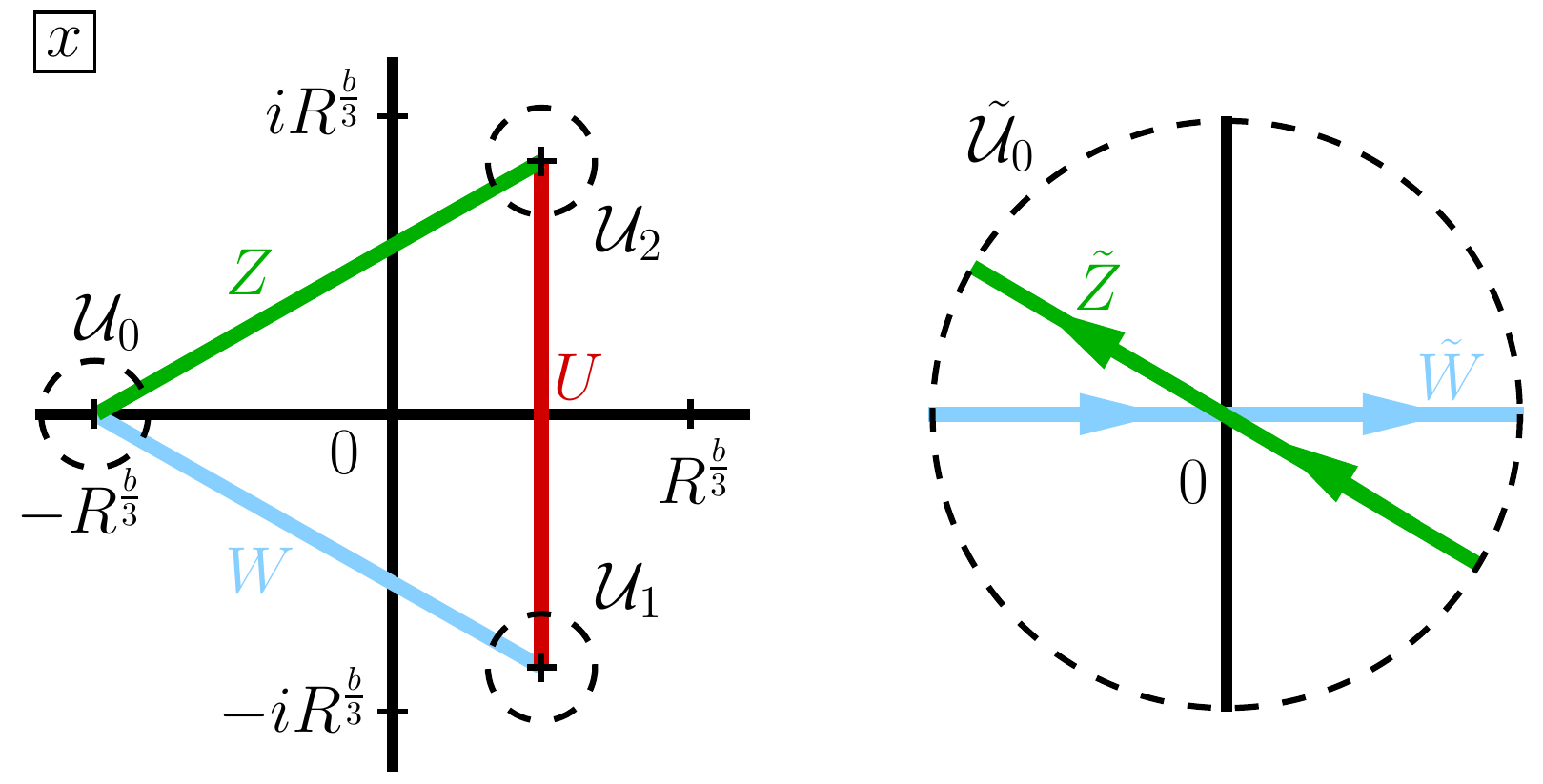}
  \caption{In this Figure we illustrate our basis choice for the cycles in the $x$- and $y$-plane for the $f$-tube. $\tilde{\mathcal{U}}_0$ is the inverse image in the double cover of $\mathcal{U}_0$ in the $x$-plane.}
  \label{fig:FTubeBasisChoice}
\end{figure}

At $p$, we have $\theta_t=0$ and the curve is a double cover of the $x$-plane, this time with the four branch points; three at $-R^{\frac{b}{3}}e^{\frac{2\pi i k }{3}}$, $k=0,1,2$, and one at $\infty$. We focus on the first three. Upon traveling from $\theta_t=0$ to $\theta_t=2\pi$, we find that these points are permuted according to $k\rightarrow (k+b)$ modulo three. Consequently, it is convenient to phrase the following discussion in terms of segments $W, U, Z$ that connected the three branch points. Let $W$ connect the $k=0,1$ branch points, $U$ connect the $k=1,2$ branch points, and $Z$ connect the $k=2,0$ branch points. Let $\mathcal{U}_k$ be local neighborhoods of the branch point $k=0,1,2$, and $\tilde{\mathcal{U}}_k$ be the inverse image of these neighborhoods in the double cover. 

Let us look at the neighborhood $\mathcal{U}_0$ and its double cover $\tilde{\mathcal{U}}_0$ in more detail. We choose local coordinates such that $W$ is oriented along the positive $\re{u_0}$-axis. The angle
of $Z$ is $\frac{2\pi}{6}$ ($\frac{\pi}{6}$) in $\mathcal{U}_0$ ($\tilde{\mathcal{U}}_0$). Note that the cycle $U$ is not visible in this local neighborhood. By a similar analysis as in the $g$-tube, $W\cdot Z =1$ and we take $W=\pi_1$ and  $Z=\pi_3$.

Upon traversing the $f$-tube from $\theta_t=0$ to $\theta_t=2\pi$, we
find a counter-clockwise rotation by $\frac{2\pi b}{3}$ in the
$x$-plane.  When $b=1$, this rotates the $Z$ segment to the $W$
segment on the left hand side of Figure~\ref{fig:FTubeBasisChoice}, and
similarly the $W$ ($U$) segment to the $U$ ($Z$) segment. This mapping
of segments determines the mapping of each associated cycle up to a
sign. Let us determine the signs, writing $Z\mapsto a_w W$, $W\mapsto
a_u U$, $U\mapsto a_z Z$ with $a_w,a_u,a_z=\pm 1$. Encircling the
origin three times via going from $\theta_t=0$ to $\theta_t=6\pi$ we have
$Z\mapsto a_w a_u a_z Z$, $U\mapsto a_z a_w a_u U$, $W\mapsto a_u a_z
a_w W$. This corresponds to a $2\pi$ rotation in $\mathcal{U}_0$, and therefore
a rotation by $\pi$ in $\tilde{\mathcal{U}_0}$, which reverses the orientation
of the cycle, requiring $a_wa_ua_z=-1$. At this point there are two possibilities: all $a_i$ negative,
or one negative. However, preserving the intersection of
the cycles under the mapping, or alternatively symmetry considerations, requires $a_w=a_u=a_z=-1$.
Thus, for $b=1$ one rotation in $\theta_t$ gives $Z\mapsto -W$, $W \mapsto -U$, and $U\mapsto -Z$. For general $b$, the braid acts as
\begin{align*}
(W,U,Z)&\mapsto (-1)^b\;(U,Z,W)~\text{for}~b=1,4,&~~ (W,U,Z)\mapsto (-1)^b\;(Z,W,U)~\text{for}~b=2,5,\\
(W,U,Z)&\mapsto -(W,U,Z)~\text{for}~b=3.
\end{align*}
Recalling $W=\pi_1$, $Z=\pi_3$ from the previous paragraph and also $U+W+Z=0$ the matrix
\begin{align}
  M=\begin{pmatrix}1 & -1 \\ 1 & 0\end{pmatrix}\,,
\end{align}
encodes the monodromy, which is given by $M^b$.

In order to determine the vanishing cycles we proceed similarly to the previous Section. The discriminant of the Weierstrass model at $\theta_t=0$, intersected with the three-sphere, yields
\begin{align}
 4r_z^{3a}e^{3 i a \theta_z}=27(R^2-r_z^2)^b\quad\Rightarrow\quad \theta_z=\frac{2\pi k}{3a},\qquad k\in \{0,1,\dots,3a-1\}\,.
\end{align}
We study again the solutions of the Weierstrass equation along straight line paths from $z=0$ to $p_k=r_z^*e^{\frac{2\pi i k}{3a}}$, which reads
\begin{align}
 y^2=x^3-r_z^{a}e^{\frac{2\pi i k}{3}}x+(R^2-r_z^2)^{\frac{b}{2}}=:v_k(x)\,.
\end{align}
As $r_z$ varies from 0 to $r_z^*$, two roots collide and we determine which ones by studying the imaginary part. Starting from $k=0$, we find the ordered set of vanishing cycles 
\begin{align}
 \{U,W,Z,U,W,Z,\ldots\}
\end{align}
such that the $3a$ vanishing cycles are given by repeating the vanishing cycles $U,W,Z$ a total of $a$ times. Finally, we find that upon traveling down the $f$-tube by varying $\theta_t$ from $0$ to $2\pi$, the $t$ term in the discriminant is rotated by a phase $e^{4\pi i b}$, such that 
\begin{align}
 3a\theta_z=2\pi k+4\pi b=2\pi\tilde{k}, \quad \tilde{k}=(k+2b)~\text{mod}~3a
\end{align}
which means that the braid induces a permutation which sends the seven-brane with index $k$ to the seven-brane with index $(k+2b)$ modulo $3a$.

\subsection{Braid action on intersection form}
\label{sec:IntersectionForm}

In the previous Sections we studied an elliptic fibration over a disc with the inverse image of the origin of the disc being
a smooth elliptic curve. For unified notation in this Section, we take the elliptic fibration to be $X\xrightarrow{\pi} D$
with $\pi^{-1}(0)=E$. String junctions are elements of two-cycles relative $E$, i.e.\ $J\in H_2(X,E)$.

\begin{figure}[t]
  \centering
  \subfloat[t][Tube and second base point at $\theta=0$.]{
    \parbox{0.45\textwidth}{\centering\includegraphics[height=0.35\textwidth]{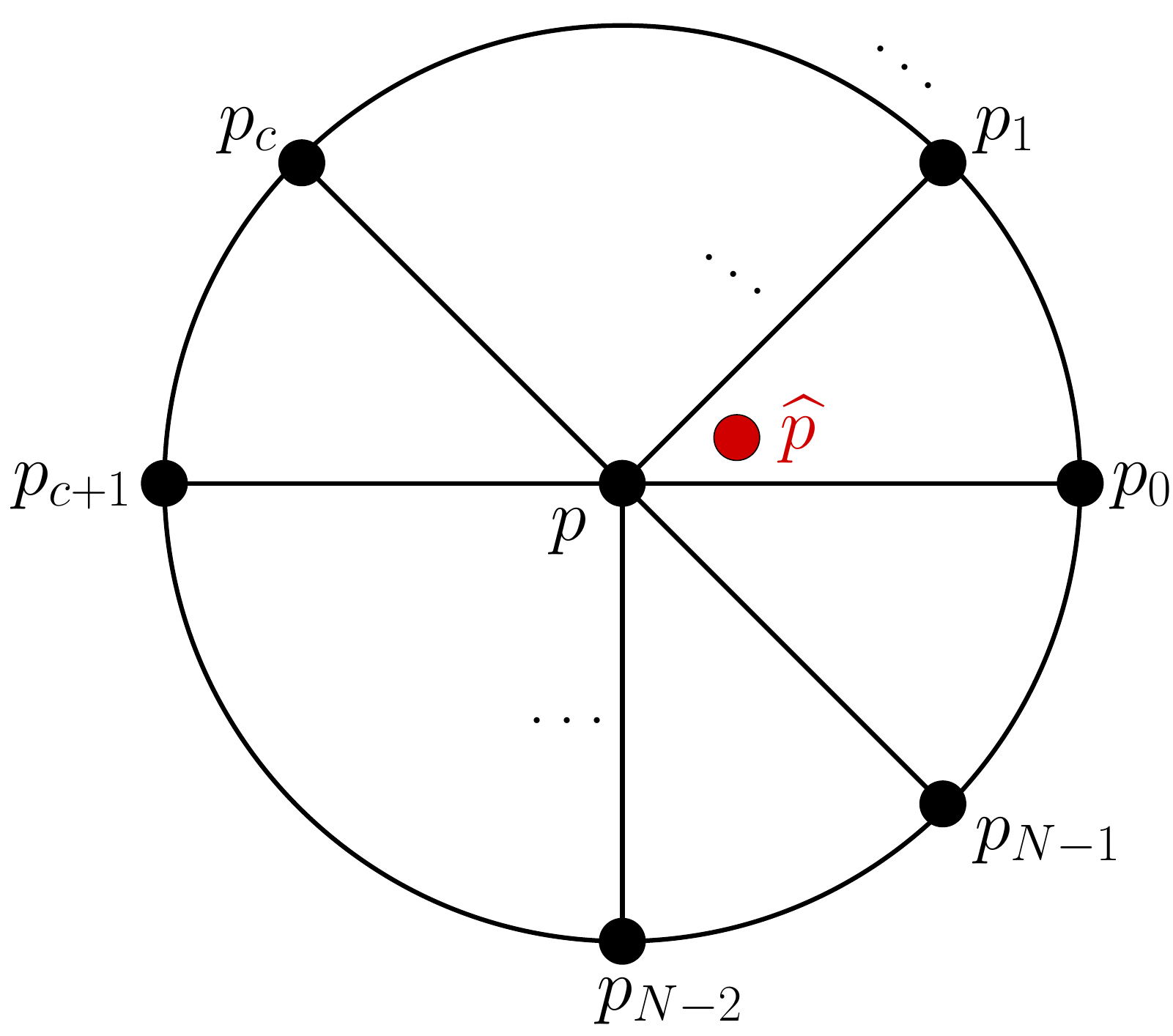}}
  }
  \qquad
  \subfloat[t][Tube and second base point at $\theta=2\pi$.]{
    \parbox{0.45\textwidth}{\centering\includegraphics[height=0.35\textwidth]{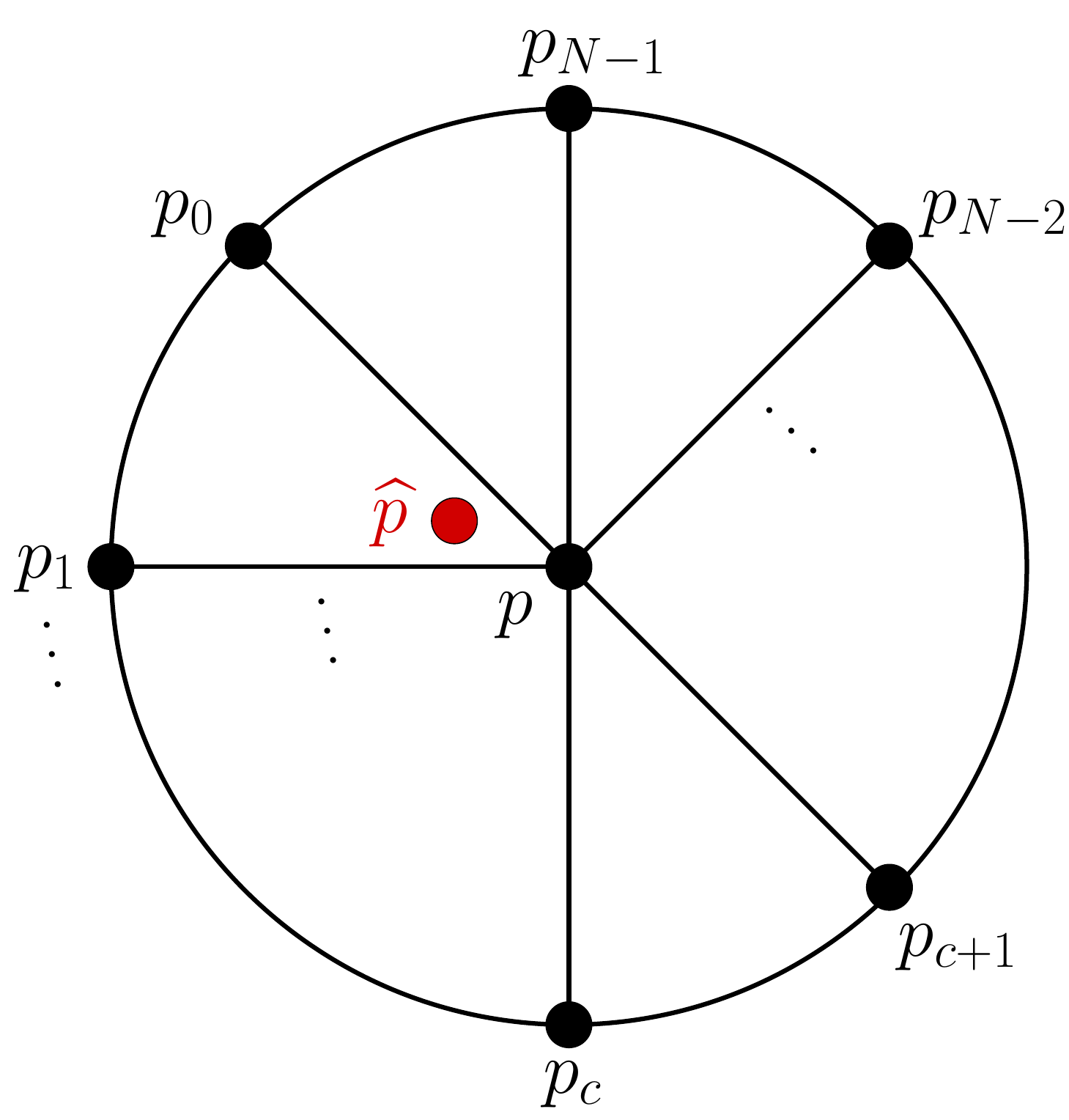}}
  } 
 \caption{This Figure illustrates the motion of the second base point under the permutation induced by the braid. The cone given by $p$, $p_0$ and $p_1$ is moved to the cone originally corresponding to $p$, $p_{c}$ and $p_{c+1}$.}
 \label{fig:ConeMotion}
\end{figure}

Following \cite{Grassi:2014ffa}, we can define a self-intersection
pairing $\langle J,J\rangle$ for a given junction $J=\sum_{i} J_i
\Gamma_i$ written in terms of a basis $\Gamma_i$ (the junction basis)
on $H_2(X,E)$, where the boundary of the junction is $\partial \Gamma_i =: \gamma_i \in H_1(E,\bZ)$.
Then the pairing is 
\begin{align}
\label{eq:IntersectionForm}
 \langle J,J\rangle_0=-\sum_{k>l\geq1}^{N-1}J_k J_l~ \gamma_k\cdot \gamma_l-\sum_{l=0}^{N-1}J_l^2.
\end{align}
Note that the first index $l=0$ is skipped in the first sum. This is the case since the pairing is a pairing on relative classes which depends on choosing a second base point $\widehat{p}$ nearby the first base point $p$. The rays that connect the base point $p$ to the points $p_k$ on the discriminant locus divide the plane into cones. In writing \eqref{eq:IntersectionForm}, we have arbitrarily put this second base point into the $0^\text{th}$ cone between $p_0$ and $p_1$. Since the intersection pairing depends on which cone the second base point $\widehat{p}$ lies in, we indicate the cone used in the pairing on relative homology with a subscript $\langle\cdot,\cdot\rangle_i$. While this choice is irrelevant for the intersection pairing on classes with vanishing asymptotic charge, it becomes relevant for junctions whose asymptotic charge is non-zero, which correspond to matter states. Consequently, if we want to compare these junctions at $\theta=0$ and $\theta=2\pi$, we need to track the motion of the second base point upon traveling down the $g$- or $f$-tube. If the braid induces a permutation $\gamma_i\mapsto \gamma_{i+c}$ where the indices are to be read mod $N-1$, the cone $\widehat{p}$ is in also moves counter-clockwise by $c$, so $\widehat{p}$ ends up in the $(c-1)^\text{th}$ cone, cf.\ Figure~\ref{fig:ConeMotion}. The new intersection form then reads
\begin{align}
\label{eq:IntersectionForm2}
 \langle J,J\rangle_{c-1}=-\sum_{k>l\geq c}^{N-1+c}J_k J_l~ \gamma_k\cdot \gamma_l-\sum_{l=0}^{N-1}J_l^2,
\end{align}
where the indices in the first sum are to be read modulo $N-1$.

\subsection{Monodromy Action on Bases of String Junctions}
\label{sec:monodromyactiononbasis}

Having performed a general analysis of the $g$-tube and $f$-tube, we are ready to state the associated action on relative homology, which will be utilized in examples to perform a map on simple roots, studying associated Lie algebraic structure in the quotient.

Let us begin with the $g$-tube. The results of \cite{Grassi:2014ffa} shows that the thimbles or prongs $\Gamma_i$ with $i=0,\cdots, 2b-1$ form a basis for the relative homology $H_2(X_g,E_g)$, that is, a basis of string junctions. Each thimble has an associated vanishing cycle $\partial \Gamma_i\in H_1(E_g,\bZ)$, and the results of the previous Section show that 
\begin{align*}
\partial \Gamma_i = W \quad \text{for $i$ even},\qquad\qquad
\partial \Gamma_i = Z \quad \text{for $i$ odd}.
\end{align*}
The combined action of the braid map and monodromy on $E_g$ induces the following map in basis~1:
\begin{align}
\label{eqn:braidactiononbasisgtube}
\Gamma_i \mapsto (-1)^{\lfloor \frac{a}{2}\rfloor} \Gamma_{(i+3a)\nicemod 2b} 		~\text{ for $i$ even}, \qquad
\Gamma_i \mapsto (-1)^{\lfloor \frac{a}{2} + \frac12\rfloor} \Gamma_{(i+3a)\nicemod 2b} ~\text{ for $i$ odd},
\end{align}
where $\lfloor\cdot\rfloor$ is the floor function, $\lfloor x\rfloor=n$ for $x\in\bR, n\in\bZ$, where $n$ is the largest integer
satisfying $n\leq x$.

We now turn to the $f$-tube. Now there are prongs $\Gamma_i$ with $i=0,\dots, 3a-1$
that form a basis on relative homology $H_2(X_f,E_f)$.
The results of the previous Section show that
\begin{align}
\label{eqn:braidactiononbasisftube}
\partial \Gamma_i &= U \quad \text{for }i\nicemod 3=0,\qquad\qquad
\partial \Gamma_i = W \quad \text{for }i\nicemod 3=1,\qquad\qquad\nonumber \\
\partial \Gamma_i &= Z \quad \text{for }i\nicemod 3=2.
\end{align}
The braid map and the monodromy on $E_f$ induce a map on the
basis
\begin{align*}
\Gamma_i \mapsto (-1)^{b}\,\Gamma_{(i+2b)\nicemod 3a} \qquad \forall i,
\end{align*}
which is simpler than that of the $g$-tube.

Summarizing, when the D3-brane traverses the $f$-tube or $g$-tube it is taking a small, closed loop
in the geometry near the seven-brane self-intersection at $z=t=0$. The seven-branes spiral around the 
D3-brane as it traverses the tube; since string junctions end on the spiraling seven-branes and the D3-brane,
this induces a monodromy on string junction states. Mathematically, in the $f$-tube and $g$-tube we
have computed the monodromy maps
\begin{align}
M_f: H_2(X_f,E_f)\to H_2(X_f,E_f) \qquad \qquad M_g: H_2(X_g, E_g)\to H_2(X_g,E_g)\,,
\end{align}
which act on the string junction spectrum ending on the D3-brane.

\newsec{D3-branes and Duality-required Monodromy Quotients}
\label{sec:D3Theories}

Let us finally study the D3-brane theory at the codimension two singularity $z=t=0$ where the seven-brane
described by the Weierstrass model
\begin{align}
y^2 = x^3 - z^a x + t^b
\label{eqn:deformedBG}
\end{align} 
self-intersects. This is the theory that we called Theory $h$ in the introduction, and to study it we will use the geometric action
of Section~\ref{sec:LinkTheory}.

Recall from the introduction that this D3-brane theory can
be naturally thought of in terms of deformations of a simpler seven-brane background, in which the D3-brane realizes
an $\cN=2$ SCFT. These theories, which we call Theory $f$ and Theory $g$ for brevity, are defined to be the D3-brane theory at $z=t=0$ in the
seven-brane background
\begin{align}
\text{Theory $f$:} \qquad y^2 &= x(x^2-z^a) \nonumber \\
\text{Theory $g$:} \qquad y^2 &= x^3 + t^b.
\end{align}
Each is an SCFT of Argyres-Douglas \cite{Argyres:1995jj,Argyres:1995xn}, Minahan-Nemeschansky \cite{Minahan:1996fg,Minahan:1996cj}, or massless $N_f=4$ Seiberg-Witten \cite{Seiberg:1994rs,Seiberg:1994aj} type,
and in general the flavor symmetries of these theories are different simple Lie groups of different rank.
The possible flavor symmetries for Theory $f$ and Theory $g$ are 
\begin{align}
G_f&\in \{SU(2),SO(8),E_7\} \nonumber \\
G_g&\in \{\emptyset,SU(3),SO(8),E_6, E_8\},
\end{align}
respectively. Deforming each of these seven-brane backgrounds to the same seven-brane background described by
(\ref{eqn:deformedBG}), keeping the D3-brane fixed at $z=t=0$, gives two different descriptions of the deformed
D3-brane theory. These dual descriptions must have the same global symmetries, and therefore the deformation
must reduce $G_f$ and $G_g$ to some common group $G_h$. Theory $f$ and Theory $g$ have massless flavors.

Alternatively, the necessary reduction to a common flavor group $G_h$ can be seen in the fixed background (\ref{eqn:deformedBG}).
In the $g$-tube the D3-brane is at $t=0$ $z=Re^{i\theta_z}$, with $R$ the three-sphere radius. The Weierstrass model over the
associated disc centered at $t=0$ is 
\begin{align}
y^2 = x^3 - R^a e^{ia \theta_z} x + t^b
\end{align}
where ordinarily $R\ne 0$ would be thought of as a mass deformation with massive flavors in representations of $G_g$.
Then the limit $R\mapsto 0$ takes the D3-brane to the singularity at $z=t=0$ and the flavors become massless.
Similar statements apply to the $f$-tube theory, which has massive flavors in representations of $G_f$ that become
massless as the D3-brane moves to $z=t=0$. But the D3-brane theory at $z=t=0$ does not care about its path to $z=t=0$,
and thus there must be something wrong with the description of that theory as the massless limit of $\cN=2$ SCFTs
with flavor symmetries $G_f$ and $G_g$ that generally differ.

The resolution is simple: $z$ and $t$ are not simply mass deformations
of an $\cN=2$ theory with a one-dimensional Coulomb branch, but are
both dimensions of space into which the D3-brane may move and the
seven-brane may extend. In particular, the deformation breaks the supersymmetry
on the D3-brane to $\cN=1$ since the seven-brane background now preserves $8$ supercharges and the
D3-brane is half BPS. So there is no paradox, as long as the
non-trivial extension of the seven-branes into both directions reduces
$G_f$ and $G_g$ to some common group $G_h$. 
 
It is natural to expect that the reduction arises from the seven-brane monodromy on string junctions.
The correct prescription is that the string junctions in the spectrum of the D3-brane at $z=t=0$
are those junctions from the $g$-tube and the $f$-tube theories that are invariant under the associated
monodromies $M_g$ and $M_f$. Thus, though the flavor symmetries away from $z=t=0$ are generally different as encoded
in the fact that generally\footnote{Using the definition from the introduction, the root junction lattice
of the flavor algebra associated with $G_{f,g}$ is $R_{f,g}=\{J\in H_2(X_{f,g},E_{f,g}) \,\, | \, \, (J,J)=-2,\,\,\ a(J)=\partial J=0\}$.} $R_g\neq R_f$ and $H_2(X_g,E_g)\neq H_2(X_f,E_f)$, if the spectrum on the D3-brane at $z=t=0$
is the monodromy-invariant spectrum, one expects an isomorphism
\begin{align}
\frac{H_2(X_f,E_f)}{M_f}\cong \frac{H_2(X_g,E_g)}{M_g}\,.
\label{eqn:iso}
\end{align}
In fact this is always the case, as computed for all fifteen examples
with $a\in\{1,2,3\}$ and $b\in\{1,2,3,4,5\}$ in Appendix~\ref{sec:Examples} using the monodromy action on the junction basis
derived in Section~\ref{sec:monodromyactiononbasis}. More specifically, quotienting
by $M_f$ and $M_g$ projects out any $U(1)$-charged junctions, so that the only 
junctions in the quotient have asymptotic charge zero. In some cases there are no
monodromy-invariant junctions, though there are in others; in the latter case there is a set
of shortest junctions\footnote{Note that if a junction $J$ is invariant so is $n J$, $n\in\bZ$. These junctions fall into higher-dimensional irreducible representations whose highest weights have Dynkin labels $(2n)$ or $(n,n)$ for $SU(2)$ or $SU(3)$, respectively.} which has the structure of either an $SU(2)$ or $SU(3)$ root lattice
that matches across the two different monodromies. This establishes the isomorphism and the existence of the duality-required common flavor group $G_h$ which in these examples is $G_h \in \{\emptyset,SU(2),SU(3)\}$. Specifically, defining the set of monodromy invariant junctions on the D3-brane to be $J_{\text{inv}}:= H_2(X_f,E_f)/M_f \cong H_2(X_g,E_g)/M_g$, the root junction lattice of $G_h$ is 
\begin{equation}
R_h=\{J\in J_\text{inv}\ \,\, | \,\, (J,J)=-2, \,\, a(J)=\partial J=0\}.
\end{equation} The data for each of the fifteen examples
is given in Table~\ref{tab:Codim2Singularities}.

\begin{table}[t]
\centering
\begin{tabular}{|ccccccccc|}
\hline
$(a,b)$ & Torus Link & $F_f$ & $G_f$ &  $F_g$ & $G_g$ & $G_h$ & One Red. & Both Red. \\ \hline
$(1,1)$ & $(3,2)$ & $\III$ & $SU(2)$ & $\II$ & $\emptyset$  & $\emptyset$ & $\checkmark$ & \\ \hline
$(1,2)$ & $(3,4)$ & $\III$ & $SU(2)$ & $\IV$ & $SU(3)$ & $SU(2)$ & $\checkmark$ & \\ \hline
$(1,3)$ & $(3,6)$ & $\III$ & $SU(2)$ & $\IN_0^*$ & $SO(8)$  & $\emptyset$ & & $\checkmark$ \\ \hline
$(1,4)$ & $(3,8)$ & $\III$ & $SU(2)$ & $\IV^*$ & $E_6$  & $SU(2)$& $\checkmark$  & \\ \hline
$(1,5)$ & $(3,10)$ & $\III$ & $SU(2)$ & $\II^*$ & $E_8$  & $\emptyset$ & & $\checkmark$  \\ \hline
$(2,1)$ & $(6,2)$ & $\IN_0^*$ & $SO(8)$ & $\II$ & $\emptyset$  & $\emptyset$ & $\checkmark$ & \\ \hline
$(2,2)$ & $(6,4)$ & $\IN_0^*$ & $SO(8)$ & $\IV$ & $SU(3)$ & $SU(3)$ & $\checkmark$  & \\ \hline
$(2,3)$ & $(6,6)$ & $\IN_0^*$ & $SO(8)$ & $\IN_0^*$ & $SO(8)$  & $\emptyset$ & & $\checkmark$   \\ \hline
$(2,4)$ & $(6,8)$ & $\IN_0^*$ & $SO(8)$ & $\IV^*$ &  $E_6$& $SU(3)$& &$\checkmark$  \\ \hline
$(2,5)$ & $(6,10)$ & $\IN_0^*$ & $SO(8)$ & $\II^*$ & $E_8$  & $\emptyset$ & &$\checkmark$  \\ \hline
$(3,1)$ & $(9,2)$ & $\III^*$ & $E_7$ & $\II$ & $\emptyset$  & $\emptyset$& $\checkmark$ & \\ \hline
$(3,2)$ & $(9,4)$ & $\III^*$ & $E_7$ & $\IV$ & $SU(3)$ & $SU(2)$& & $\checkmark$ \\ \hline
$(3,3)$ & $(9,6)$ & $\III^*$ & $E_7$ & $\IN_0^*$ & $SO(8)$  & $\emptyset$ & & $\checkmark$  \\ \hline
$(3,4)$ & $(9,8)$ & $\III^*$ & $E_7$ & $\IV^*$ &  $E_6$ & $SU(2)$& & $\checkmark$ \\ \hline
$(3,5)$ & $(9,10)$ & $\III^*$ & $E_7$ & $\II^*$ & $E_8$  & $\emptyset$ & & $\checkmark$ \\ \hline
\end{tabular}
\caption{The data associated to the theories $f$, $g$ and $h$ that we study. The last two columns indicate whether either $G_f$ or $G_g$ is reduced to obtain $G_h$ (second-to-last), or both (last).}
\label{tab:Codim2Singularities}
\end{table}

In detail, the computational steps are as follows: For a given set $(a,b)$, the vanishing cycles of theory $f$ and $g$ are determined, as is the braid action $B$. In order to analyze the theory in the $f$- or $g$-tube, we first construct the root junctions (i.e.\ those junctions $J$ with $a(J)=0$ and $(J,J)=-2$) and from them the simple root junctions. We then determine the map of the asymptotic charges under the braid action and find that asymptotic charge zero states are mapped to asymptotic charge zero states. This establishes that the braid action is an automorphism on the root lattice. In order to find out whether the automorphism is inner or outer, we construct the Weyl group and check whether the action induced by the braid on the simple roots is a Weyl group element. In all cases we only find inner automorphisms. Subsequently we construct the monodromy-invariant root junction, i.e. those in $R_h$. The simplest way to do so is to find the eigenspace of the braid map $B$ with eigenvalue one. The eigenvectors can then be expressed in terms of the simple roots of the original algebra. 
Since in all examples the only invariant states have asymptotic charge zero, we find that there are no monodromy-invariant charged states. All these steps are automated in a Mathematica notebook which we provide in \cite{MathematicaProgram:2016mat}.

\vspace{.5cm}

We would like to understand more about the physics of the strongly
coupled theory of the D3-brane at $z=t=0$, which we call Theory $h$,
based on the geometry. Henceforth we will denote it with the
superscript $(a,b)$ in order to talk about the D3-brane theory at
$z=t=0$ in a fixed seven-brane background defined by $a$ and
$b$. Similarly, the associated $\cN=2$ theories that deform to the
$\cN=1$ theory $h^{(a,b)}$ will be denoted $f^a$, $g^b$. $h^{(a,b)}$
is a deformation of two different theories that generally do not have
a Lagrangian description; we will therefore make conjectural
statements about the quantum D3-brane theory $h^{(a,b)}$ purely from
the geometry, attempting to find a unified description of the physics.

For every theory $h^{(a,b)}$ the geometry implies some common features:
\begin{itemize}
\item Two dual descriptions of $h^{(a,b)}$ in terms of distinct deformations of distinct $\cN=2$ SCFTs $f^a$ and $g^b$,
  where $f^a$ and $g^b$ have holomorphic gauge coupling $\tau = i$ and $\tau = e^{2\pi i/3}$, respectively. Interestingly, these are not $SL(2,\bZ)$ equivalent.
\item $f^a$ and $g^b$ are well known theories with massless charged monopoles and dyons:
  \begin{center}
    \begin{tabular}{c|c|c}
      & Theory & Flavor Symmetry \\ \hline
      $f^1$ & $H_1$ Argyres-Douglas & $SU(2)$ \\
      $f^2$ & $N_f=4$ Seiberg-Witten & $SO(8)$ \\
      $f^3$ & $E_7$ Minahan-Nemeschansky & $E_7$\\
      $g^1$ & $H_0$ Argyres-Douglas & $\emptyset$ \\
      $g^2$ & $H_2$ Argyres-Douglas & $SU(3)$\\
      $g^3$ & $N_f=4$ Seiberg-Witten & $SO(8)$ \\
      $g^4$ & $E_6$ Minahan-Nemeschansky & $E_6$ \\
      $g^5$ & $E_8$ Minahan-Nemeschansky & $E_8$ \\
    \end{tabular}
  \end{center}
\item The deformations of $f^a$ and $g^b$ give the same D3-brane theory $h^{(a,b)}$ with no
  massless charged particles, which stems from the fact that the monodromy invariant
junction lattice $J_{\text{inv}}$ defined by
  (\ref{eqn:iso}) consists only of junctions with asymptotic charge zero.
\item At least one of the flavor symmetries of $f^a$ or $g^b$ is reduced by the deformation.
\end{itemize}
Given the last two bullet points, it is tempting to interpret the
physics as deformation-induced condensation of some number of
monopoles or dyons, which may or may not have been in non-trivial
flavor representations according to whether or not the flavor symmetry
is reduced by the deformation.  Dually, one might interpret this as
confinement since the monodromy reduction removes the charged states,
leaving behind charge neutral string junctions that are topologically
comprised of charged string junctions that no longer exist
individually in the spectrum. This interpretation is further supported
by the fact that the charge neutral string junctions are in higher
dimensional flavor representations than their charged constituents.
Note that this interpretation in terms of confinement would
be unconventional, however, since the monodromy-invariant charge
neutral string junction for the D3-brane theory at the origin
corresponds to a collapsed cycle and therefore a massless state,
rather than having a confinement scale mass. Nevertheless, such a massless
junction does exist.

The mass issue associated with the conventional confinement interpretation disappears, however, if the D3-brane is moved away from
$z=t=0$ along the $z$- or $t$-axis. Then the monodromy invariant
charge neutral string junction that passes through the D3-brane has
finite size, and therefore a mass, in which case it may be interpreted
as confined state comprised of an electron, monopole, and dyon. This interpretation requires that the charged states are
also projected out for the D3-brane away from the origin, i.e. that
the monodromy reduction also occurs in that theory. Instead of presenting a detailed study of metric data, string
junction masses, and identification of states to motivate the monodromy reduction, we would like to
again argue from duality. 

Consider a D3-brane at a fixed point $p_\text{away}\in \bC^2$ where
$z=0$, $t=t^*\neq 0$. In the seven-brane background $y^2=x^3-z^a x$
the D3-brane at $p_\text{away}$ is one of the $\cN=2$ SCFTs $f^a$ with
flavor symmetry $G_f$, whereas in the seven-brane background
$y^2=x^3+t^b$ it is a massive $\cN=2$ quantum field theory with flavor symmetry $G_g$ determined by $b$;
the latter is simply movement of the D3-brane on its Coulomb branch away from its SCFT point. 
One may describe the D3-brane at $p_\text{away}$ in the deformed background $y^2=x^3-z^a x + t^b$
as a deformation of either of these theories, and therefore again the differing flavor symmetries
$G_f$ and $G_g$ must be reduced to some common flavor symmetry $G_h$ on the D3-brane at $p_\text{away}$
in the deformed seven-brane background. The only way that we know for this duality-required reduction
to occur is by quotienting by the monodromy of Section \ref{sec:monodromyactiononbasis}. A similar argument holds if
$p_{away}\in \bC^2$ is instead $z=z^*\neq 0$, $t=0$.

Doing so, the $\cN=1$
D3-brane theory at $p_\text{away}$ in the deformed seven-brane background can exhibit (depending on
$a$ and $b$) a charge neutral monodromy invariant massive string junction that is comprised of 
charged string junctions that do not exist in the spectrum themselves. In one example
that we will discuss, such a monodromy invariant junction is
\begin{equation}
J=\Gamma_1 + \Gamma_2 + \Gamma_3
\end{equation}
where $\Gamma_1$, $\Gamma_2$, and $\Gamma_3$ carry electric, magnetic, and dyonic charges
but themselves do not exist in the spectrum due to the monodromy projection.
It is natural to this phenomenon as electron-monopole-dyon confinement.

We will discuss this physics in the two simplest cases, $h^{(1,1)}$ in
Section \ref{sec:ExampleMC} and $h^{(1,2)}$ in Section
\ref{sec:ExampleEDMC}. All of the details presented here are
explicitly computed in Appendix~\ref{sec:Examples}. However, before we
do so, we will look at the scaling dimensions and the flow of the
corresponding SCFTs.

\subsec{Scaling dimensions and RGE flow of the \texorpdfstring{$\mathcal{N}=2$}{N=2} SCFTs}
\label{sec:ScalingDimensions}
Though we have focused primarily on geometric issues in this paper,
would would also like to study the resulting SCFTs on the D3-brane
along the lines of
\cite{Aharony:1996bi,Grassi:2001xu,Heckman:2010qv}. To this end we
study the scaling dimensions of the deformation operators, and study
how their properties correlate with the reduction properties of $G_f$
and $G_g$ that we have derived geometrically. 

Using the procedure of \cite{Witten:1985xc}, we construct the holomorphic $(3,0)$ form of the elliptically fibered (local) CY-threefold. Starting from a Weierstrass model, we can write
\begin{align}
 \Omega=\frac{dz\wedge dx\wedge dt}{\partial V/\partial y}=\frac{dz\wedge dx\wedge dt}{2y},
\end{align}
Since the torus fiber volume is unphysical we demand $[\Omega]=2$. Homogeneity of the Weierstrass model then allows to compute the scaling dimensions in terms of $a$ and $b$ from an associated linear system of equations.

For computing the scaling dimensions one would naively start with the Weierstrass model
\begin{equation}
\label{eq:WeierstrassScalingNaive}
y^2 = x^3 - z^a x + t^b.
\end{equation}
There are four unknowns, $[x],[y],[t],[z]$, as well as three equations from homogeneity, and one from the normalization\footnote{Note that the authors of \cite{Heckman:2010qv} remark that this method, which is based on the Gukov-Vafa-Witten expression for the superpotential \cite{Gukov:1999ya}, might not be valid if the theory is altered by inclusion of the latter.} $[\Omega]=2$. This fixes the scaling dimensions to  
\begin{align}
 \label{eq:ScalingDimsAll}
  [x]=\frac{4ab}{d},\quad [y]=\frac{6ab}{d},\quad [z]=\frac{8b}{d},\quad [t]=\frac{12a}{d}
 \end{align}
 where we have defined
 \begin{align}
  d:=6a+4b-ab=a(6-b)+4b=b(4-a)+6a.
 \end{align}
 Note that for the minimal singularities $a<4$, $b<6$ and thus $d>0$. In order to study which of these theories can flow to a unitary $\mathcal{N}=1$ theory we check for which the scaling dimensions are larger than one. From \eqref{eq:ScalingDimsAll} we find
 \begin{align}
  \label{eq:scalingDimInequality}
  \frac{6a}{4+a}\leq b\leq\frac{6a}{4-a}.
 \end{align}
 These inequalities are satisfied for
 \begin{align}
  \{a=1,b=2\},\quad \{a=2,b=2,3,4,5\},\quad \{a=3,b=3,4,5\}.
 \end{align}
In the other cases we have either $[z]<1$ or $[t]<1$, see Table~\ref{tab:SCFTOverview}. 
This is typically interpreted as a decoupling of the non-unitary operator, such that its scaling dimension is set to one, and the occurrence of an accidental $U(1)$. While this is necessary in order to make sense out of an otherwise unphysical, non-unitary theory, we could use \eqref{eq:WeierstrassScalingNaive} for those cases where the scalings of $z$ and $t$ do not violate the unitarity bound. However, we are viewing Theory $h$ as a deformation of either Theory $f$ or $g$, and this is made explicit in our next approach, the results of which fits nicely with the monodromy reduction, as we shall discuss now.

Having in mind that Theory $h$ arises as a deformation of two different $\mathcal{N}=2$ theories, the natural starting point for the scaling dimensions is the Weierstrass model
\begin{equation}
y^2=x^3-\epsilon_z z^a x + \epsilon_t t^b,
\end{equation}
where an appropriate rescaling of the Weierstrass model and change of variables allows for the elimination of
$\epsilon_z$ or $\epsilon_t$, yielding
\begin{align}
\begin{split}
\text{case i):}\qquad y^2&=x^3-\epsilon z^a x + t^b\,, \\
\text{case ii):}\qquad y^2&=x^3-z^a x+ \epsilon t^b\,, 
\end{split}
\end{align}
in which case turning on $\epsilon\ne 0$ can be thought of as deforming the $\cN=2$ theories $g$ and $f$, respectively. The equations for the scaling dimensions are then given by
\begin{subequations}
\begin{align}
\text{case i):}\qquad 3[x]&=2[y]=[\epsilon]+a[z]+[x]=b[t], \qquad [\Omega]=2=[z]+[x]+[t]-[y]\,, \label{eqn:N2epz}\\
\text{case ii):}\qquad 3[x]&=2[y]=a[z]+[x]=[\epsilon]+b[t], \qquad [\Omega]=2=[z]+[x]+[t]-[y]\,.\label{eqn:N2ept}
\end{align}
\end{subequations}
In both cases we have the same number of equations from homogeneity and normalization as
compared to the naive approach, but one more variable $[\epsilon]$,
and therefore one needs additional physical input to fix the
scaling dimensions.

This additional physical input is the following. Consider,
for example, the case of a D3-brane at $t=0$ in the background of case $i)$,
\begin{equation}
y^2 = x^3 -\epsilon z^a x + t^b.
\end{equation}
In the limit $\epsilon \to 0$ this recovers the $\cN=2$ theory,
and movement of the D3-brane in the $z$-direction simply moves
it along the seven-brane at $t=0$ without changing the $\cN=2$
physics on the D3-brane. This implies that in the limit $\epsilon\to0$,
$z$ is a free field in a hypermultiplet and therefore should have scaling
dimension $[z]=1$. Setting $[z]=1$, the equations \eqref{eqn:N2epz}
become
\begin{equation}
3[x]=2[y]=[\epsilon]+a+[x]=b[t], \qquad 1=[x]+[t]-[y],
\end{equation}
and we note that the three equations that do not involve the
scaling dimension $[\epsilon]$ of the parameter $\epsilon$ are
precisely the equations that determine the scaling dimensions
of the $\cN=2$ theory, with normalization set by the Seiberg-Witten
differential, so $[x],[y],[t]$ are that of the $\cN=2$ theory,
and $t$ is the Coulomb branch operator.
This fixes $[\epsilon]$ for any fixed $(a,b)$, and one could perform
a similar calculation for $y^2=x^3-z^a x +\epsilon t^b$
that deforms the other $\cN=2$ theory.
For each of these types of deformations and every
$(a,b)$, $[\epsilon]$ is computed in Table \ref{tab:SCFTOverview}. We find that $\epsilon$ irrelevant if
\begin{align}
  b<\frac{6a}{4+a}~~\text{in case i)},\qquad\qquad b>\frac{6a}{4-a}~~\text{in case ii)}
 \end{align}
 We thus see that the cases are mutually exclusive and that they cover the rest of the models which do not satisfy \eqref{eq:scalingDimInequality}.

Comparing the different approaches, we make the following observations:
\begin{itemize}
\item Though the deformation is irrelevant in some cases,
  i.e. $[\epsilon] < 0$, it still affects the infrared physics
  since $\epsilon$ enters into the $J$-invariant of the elliptic curve
  \begin{equation}
    J = \frac{4f^3}{4f^3+27g^2},
  \end{equation}
  and therefore affects the holomorphic gauge coupling $\tau$
  on the D3-brane. Such irrelevant operators that affect low
  energy physics are often called dangerously irrelevant operators.
\item In those cases where the naive analysis gives rise to a non-unitary field and it is thus interpreted as a free field, it appears in a deformation with $\epsilon$ which is irrelevant.
\item Using the approach where we start with an explicit deformation parameter $\epsilon$, we observe
  \begin{enumerate}
  \item[a)] $[\epsilon]>0$ if and only if the flavor symmetry of the $\cN=2$ theory is reduced by the deformation, 
  \item[b)] $[\epsilon]=0$ if and only if the associated flavor symmetry is preserved,
  \item[c)] if $[\epsilon] < 0$ then the flavor symmetry may or may not be reduced.
  \end{enumerate}
\end{itemize}
We hope to return to a more detailed treatment of these operator analyses in the future.

\begin{table}[t]%
\centering
 \begin{tabular}{|c||c|c|c|c|}
  \hline
  \multirow{2}{*}{(a,b)}& Which flavor & Non-unitary & \multicolumn{2}{c|}{$[\epsilon]$ in deformed $\mathcal{N}=2$ theory}\\ 
  &is reduced&field& ~~~~case $i)$~~~~&case $ii)$\\
  \hline
  $(1,1)$ & $G_f$ & $z$  & $-\frac15$ & $1$\\
  $(1,2)$ & $G_g$ & $-$ & $1$ & $0$\\
  $(1,3)$ & $G_f,G_g$ & $t$ & $3$ & $-1$\\
  $(1,4)$ & $G_g$ & $t$ & $7$ & $-2$\\
  $(1,5)$ & $G_f,G_g$ & $t$ & $19$ & $-3$\\
  \hline
  $(2,1)$ & $G_f$ & $z$ & $-\frac65$ & $5$\\
  $(2,2)$ & $G_f$ & $-$ & $0$ & $4$\\
  $(2,3)$ & $G_f,G_g$ & $-$ & $2$ & $3$\\
  $(2,4)$ & $G_f,G_g$ & $-$ & $6$ & $2$\\
  $(2,5)$ & $G_f,G_g$ & $-$ & $18$ & $1$\\
  \hline
  $(3,1)$ & $G_f$ & $z$ & $-\frac{11}{5}$ & $17$\\
  $(3,2)$ & $G_f,G_g$ & $z$ & $-1$ & $16$\\
  $(3,3)$ & $G_f,G_g$ & $-$ & $1$ & $15$\\
  $(3,4)$ & $G_f,G_g$ & $-$ & $5$ & $14$\\
  $(3,5)$ & $G_f,G_g$ & $-$ & $17$ & $13$\\
  \hline 
 \end{tabular}
 \caption{Overview table of SCFT scaling dimensions. For fixed 
 $(a,b)$ the flavor symmetry reduction is listed, as is the
 non-unitary field of method one, and the scaling dimension
 of the deformation parameter $\epsilon$ of the $\cN=2$ theories using
 method two.}
 \label{tab:SCFTOverview}
\end{table}%

\subsec{Example: A Flavor-Breaking Deformation of \texorpdfstring{$H_1$}{H\_1} Argyres-Douglas Theory}\label{sec:ExampleMC}
The theory $h^{(1,1)}$ can be described as deformed $H_0$ or $H_1$ Argyres-Douglas theory. This
was also an example studied in \cite{Grassi:2001xu}. The theory $f^1$, which is $H_1$ Argyres-Douglas, has massless monopoles, electrons,
and dyons with an $SU(2)$ flavor algebra encoded in a three-pronged string junction $J=\Gamma_0 + \Gamma_1 + \Gamma_2$ that 
has asymptotic charge zero. An arbitrary string junction is written as $J=\sum_{i=1}^3 J_i \Gamma_i$ and the deformation induces
an action according to \eqref{eq:ab11BraidActionFTube}, 
\begin{align}
(\Gamma_1,\Gamma_2,\Gamma_3) \mapsto (-\Gamma_3,-\Gamma_1,-\Gamma_2)\,.
\end{align} 
Quotienting by this action gives a trivial lattice of string junctions for
the theory $h^{(1,1)}$ as viewed from the point of view of the deformation of theory $f^1$. That is, both the charged and charge neutral
string junctions are projected out in the quotient.  

On the other hand, one can also consider $h^{(1,1)}$ as a deformation
of another $\cN=2$ SCFT $g_1$, which is $H_0$ Argyres-Douglas
theory. There is no flavor algebra, but there are massless charged
monopoles and dyons, cf.\ Table~\ref{tab:IICycles}. Deforming this theory to obtain $h^{(1,1)}$ gives
an action on the basis of string junctions of the $H_0$
theory, cf.\ \eqref{eq:ab11BraidActionGTube}
\begin{align}
(\Gamma_1,\Gamma_2)\mapsto  (-\Gamma_2,\Gamma_1)\,.
\end{align} 
Quotienting by this action (which is a different quotient and initial
junction lattice than from the $f^1$ point of view) also gives the
lattice of string junctions for the theory $h^{(1,1)}$, which is again
trivial. Here there is no flavor symmetry to break, but the
spectrum after deformation no longer has charged states.

We will say more in the next example, since the most interesting features occur when there are string junctions
that survive the monodromy projection. We simply conclude here that from both points of view
the theory $h^{(1,1)}$ has no invariant string junction states.

\subsec{Example: Another \texorpdfstring{$H_1$}{H\_1} Deformation with Electron-Dyon-Monopole Bound States}\label{sec:ExampleEDMC}
Let us turn to $h^{(1,2)}$, which is more interesting since the residual flavor symmetry after deformation
is $SU(2)$, as can be seen from Table \ref{tab:Codim2Singularities}. $h^{(1,2)}$ can be described as a deformation
of $f^1$ or $g^2$, which are the $H_1$ and $H_2$ Argyres-Douglas theories.

Thought of as a deformation of $f^1$, which has $SU(2)$ flavor symmetry, the deformation induces the action \eqref{eq:ab12BraidActionFTube} on the basis of string junctions
\begin{align}
(\Gamma_1,\Gamma_2,\Gamma_3) \mapsto (\Gamma_2,\Gamma_3,\Gamma_1)\,.
\end{align} 
Unlike the deformation of $f^1$ associated with the $h^{(1,1)}$ theory
(where the Weierstrass model is deformed by $t$ instead of $t^2$) in
this case the junction $J=\Gamma_1+\Gamma_2+\Gamma_2$ is left
invariant by the deformation-induced monodromy. Together with $-J$,
this leaves an $SU(2)$ flavor algebra intact on the D3-brane theory
$h^{(1,2)}$. However, there are no monodromy-invariant charged
junctions.  The $U(1)$ gauge symmetry has been broken, but the $SU(2)$
flavor symmetry remains intact.

Moving the D3-brane away from $z=0$ along the $z$-axis to $z=z^*$
as prescribed previously, the
string junctions $J$ and $-J$ become finite size. This implies that
for the D3-brane at $z=z^*$, $t=0$, the seven-brane deformation
induces the condensation of a monopole in a trivial flavor
representation of the $H_1$ Argyres-Douglas theory.
The displaced D3-brane theory has no charged states, but
does exhibit a massive charge neutral flavor adjoint of $SU(2)$ that
becomes massless as it is moved back to the SCFT point. Since charged
junctions are absent, it is natural to interpret the massive charge neutral
flavor adjoint with root junctions $(J,0,-J)$, where
$J=\Gamma_1+\Gamma_2+\Gamma_3$, as a confined state made of the
(anti-)dyon $\Gamma_1=\pi_2$, electron $\Gamma_2=\pi_1$, and monopole
$\Gamma_3=\pi_3$.

Let us see whether a similar picture emerges from the point of view of
the deformation of theory $g^2$, which has $SU(3)$ flavor symmetry. The
ordered set of vanishing cycles associated to theory $g^2$ (which are that of
a deformed type $IV$ fiber) are $\{\pi_1,\pi_3,\pi_1,\pi_3\}$. So with
our conventions where $\pi_1$ and $\pi_3$ define electric and magnetic
charge for the D3-brane theory, $\Gamma_1$ and $\Gamma_3$ are
electrons, $\Gamma_2$ and $\Gamma_4$ are monopoles, and appropriate
sums are (anti-)dyons with electric and magnetic charge $-(1,1)$. The
deformation of this theory to arrive at $h^{(1,2)}$ (which is to
deform the Weierstrass model $y^2=x^3+t^2$ by $z x$) induces the
action \eqref{eq:ab12BraidActionGTube} on the junction basis
\begin{align}
(\Gamma_1,\Gamma_2,\Gamma_3,\Gamma_4) \mapsto (\Gamma_4,-\Gamma_1,\Gamma_2,-\Gamma_3).
\end{align}
Prior to deformation, the junction lattice has dimension four, but
after deformation one must quotient by this action to obtain the
junctions of the $h^{(1,2)}$ theory. This quotient lattice has
dimension one and is an $SU(2)$ generated by
$J=\Gamma_1-\Gamma_2-\Gamma_3+\Gamma_4$. Written this way, $J$ is
comprised of an electric junction $\Gamma_1$, an (anti-)dyonic
junction $-\Gamma_2-\Gamma_3$, and a monopole junction
$\Gamma_4$. Alternatively this can be seen by constructing the
decuplet of the original $SU(3)$ forms from an electric quark, dyonic
quark, and monopole quark; this decuplet contains $J$. 

Moving the D3-brane
away from $z=t=0$ to $z=z^*$, $t=0$ this junction becomes massive, and
it is again natural to interpret it as a confined object comprised of an electron, dyon, and monopole string junction.
We see again, this time from the deformation of theory $g^2$, that
$h^{(1,2)}$ has a charge neutral flavor adjoint of $SU(2)$ that can be
interpreted as a confinement of an electron, dyon, and monopole 
when the D3-brane is placed away
from the SCFT point at $z=t=0$.

\newsec{Conclusions and Outlook}

In this paper we studied  $\cN=1$ D3-branes in non-trivial seven-brane
backgrounds that have dual descriptions as deformations of two
 $\cN=2$ SCFTs or QFTs with different flavor symmetries.  Via a
geometric analysis involving string junctions and seven-brane link
induced monodromy, we demonstrated that the dual descriptions have a
common reduced flavor symmetry (as they must) and the deformation of
the $\cN=2$ theories removes their charged states. 
Duality also requires that the monodromy be imposed on the D3-brane theory away from the origin in the deformed seven-brane background,
which is the SCFT point.
After imposing the monodromy quotient, the 
states of the deformed theory (both at and away from the SCFT point) are charge neutral and in higher dimensional
representations of the flavor symmetry, if it exists. 

If the charge neutral states are also massive,
this is the behavior of quark confinement into baryons and
mesons, or alternatively of monopole condensation under electric-magnetic
duality; the charge neutral junctions of the displaced D3-brane theory
in the deformed seven-brane background are indeed such massive states.  
Compared to the conventional case, though, the charge neutral states in
the theories we study are not comprised
purely of confining electric states (e.g. of the Cartan subalgebra of QCD),
but instead arise from electric, magnetic, and dyonic
states. 

\vspace{1cm}
\noindent Having summarized the physics of our results, let us give more details.

The D3-branes are located at or near the non-trivial
self-intersection of a seven-brane background described by a
Weierstrass model. The associated elliptic fibration is smooth at complex
codimension one in its base, but the seven-brane self-intersects in codimension two near the D3-brane. Albeit occurring rather frequently, such
seven-brane configurations (and corresponding singularities) have not
received much attention in the past, and we study them via torus knots
(or links) and string junctions that appear naturally in the geometry.

In more detail, there are apparent
discrepancies when studying the codimension two singularity from
different points of view in codimension one, which we call Theories
$f$ and $g$, and it is critical to resolve these discrepancies. Applying and extending the techniques for studying
singularities via deformations and string junctions introduced in
\cite{Grassi:2013kha,Grassi:2014ffa,Halverson:2016vwx}, we construct
the flavor algebra and representations of Theories $f$ and $g$, but
find that the deformed seven-brane background reduces or completely
breaks the flavor algebra and the charged states are projected out,
giving rise to a Theory $h$ (a D3 theory at the SCFT point where the seven-brane self-intersects) consistent with
approaching it from either Theory $f$ or $g$ (i.e. approaching the
codimension two singularity along two different codimension one
loci). The reduction is obtained from a monodromy induced by the
seven-branes wrapped on a torus knot or link. We argued that the monodromy
quotient should also be imposed on the spectrum of the D3-brane
displaced from SCFT point, since duality still requires flavor
symmetry reduction to a common group.

We interpret the physical meaning of our results based on our
geometric analysis, since there is no known Lagrangian description for
the theories we discuss. The two theories $f$ and $g$ by themselves
correspond to well-known $\mathcal{N}=2$ theories whose BPS spectra
can be constructed; they are theories of Seiberg-Witten, Argyres-Douglas,
or Minahan-Nemeschansky. Theory $h$ is dually described as a deformation of
Theory $f$ or $g$, and these deformations break the
supersymmetry to $\mathcal{N}=1$. The geometry implies that Theory $h$ never has
charged states, but does have charge neutral states in higher
dimensional flavor representations. These gain a mass when the D3-brane
associated with Theory $h$ is displaced from the origin,
which we interpret as deformation-induced monopole condensation or (dually) confinement; it is deformation-induced since charged states of the $\cN=2$
theories emerge in the limit of the undeformed seven-branes. The flavor symmetry of at
least one of the $\cN=2$ theories is broken, giving a common flavor
symmetry group for Theory $h$, $G_h\in \{\emptyset, SU(2), SU(3)\}$.
In a few cases, the deformation does not break the flavor symmetry,
which can be interpreted and as the condensation of flavor-neutral
charged BPS states of one of the underlying $\mathcal{N}=2$
theories. We will discuss the mathematical aspects of $G_h$ in another publication \cite{Grassi:2017xxx}.

In one of the examples, we compared the junctions invariant under the
braid action associated with the link, and find that in the monodromy
invariant junction of Theories $f$ or $g$ that survives in Theory
$h$ is built from three junctions of the $\cN=2$ theory that carry one
unit of electric, magnetic, and dyonic charge, which we interpret as a
confined state of these when the D3-brane is moved away from the origin. To the best of our knowledge, such theories have
not been described previously.

\vspace{5mm}

While the examples we present in this paper is exhaustive for the
specific type of codimension two singularity we look at (i.e.\
$a=1,2,3$ and $b=1,2,3,4,5$ are the only possibilities leading to
minimal models in codimension two), many points remain open. First, it
would be interesting to study and interpret our results in terms of
five-branes as done in \cite{Grassi:2001xu}. Second, it would be
interesting to apply the same techniques to other types of
singularities that could potentially give rise to iterated torus
knots.  From the CFT point of view it would be very interesting to
study these new theories $h$ more in-depth.  From the mathematical
point of view it would be worthwhile to establish a connection between
the intersection pairing and monodromy used in knot theory and the
techniques we have utilized, which involve monodromy on the second
homology of a one parameter family of elliptic surfaces.  It would
also be interesting to study other aspects of the mathematics; e.g.
for $b >1$ the elliptic threefold defined by \eqref{eqn:introWeier}
has an isolated singularity at $z=t=x=y=0$, which we will study in
this context in a sequel mathematics paper
\cite{Grassi:2017xxx}. Furthermore, the monodromy reduction of the
theories and the mass of the string junctions associated with it
should be reflected in properties of the metric. It would therefore be
worthwhile to study metric data in such setups.

\section*{Acknowledgments}
We thank I\~naki Garc\'ia-Etxebarria, Cody Long, Joe Minahan, Brent Nelson, Diego Regalado, Mauricio Romo and Timo Weigand for useful discussions. We are particularly thankful for very helpful comments by Philip Argyres and Mario Martone. JH is supported by NSF Grant PHY-1620526 and startup funding from Northeastern University. He thanks the Center for Theoretical Physics at MIT and the University of Pennsylvania for hospitality at various stages of this project. This work was in part performed during a visit of AG, JH and JS to the Aspen Center for Physics, which is supported by National Science Foundation grant PHY-1066293. The work of FR is supported by by the EPSRC grant EP/N007158/1 ``Geometry for String Model Building''.

\begin{appendices}

\newsec{Examples}
\label{sec:Examples}
In this Section we apply the general discussion of Sections~\ref{sec:LinkTheory} and \ref{sec:D3Theories} to concrete examples. We limit our discussion to minimal Weierstrass fibrations, which means $1 \leq a \leq 3$ and $1\leq b \leq 5$, so there are $15$ cases to be studied.

Let us fix some notation to help us be brief in each example. Let $F_f$ and $F_g$ be the Kodaira fibers in the $f$-tube and $g$-tube, respectively, with associated ADE singularities (in those surfaces) $G_f$ and $G_g$. An overview over all 15 cases can be found in Table~\ref{tab:Codim2Singularities}. For these cases $L_\Delta$ is a $(3a,2b)$ torus knot or link. If $gcd(3a,2b)>1$ then these are not coprime and the link has multiple components, which happens if and only if either $F_f$ or $F_g$ are $I_0^*$ fibers. 

\begin{table}[t]
\centering
\begin{tabular}{|ccc||c|c|}
\hline
\multirow{2}{*}{Knot/Link} & \multirow{2}{*}{$F_f$} & \multirow{2}{*}{$F_g$} & \multicolumn{2}{c|}{Vanishing cycles} \\
     &       &       & $g$-tube ($t=0$) & $f$-tube ($z=0$) \\ \hline
$(3,2)$ & $\III$ & $\II$ & $\{\pi_1,\pi_3\}$ & $\{\pi_2,\pi_1,\pi_3\}$  \\ \hline
$(3,4)$ & $\III$ & $\IV$ & $\{\pi_1,\pi_3,\pi_1,\pi_3\}$ & $\{\pi_2,\pi_1,\pi_3\}$ \\ \hline
$(3,6)$ & $\III$ & $\IN_0^*$ & $\{\pi_1,\pi_3,\pi_1,\pi_3,\pi_1,\pi_3\}$ & $\{\pi_2,\pi_1,\pi_3\}$ \\ \hline
$(3,8)$ & $\III$ & $\IV^*$ & $\{\pi_1,\pi_3,\pi_1,\pi_3,\pi_1,\pi_3,\pi_1,\pi_3\}$ & $\{\pi_2,\pi_1,\pi_3\}$ \\ \hline
$(3,10)$ & $\III$ & $\II^*$ & $\{\pi_1,\pi_3,\pi_1,\pi_3,\pi_1,\pi_3,\pi_1,\pi_3,\pi_1,\pi_3\}$  & $\{\pi_2,\pi_1,\pi_3\}$ \\ \hline
$(6,2)$ & $\IN_0^*$ & $\II$ & $\{\pi_1,\pi_3\}$ & $\{\pi_2,\pi_1,\pi_3,\pi_2,\pi_1,\pi_3\}$ \\ \hline
$(6,4)$ & $\IN_0^*$ & $\IV$ & $\{\pi_1,\pi_3,\pi_1,\pi_3\}$ & $\{\pi_2,\pi_1,\pi_3,\pi_2,\pi_1,\pi_3\}$ \\ \hline
$(6,6)$ & $\IN_0^*$ & $\IN_0^*$ & $\{\pi_1,\pi_3,\pi_1,\pi_3,\pi_1,\pi_3\}$ & $\{\pi_2,\pi_1,\pi_3,\pi_2,\pi_1,\pi_3\}$ \\ \hline
$(6,8)$ & $\IN_0^*$ & $\IV^*$ & $\{\pi_1,\pi_3,\pi_1,\pi_3,\pi_1,\pi_3,\pi_1,\pi_3\}$ & $\{\pi_2,\pi_1,\pi_3,\pi_2,\pi_1,\pi_3\}$\\ \hline
$(6,10)$ & $\IN_0^*$ & $\II^*$ & $\{\pi_1,\pi_3,\pi_1,\pi_3,\pi_1,\pi_3,\pi_1,\pi_3,\pi_1,\pi_3\}$ & $\{\pi_2,\pi_1,\pi_3,\pi_2,\pi_1,\pi_3\}$\\ \hline
$(9,2)$ & $\III^*$ & $\II$ & $\{\pi_1,\pi_3\}$ & $\{\pi_2,\pi_1,\pi_3,\pi_2,\pi_1,\pi_3,\pi_2,\pi_1,\pi_3\}$ \\ \hline
$(9,4)$ & $\III^*$ & $\IV$ & $\{\pi_1,\pi_3,\pi_1,\pi_3\}$ & $\{\pi_2,\pi_1,\pi_3,\pi_2,\pi_1,\pi_3,\pi_2,\pi_1,\pi_3\}$ \\ \hline
$(9,6)$ & $\III^*$ & $\IN_0^*$ & $\{\pi_1,\pi_3,\pi_1,\pi_3,\pi_1,\pi_3\}$ & $\{\pi_2,\pi_1,\pi_3,\pi_2,\pi_1,\pi_3,\pi_2,\pi_1,\pi_3\}$\\ \hline
$(9,8)$ & $\III^*$ & $\IV^*$ & $\{\pi_1,\pi_3,\pi_1,\pi_3,\pi_1,\pi_3,\pi_1,\pi_3\}$ & $\{\pi_2,\pi_1,\pi_3,\pi_2,\pi_1,\pi_3,\pi_2,\pi_1,\pi_3\}$\\ \hline
$(9,10)$ & $\III^*$ & $\II^*$ & $\{\pi_1,\pi_3,\pi_1,\pi_3,\pi_1,\pi_3,\pi_1,\pi_3,\pi_1,\pi_3\}$ & $\{\pi_2,\pi_1,\pi_3,\pi_2,\pi_1,\pi_3,\pi_2,\pi_1,\pi_3\}$\\ \hline
\end{tabular}
\caption{Vanishing cycles for the various torus knots or links.}
\label{tab:VanishingCycles}
\end{table}

We have collected the vanishing cycles for the gauge groups relevant
to our study in Table~\ref{tab:VanishingCycles}. Each of the strands
$s_j$ of the torus knot or link is associated with a vanishing cycles $\pi_i$, $i =1,2,3$. Upon
traversing the knot or link in the $f$-tube we find that a $(3a,2b)$ torus knot or
link can be described as a braid with $s_j$ strands, $j=1,\ldots,3a$,
which induces a permutation $s_j\rightarrow s_{j+2b}$ modulo
$3a$. Similarly, in the $g$-tube the $(3a,2b)$ torus knot or link can be
described as a braid with $2b$ strands, which induces a permutation
$s_j\rightarrow s_{j+3a}$ modulo $2b$.

Next we address the string junctions following the conventions of
\cite{Grassi:2014ffa}. As we have discussed in detail in the previous
section, traveling down the $g$- or $f$-tube rotates the straight line
paths that start at $p$ end end on a seven-brane at $p_k$. However, in
each slice all $p_k$ are rotated by the same amount, which means that
the paths $\delta_k$ that connect $p$ with $p_k$ never cross each
other. In particular, this means that no Hanany-Witten moves \cite{Hanany:1996ie} are
introduced. Thus, the only effect of traversing the torus knot or link is a
permutation of the strands, which leads to a permutation plus a
possible sign flip of the vanishing cycles associated with them.
This gives the action on relative homology (string junctions)
presented in Section~\ref{sec:monodromyactiononbasis}.

Let us now go through each of the 15 examples. We group the discussion according to the 7 distinct Kodaira singular fibers that occur
in the slice of the $f$-tube ($g$-tube) at $\theta_t=0$ ($\theta_z=0$); they are $\{II,III,IV, I_0^*, IV^*, III^*,II^*\}$.

\subsection{Type \II fibers}

\begin{figure}[t]
  \centering
  \subfloat[t][Braid of $(3,2)$ torus knot.\label{fig:IITorusLink11}]{
    \parbox{.45\textwidth}{\centering\minibraidc{.5}{s_1s_1s_1}{red}{cyan}}
  }
  \qquad
  \subfloat[t][Braid of $(6,2)$ torus link.\label{fig:IITorusLink21}]{
    \parbox{.45\textwidth}{\centering\minibraid{.5}{s_1s_1s_1s_1s_1s_1}}
  }
  \newline
  \subfloat[t][Braid of $(9,2)$ torus knot.\label{fig:IITorusLink31}]{
    \parbox{.45\textwidth}{\centering\minibraid{.5}{s_1s_1s_1s_1s_1s_1s_1s_1s_1}}
  }
  \qquad
  \parbox{.45\textwidth}{~}
  \caption{The braids corresponding to the type \II fibers in the $g$-tube, which have $b=1$.}
  \label{fig:IITorusLinks}
\end{figure}

\begin{figure}[t]
  \centering
  \includegraphics[height=20ex]{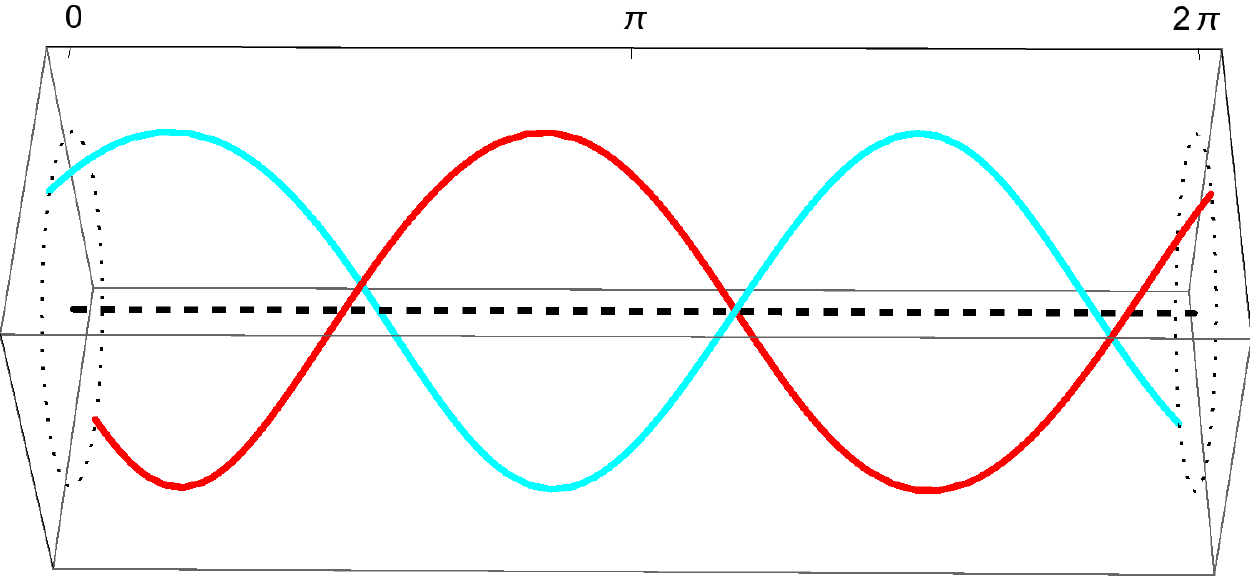}
  \qquad
  \includegraphics[height=20ex]{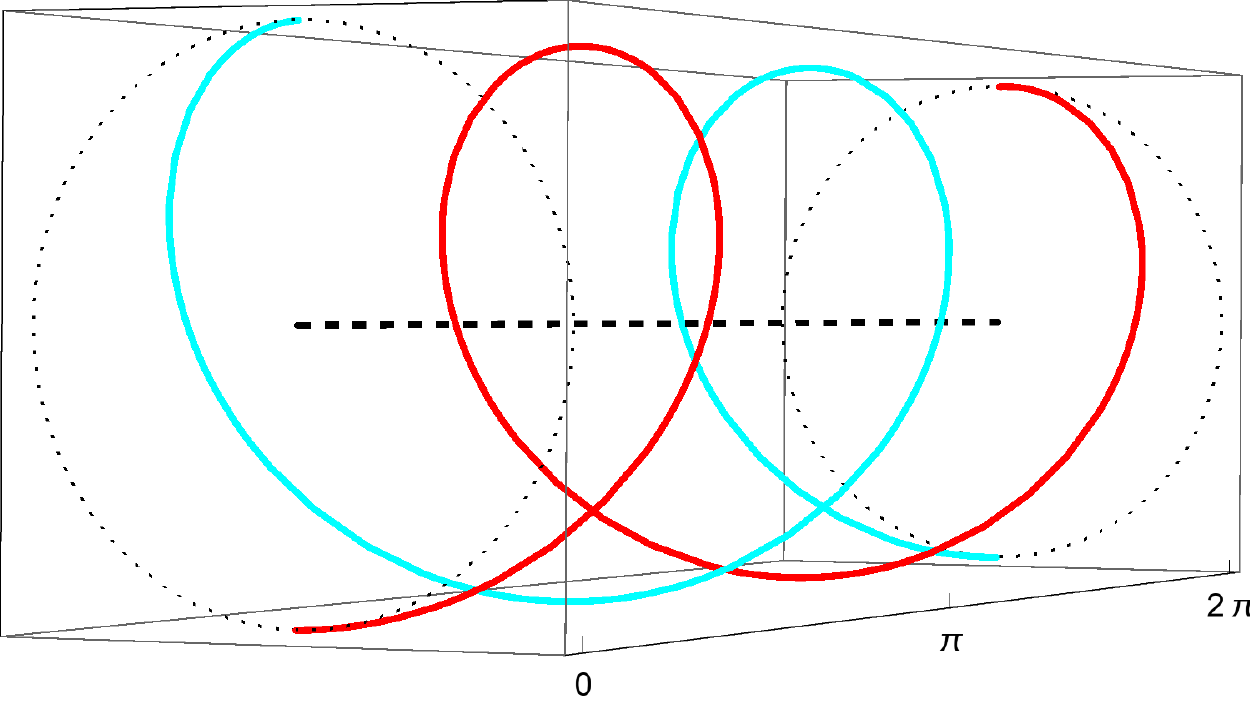}
  \caption{A visualization of the $g$-tube for \ab11 as $\theta_z$ is varied from $0$ to $2\pi$. The dashed line represents the path traveled  by the base point $p$ at the center $D_t(0)$. The strands are seven-branes and the color coding matches that of Figure~\ref{fig:IITorusLink11}.}
  \label{fig:IITorusLinkTube}
\end{figure}

The type \II singularities occur in the $g$-tube for $b=1$. We thus have to look at the $(3a,2)$ torus knot or links, cf.\ Figure \ref{fig:IITorusLinks} for the braid representation and Figure \ref{fig:IITorusLinkTube} for the tube representation for $a=1$. 
As discussed in Section~\ref{sec:GTube} the vanishing cycles are
\begin{align}
\{W,Z\}\,.
\end{align}
This example is rather special in that this cuspidal curve does not
give rise to a non-trivial Lie algebra. In terms of string
junctions, this means that there are no string junctions corresponding
to simple roots, i.e.\ no junctions $J$ with $(J,J)=-2$ and $a(J)=0$.

In basis one we have $W=\pi_1$ and $Z=\pi_3$, using the cycle labels of \eqref{eq:defPis} and the usual intersection product of one-cycles $(p,q)\cdot (r,s)=ps-rq$. Written this way,
the vanishing cycles are thus
\begin{align}
\{\pi_1,\pi_3\}\,.
\end{align}
 Using \eqref{eq:IntersectionForm}, the $I$-matrix reads
\begin{align}
I=
\begin{pmatrix}
 -1 & \frac{1}{2} \\
 \frac{1}{2} & -1 \\
\end{pmatrix}\,,
\end{align}
from which we find that there are no junctions $J$ of $(J,J)=-2$ and $a(J)=(0,0)$. There are, however, BPS junctions with $(J,J)=-1$ and non-vanishing asymptotic charge, cf.\ Table~\ref{tab:IIBasis1Cycles}. Note that if a junction $J$ is BPS, then $-J$ will also be BPS and its asymptotic charge will be the negative of the former.

\vspace{.5cm}
\noindent Let us turn to study the cases associated with differing values of $a$.

\textbf{$\boldsymbol{a=1}$:\quad} 
From (\ref{eqn:braidactiononbasisgtube}) the braid induces an action on the junction basis $\Gamma_i=e_i\in \bZ^{2}$ given by 
 \begin{align}
 \label{eq:ab11BraidActionGTube}
 B=
\begin{pmatrix}
 0 & -1 \\
 1 & 0
\end{pmatrix}
\,, 
\end{align}
which leaves no junctions invariant; the invariant sublattice has dimension $0$.

\begin{table}[t]
\centering
\subfloat[t][Junctions with $(J,J)=-1$ in basis 1.\label{tab:IIBasis1Cycles}]{
\begin{tabular}[h]{|c|c||c|c|}
\hline
\multicolumn{2}{|c||}{$\theta_z=0$} &\multicolumn{2}{|c|}{$\theta_z=2\pi$}\\
$a(J)$ & $J$&  $a(J')$ & $J'$  \\
\hline
$(1,0)$&$(1,0)$&$(0,1)$&$(0,1)$\\
\hline
$(0,1)$&$(0,1)$&$(-1,0)$&$(-1,0)$\\
\hline
$(-1,-1)$&$(-1,-1)$&$(1,-1)$&$(1,-1)$\\
\hline
\end{tabular}
}
\qquad
\subfloat[t][Junctions with $(J,J)=-1$ in basis 2.\label{tab:IIBasis2Cycles}]{
\begin{tabular}[h]{|c|c||c|c|}
\hline
\multicolumn{2}{|c||}{$\theta_z=0$} &\multicolumn{2}{|c|}{$\theta_z=2\pi$}\\
$a(J)$ & $J$&  $a(J')$ & $J'$  \\
\hline
$(1,0)$&$(0,1)$&$(0,-1)$&$(-1,0)$\\
\hline
$(0,1)$&$(1,0)$&$(1,0)$&$(0,1)$\\
\hline
$(1,-1)$&$(-1,1)$&$(-1,-1)$&$(-1,-1)$\\
\hline
\end{tabular}
}
\caption{Junctions with self-intersection $-1$ in bases one and two for $b=1$ in the $g$-tube. As can be seen, the sign of the second entry of the junction vector is flipped after the braid action.}
\label{tab:IICycles}
\end{table}

In more detail, the braid action permutes the junction entry vectors $j_0$ and $j_1$ and subsequently flips the sign of the (now) first entry $j_1$. As a consequence, this means that in this basis states with asymptotic charge $a=(a_1,a_2)$ are mapped to states $a'=(-a_2,a_1)$. We have collected the string junctions before and after the braid action in Table \ref{tab:IIBasis1Cycles}. As explained in Section~\ref{sec:IntersectionForm}, in order to check the self-intersection of the states at $\theta_t=2\pi$, we have to use a new $I$-matrix that takes the motion of the base point $\widehat{p}$ into account. With respect to this new intersection form
\begin{align}
 I'=\begin{pmatrix}
 -1 & -\frac{1}{2} \\
 -\frac{1}{2} & -1 \\
\end{pmatrix},
\end{align}
we find that the self-intersection $(J,J)=-1$ is preserved throughout the $g$-tube, as it should be since it is a topological quantity. Indeed, the self-intersection of the third junction\footnote{We do not write the cycles explicitly in $J$ when the choice of cycles is clear; here we work in basis one, so the first entry of $J$ comes with $\pi_1$ and the second with $\pi_3$.} at $\theta_z=2\pi$, $J=\{-1,1\}$, would have self-intersection $(J,J)=-3$ if we were to use $I$ instead of $I'$.

Before moving on to the cases $a=2,3$ let us compare our results with those obtained from using basis 2. If we have done everything correctly we must reach the same conclusions. In basis two we have $W=(0,1)=\pi_3$ and $Z=(1,0)=\pi_1$. The vanishing cycles $\{W,Z\}$ are therefore 
\begin{align}
\{\pi_3,\pi_1\}\,.
\end{align}
The monodromy on $E_g$ induced by traversing the $g$-tube acts as $W\mapsto -Z$ and $Z\mapsto W$, i.e.\ $\pi_3\mapsto -\pi_1$ and $\pi_1\mapsto \pi_3$. Note, written as an action on $\pi_1$ and $\pi_3$ this monodromy is the same as that of the previous basis. The difference is that the vanishing cycles have changed, as has the $I$-matrix,
\begin{align}
I=
\begin{pmatrix}
 -1 & -\frac{1}{2} \\
 -\frac{1}{2} & -1 \\
\end{pmatrix}
\,.
\end{align}
The braid and the mapping of the cycles induces the following action on the junctions:
\begin{align}
 \label{eq:ab11BraidActionGTube2}
 B=
\begin{pmatrix}
 0 & 1 \\
 -1 & 0
\end{pmatrix}
\,, 
\end{align}
Thus, the action in this basis corresponds again to permuting the two junction vector entries, but this time it is followed by a sign flip of the second entry. Since the first junction vector entry comes with $\pi_3=(0,1)$, this means, however, that again the sign of the second entry of the asymptotic charge is flipped. Hence the asymptotic charges of the BPS string junctions after traveling down the $g$-tube are independent of the base choice.

We have collected the BPS junctions with $(J,J)=-1$ in this basis in Table~\ref{tab:IIBasis2Cycles}. We find again that with respect to the new intersection form
\begin{align}
 I'=\begin{pmatrix}
 -1 & \frac{1}{2} \\
 \frac{1}{2} & -1 \\
\end{pmatrix},
\end{align}
the self-intersection is left unchanged. Since this analysis carries over to the other cases as well, we will use basis one throughout the rest of the paper.

\textbf{$\boldsymbol{a=2}$:\quad} 
From (\ref{eqn:braidactiononbasisgtube}) the braid induces an action on the junction basis $\Gamma_i=e_i\in \bZ^{2}$ given by 
 \begin{align}
 B=
\begin{pmatrix}
 -1 & 0 \\
 0 & -1
\end{pmatrix}
\,, 
\end{align}
which leaves no junctions invariant; the invariant sublattice has dimension $0$.

\textbf{$\boldsymbol{a=3}$:\quad} 
From (\ref{eqn:braidactiononbasisgtube}) the braid induces an action on the junction basis $\Gamma_i=e_i\in \bZ^{2}$ given by 
 \begin{align} 
 B=
\begin{pmatrix}
 0 & 1 \\
 -1 & 0
\end{pmatrix}
\,, 
\end{align}
which leaves no junctions invariant; the invariant sublattice has dimension $0$.

\subsection{Type \III fibers}
The type \III singularities occur in the $f$-tube for $a=1$. We thus
look at the $(3,2b)$ torus knots or links with $1\leq b \leq 5$. In this case
we find an $SU(2)$ gauge algebra with states transforming in the
fundamental representation. However, again the braid cannot induce
non-trivial outer automorphisms since all automorphisms are inner for
$SU(2)$.  The vanishing cycles are given in
table~\ref{tab:VanishingCycles}, from which we find the intersection
matrix
\begin{align}
 I=
 \begin{pmatrix}
 -1 & \frac{1}{2} & -\frac{1}{2} \\
 \frac{1}{2} & -1 & \frac{1}{2} \\
 -\frac{1}{2} & \frac{1}{2} & -1 \\
\end{pmatrix}.
\end{align}
The simple root junction is given by $\alpha_1=(1,1,1)$, which can be checked to have asymptotic charge $(0,0)$ and self-intersection $-2$. 

\vspace{.5cm}
\noindent Let us turn to study the cases associated with differing values of $b$.

\begin{figure}[t]
  \centering
  \subfloat[t][Braid of $(3,2)$ torus knot.\label{fig:IIITorusLink11}]{
    \parbox{.45\textwidth}{\centering\minibraidd{1.4}{s_1s_2s_1s_2}{red}{blue}{green}}
  }
  \qquad
  \subfloat[t][$f$-tube of $(3,2)$ torus knot.\label{fig:IIITorusLinkTube}]{
    \parbox{.45\textwidth}{\centering\includegraphics[height=17ex]{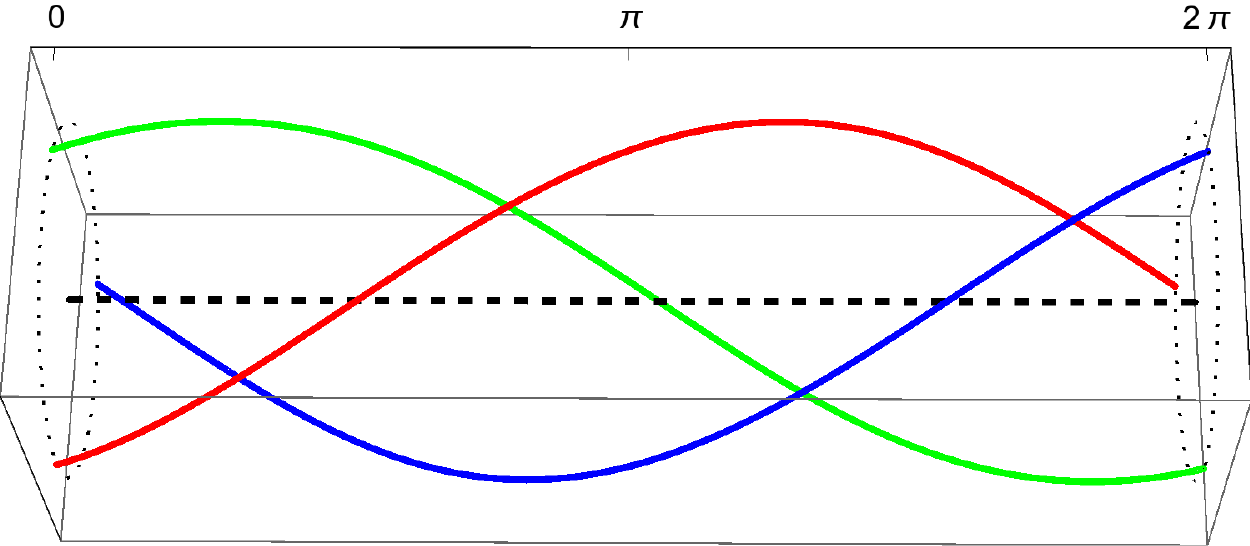}}
  }
  \newline
  \subfloat[t][Braid of $(3,4)$ torus knot.\label{fig:IIITorusLink34}]{
    \parbox{.45\textwidth}{\centering\minibraid{.59}{s_1s_2s_1s_2s_1s_2s_1s_2}}
  }
  \subfloat[t][Braid of $(3,6)$ torus link.\label{fig:IIITorusLink36}]{
    \parbox{.45\textwidth}{\centering\minibraid{.59}{s_1s_2s_1s_2s_1s_2s_1s_2s_1s_2s_1s_2}}
  }
  \newline
  \subfloat[t][Braid of $(3,8)$ torus knot.\label{fig:IIITorusLink38}]{
    \parbox{.45\textwidth}{\centering\minibraid{.36}{s_1s_2s_1s_2s_1s_2s_1s_2s_1s_2s_1s_2s_1s_2s_1s_2}}
  }
  \subfloat[t][Braid of $(3,10)$ torus knot.\label{fig:IIITorusLink310}]{
    \parbox{.45\textwidth}{\centering\minibraid{.36}{s_1s_2s_1s_2s_1s_2s_1s_2s_1s_2s_1s_2s_1s_2s_1s_2s_1s_2s_1s_2}}
  }
  \caption{The braid and the $f$-tube presentation for the type \III fiber with $a=1$.}
  \label{fig:IIITorusLinks}
\end{figure}

\textbf{$\boldsymbol{b=1}$:\quad} 
The braid and the tube action for $\ab11$ in the $f$-tube are given in Figure~\ref{fig:IIITorusLinks}.
From (\ref{eqn:braidactiononbasisftube}) the braid induces an action on the junction basis $\Gamma_i=e_i\in \bZ^{3}$ given by 
 \begin{align}
 \label{eq:ab11BraidActionFTube}
 B=
\begin{pmatrix}
 0 & -1 & 0 \\
 0 & 0 & -1 \\
 -1 & 0 & 0
\end{pmatrix}
\,, 
\end{align}
which leaves no junctions invariant; the invariant sublattice has dimension $0$.
The asymptotic charge of a junction $J$ at $\theta_t=0$ is given by $a(J)=(J_1-J_0,J_2-J_0)$. At $\theta_t=2\pi$, we find $a'(J)=(J_1-J_2,J_1-J_0)$. Consequently $a(J)=(a_1,a_2)\mapsto a'(J)=(a_1-a_2,a_1)$, from which we see again that states of asymptotic charge $(0,0)$ are mapped to states with asymptotic charge $(0,0)$, i.e.\ the braid acts as an automorphism on the root junctions. 

We list the junctions $J$ with $(J,J)=-1$ in Table~\ref{tab:Spectrum11FTube}. There are two singlets and two doublets (plus their conjugates) of the flavor group $SU(2)$. They correspond to the BPS states of the undeformed $\mathcal{N}=2$ theory. Note that they are projected out under the braid action, since they carry non-vanishing asymptotic charge.

\begin{table}
 \centering
 \begin{tabular}{|c|c|c|}
 \hline
 Irrep & asymp.\ charge & weight junctions \\
 \hline
 \rep{1} & (1,0) & \{(0,1,0)\}\\
 \rep{1} & (1,2) & \{(-1,0,1)\}\\
 \rep{2} & (0,1) & \{(-1,-1,0),(0,0,1)\}\\
 \rep{2} & (1,1) & \{(-1,0,0),(0,1,1)\}\\
 \hline
 \end{tabular}
 \caption{Junction vectors with self-intersection $-1$ in the $f$-tube for $a=1$. We give their irreducible representations under the flavor $SU(2)$, their asymptotic charges, and the corresponding junctions. Since the \rep{1} is real and the $\rep{2}$ is pseudo-real, the negatives of these junctions are in the spectrum as well.}
 \label{tab:Spectrum11FTube}
\end{table}

\textbf{$\boldsymbol{b=2}$:\quad}
From (\ref{eqn:braidactiononbasisftube}) the braid induces an action on the junction basis $\Gamma_i=e_i\in \bZ^{3}$ given by 
 \begin{align}
  \label{eq:ab12BraidActionFTube}
 B=
\begin{pmatrix}
 0 & 0 & 1 \\
 1 & 0 & 0 \\
 0 & 1 & 0
\end{pmatrix}
\,, 
\end{align}
which determines a non-trivial braid invariant sublattice of the junction lattice generated by
\begin{center}
\begin{tabular}{c|c|c} 
 Junction $J$ & $(J,J)$ & $a(J)$ \\ \hline
$(1, 1, 1)$ & $-2$ & $(0, 0)$
\end{tabular}.
\end{center}
We see that the generator of braid invariant junctions is the simple root $\alpha_1$ of $G_f=SU(2)$, so $G_h=SU(2)$. 
The asymptotic charges map as $a(J)=(a_1,a_2)\mapsto a'(J)=(-a_2,a_1-a_2)$, which shows again that the braid acts as an automorphism on the root junctions. 

\textbf{$\boldsymbol{b=3}$:\quad}
From (\ref{eqn:braidactiononbasisftube}) the braid induces an action on the junction basis $\Gamma_i=e_i\in \bZ^{3}$ given by 
 \begin{align} 
 B=
\begin{pmatrix}
 -1 & 0 & 0 \\
 0 & -1 & 0 \\
 0 & 0 & -1
\end{pmatrix}
\,, 
\end{align}
which leaves no junctions invariant; the invariant sublattice has dimension $0$, so $G_h=\emptyset$.
The braid induces a map on asymptotic charges map as $a(J)=(a_1,a_2)\mapsto a'(J)=(-a_1,-a_2)$, which shows again that the braid acts as an automorphism on the root junctions. The braid maps $\alpha_1\mapsto-\alpha_1$, as in the $b=1$ case.

\textbf{$\boldsymbol{b=4}$:\quad}
From (\ref{eqn:braidactiononbasisftube}) the braid induces an action on the junction basis $\Gamma_i=e_i\in \bZ^{3}$ given by 
 \begin{align} 
 B=
\begin{pmatrix}
 0 & 1 & 0 \\
 0 & 0 & 1 \\
 1 & 0 & 0
\end{pmatrix}
\,, 
\end{align}
which determines a non-trivial braid invariant sublattice of the junction lattice generated by
\begin{center}
\begin{tabular}{c|c|c} 
 Junction $J$ & $(J,J)$ & $a(J)$ \\ \hline
$(1, 1, 1)$ & $-2$ & $(0, 0)$
\end{tabular}.
\end{center}
We see that the invariant junctions are generated by the root $\alpha_1$ of $G_f=SU(2)$, 
and therefore the braid invariant subalgebra of $G_f$ is $G_h=SU(2)$.
Again it can be shown that the braid gives an automorphism on roots.

\textbf{$\boldsymbol{b=5}$:\quad}
From (\ref{eqn:braidactiononbasisftube}) the braid induces an action on the junction basis $\Gamma_i=e_i\in \bZ^{3}$ given by 
 \begin{align} 
 B=
\begin{pmatrix}
 0 & 0 & -1 \\
 -1 & 0 & 0 \\
 0 & -1 & 0
\end{pmatrix}
\,, 
\end{align}
which leaves no junctions invariant; the invariant sublattice has dimension $0$, so that $G_h=\emptyset$.
The braid is given in Figure \ref{fig:IIITorusLink310}. 
The asymptotic charges map as $a(J)=(a_1,a_2)\mapsto a'(J)=(a_2,a_2-a_1)$, and we see that the asymptotic charge zero states are mapped to asymptotic charge zero states. The braid induces a Weyl reflection of the (simple) root junction, $\alpha_1\mapsto-\alpha_1$.

\subsection{Type \IV fibers}
Type \IV knots or links are obtained from the $g$-tube for $b=2$, i.e.\ from \ab{$3a$}{4} torus knots or links. The $I$-matrix in basis one is given by
\begin{align}
I=\left(
\begin{array}{cccc}
 -1 & \frac{1}{2} & 0 & \frac{1}{2} \\
 \frac{1}{2} & -1 & -\frac{1}{2} & 0 \\
 0 & -\frac{1}{2} & -1 & \frac{1}{2} \\
 \frac{1}{2} & 0 & \frac{1}{2} & -1 \\
\end{array}
\right)\,.
\end{align} 
We can choose a set of simple root junctions given by
\begin{align}
\alpha_1=\{0, 1, 0, -1\}\,,\qquad \alpha_2=\{1, 0, -1, 0\}\,.
\end{align}
While there exists an outer automorphism folding $SU(3)$ to $Sp(1)$, we find a Weyl group element that corresponds to the braid action and consequently the induced automorphism is inner. Again we have to quotient by this action, which changes the rank of the resulting flavor group as discussed in the following.

\textbf{$\boldsymbol{a=1}$:\quad}
\begin{table}
 \centering
 \begin{tabular}{|c|c|c|}
 \hline
 Irrep & asymp.\ charge & weight junctions \\
 \hline
 \rep{1} & $(1,-1)$ & $\{(0,-1,1,0)\}$\\
 \rep{1} & $(2,1)$ & $\{(1,0,1,1)\}$\\
 \rep{1} & $(1,2)$ & $\{(1,1,0,1)\}$\\
 \rep{3} & $(1,0)$ & $\{(0,-1,1,1),(0,0,1,0),(1,0,0,0)\}$\\
 \rep{3} & $(0,1)$ & $\{(0,0,0,1),(0,1,0,0),(1,1,-1,0)\}$\\
 \rep{3} & $(-1,-1)$ & $\{(-1,-1,0,0),(-1,0,0,-1),(0,0,-1,-1)\}$\\
 \hline
 \end{tabular}
 \caption{Junction vectors with self-intersection $-1$ in the $g$-tube for $b=2$. We give their irreducible representations under the flavor group $SU(3)$, their asymptotic charges, and the corresponding junctions. Again, the negatives of these junctions are in the spectrum as well and correspond to the respective conjugate irreps.}
 \label{tab:Spectrum12}
\end{table}
From (\ref{eqn:braidactiononbasisgtube}) the braid induces an action on the junction basis $\Gamma_i=e_i\in \bZ^{4}$ given by 
 \begin{align}
  \label{eq:ab12BraidActionGTube}
 B=
\begin{pmatrix}
 0 & -1 & 0 & 0 \\
 0 & 0 & 1 & 0 \\
 0 & 0 & 0 & -1 \\
 1 & 0 & 0 & 0
\end{pmatrix}
\,, 
\end{align}
which determines a non-trivial braid invariant sublattice of the junction lattice generated by
\begin{center}
\begin{tabular}{c|c|c} 
 Junction $J$ & $(J,J)$ & $a(J)$ \\ \hline
$(1, -1, -1, 1)$ & $-6$ & $(0, 0)$
\end{tabular}.
\end{center}
We see that the generator of the invariant junction $\beta_1=(1,-1,-1,1)=\alpha_2-\alpha_1$ is a linear combination of the simple roots $\alpha_1,\alpha_2$ of $G_g=SU(3)$. This junction $\beta_1$ is the simple root for the reduced algebra $G_h=SU(2)$.

Note that we have now seen that $G_h=SU(2)$ for the case $(a,b)=(1,2)$ in both the $f$-tube and $g$-tube. For this $(a,b)$ $G_f=SU(2)$
and is not reduced to obtain $G_h$, but $G_g=SU(3)$ and is reduced to obtain the same $G_h$. That is, we have derived $G_h$ using two
different points of view, the $f$-tube and $g$-tube.

\textbf{$\boldsymbol{a=2}$:\quad}
From (\ref{eqn:braidactiononbasisgtube}) the braid induces an action on the junction basis $\Gamma_i=e_i\in \bZ^{4}$ given by 
 \begin{align} 
 B=
\begin{pmatrix}
 0 & 0 & -1 & 0 \\
 0 & 0 & 0 & -1 \\
 -1 & 0 & 0 & 0 \\
 0 & -1 & 0 & 0
\end{pmatrix}
\, 
\end{align}
which determines a non-trivial braid invariant sublattice of the junction lattice generated by
\begin{center}
\begin{tabular}{c|c|c} 
 Junction $J$ & $(J,J)$ & $a(J)$ \\ \hline
$(1, 0, -1, 0)$ & $-2$ & $(0, 0)$ \\
$(0, 1, 0, -1)$ & $-2$ & $(0, 0)$
\end{tabular}.
\end{center}
We see that the generators of the invariant junctions are $\beta_i=\alpha_i$, i.e.\ precisely the simple roots of
 $G_g=SU(3)$, and therefore $G_h=SU(3)$.

Note that in this case $a$ and $b$ are not co-prime and the braid is actually a link with two components. The asymptotic charges map as $(a_1,a_2)\mapsto(-a_1,-a_2)$ so that junctions $J$ with $a(J)=0$ map to junctions $J'$ with $a(J)=0$; the braid gives an
automorphism of the roots, which in this case is trivial. As we shall see when we next analyze the $\IN_0^*$ cases, the result $G=SU(3)$ matches again perfectly with the result from the $f$- tube in the \ab22 case.

\textbf{$\boldsymbol{a=3}$:\quad}
From (\ref{eqn:braidactiononbasisgtube}) the braid induces an action on the junction basis $\Gamma_i=e_i\in \bZ^{4}$ given by 
 \begin{align} 
 B=
\begin{pmatrix}
 0 & 0 & 0 & 1 \\
 -1 & 0 & 0 & 0 \\
 0 & 1 & 0 & 0 \\
 0 & 0 & -1 & 0
\end{pmatrix}
\,, 
\end{align}
which determines a non-trivial braid invariant sublattice of the junction lattice generated by
\begin{center}
\begin{tabular}{c|c|c} 
 Junction $J$ & $(J,J)$ & $a(J)$ \\ \hline
$(1, -1, -1, 1)$ & $-6$ & $(0, 0)$
\end{tabular}.
\end{center}
We see that the generator of invariant junctions is $\beta_1=(1,-1,-1,1)=\alpha_1-\alpha_2$, a linear combination of the simple roots of $G_g=SU(3)$ that itself generates an algebra $G_h=SU(2)$.

In this case the asymptotic charges are mapped as $(a_1,a_2)\mapsto(a_2,-a_1)$, therefore preserving the asymptotic charge of junctions
$J$ with $a(J)=0$, so the braid induces an automorphism of roots.

\subsection{Type \texorpdfstring{$\IN_0^*$}{I*\_0} fibers}
This case is special since the fiber type $I_0^*$ can occur both in the $f$- and the $g$-tube. In the former, the corresponding torus knot or link is given by $a=2$ for arbitrary $b$ and in the latter by $b=3$ for arbitrary $a$. Note that the $D_4$ Dynkin diagram allows for outer automorphisms that fold it to either $B_3$ or $G_2$. However, we find that all braid actions induce inner automorphisms.

\subsubsection*{The \texorpdfstring{$f$}{f}-tube analysis}
An $I_0^*$ fiber is obtained in the $f$-tube when $a=2$. In that case the $I$-matrix reads
\begin{align}
 I=\left(
\begin{array}{cccccc}
 -1 & \frac{1}{2} & -\frac{1}{2} & 0 & \frac{1}{2} & -\frac{1}{2} \\
 \frac{1}{2} & -1 & \frac{1}{2} & -\frac{1}{2} & 0 & \frac{1}{2} \\
 -\frac{1}{2} & \frac{1}{2} & -1 & \frac{1}{2} & -\frac{1}{2} & 0 \\
 0 & -\frac{1}{2} & \frac{1}{2} & -1 & \frac{1}{2} & -\frac{1}{2} \\
 \frac{1}{2} & 0 & -\frac{1}{2} & \frac{1}{2} & -1 & \frac{1}{2} \\
 -\frac{1}{2} & \frac{1}{2} & 0 & -\frac{1}{2} & \frac{1}{2} & -1 \\
\end{array}
\right)\,.
\end{align}
which leads to a set of simple roots that in the junction basis $\Gamma_i$ are 
\begin{align}
\begin{split}
 \alpha_1&=\{0, 0, 0, 1, 1, 1\}\,,\qquad~~ \alpha_2=\{0, 0, 1, 0, 0, -1\}\,,\\ 
 \alpha_3&=\{0, 1, 0, 0, -1, 0\}\,,\qquad \alpha_4=\{1, 0, -1, -1, 0, 1\}\,. 
\end{split}
\end{align}

\textbf{$\boldsymbol{b=1}$:\quad}
From (\ref{eqn:braidactiononbasisftube}) the braid induces an action on the junction basis $\Gamma_i=e_i\in \bZ^{6}$ given by 
 \begin{align} 
 B=
\begin{pmatrix}
 0 & 0 & 0 & 0 & -1 & 0 \\
 0 & 0 & 0 & 0 & 0 & -1 \\
 -1 & 0 & 0 & 0 & 0 & 0 \\
 0 & -1 & 0 & 0 & 0 & 0 \\
 0 & 0 & -1 & 0 & 0 & 0 \\
 0 & 0 & 0 & -1 & 0 & 0
\end{pmatrix}
\,, 
\end{align}
which leaves no junctions invariant; the invariant sublattice has dimension $0$; thus $G_h=\emptyset$.

The asymptotic charges are mapped as $(a_1,a_2)\mapsto(a_1-a_2,a_1)$ which preserves $a(J)$ for junctions
with $a(J)=0$, and therefore the braid induces an automorphism on the roots. The action on the simple roots reads
\begin{align}
\alpha_1 \mapsto -\alpha_1-\alpha_2-\alpha_3-\alpha_4\,,\quad 
\alpha_2 \mapsto \alpha_3\,,\quad 
\alpha_3 \mapsto \alpha_2+\alpha_4\,,\quad 
\alpha_4 \mapsto -\alpha_2-\alpha_3\,.
\end{align}
and there is no invariant subalgebra, as determined also directly from $B$.

\textbf{$\boldsymbol{b=2}$:\quad}
From (\ref{eqn:braidactiononbasisftube}) the braid induces an action on the junction basis $\Gamma_i=e_i\in \bZ^{6}$ given by 
 \begin{align} 
 B=
\begin{pmatrix}
 0 & 0 & 1 & 0 & 0 & 0 \\
 0 & 0 & 0 & 1 & 0 & 0 \\
 0 & 0 & 0 & 0 & 1 & 0 \\
 0 & 0 & 0 & 0 & 0 & 1 \\
 1 & 0 & 0 & 0 & 0 & 0 \\
 0 & 1 & 0 & 0 & 0 & 0
\end{pmatrix}
\,, 
\end{align}
which determines a non-trivial braid invariant sublattice of the junction lattice generated by
\begin{center}
\begin{tabular}{c|c|c} 
 Junction $J$ & $(J,J)$ & $a(J)$ \\ \hline
$(1, 0, 1, 0, 1, 0)$ & $-4$ & $(0, 0)$ \\
$(0, 1, 0, 1, 0, 1)$ & $-4$ & $(0, 0)$
\end{tabular}.
\end{center}
We see that the invariant junctions $\beta_i$ are a linear combination of the simple roots 
\begin{align}
 \beta_1=\alpha_1+2 \alpha_2+\alpha_4\,,\qquad\beta_2=\alpha_1+\alpha_3\,.
\end{align}
of $G_f=SO(8)$. These generate a braid invariant algebra $G_h=SU(3)$.

The asymptotic charges in this case are mapped as $(a_1,a_2)\mapsto(-a_2,a_1-a_2)$. The map on the simple roots reads
\begin{align}
\alpha_1 \mapsto \alpha_1+\alpha_2+\alpha_3\,,\quad
\alpha_2 \mapsto \alpha_2+\alpha_4\,,\quad
\alpha_3 \mapsto -\alpha_2\,,\quad
\alpha_4 \mapsto -\alpha_2-\alpha_3-\alpha_4\,.
\end{align}
From the Cartan matrix \eqref{eq:CartanMatrix} we see that the two simple roots $\beta_i$ correspond to an $SU(3)$ flavor algebra. As alluded to above, this matches the $g$-tube result from the type \IV fiber.

\textbf{$\boldsymbol{b=3}$:\quad}
From (\ref{eqn:braidactiononbasisftube}) the braid induces an action on the junction basis $\Gamma_i=e_i\in \bZ^{6}$ given by 
 \begin{align} 
 B=
\begin{pmatrix}
 -1 & 0 & 0 & 0 & 0 & 0 \\
 0 & -1 & 0 & 0 & 0 & 0 \\
 0 & 0 & -1 & 0 & 0 & 0 \\
 0 & 0 & 0 & -1 & 0 & 0 \\
 0 & 0 & 0 & 0 & -1 & 0 \\
 0 & 0 & 0 & 0 & 0 & -1
\end{pmatrix}
\,, 
\end{align}
which leaves no junctions invariant; the invariant sublattice has dimension $0$; thus $G_h=\emptyset$.

The braid map in this case is minus the identity. Hence all junctions, and consequently their associated asymptotic charges are mapped to their negatives. The map induces an automorphism on roots, but there is clearly no invariant subalgebra: any junction is an eigenvector
of the braid map with eigenvalue $-1$. 

\textbf{$\boldsymbol{b=4}$:\quad}
From (\ref{eqn:braidactiononbasisftube}) the braid induces an action on the junction basis $\Gamma_i=e_i\in \bZ^{6}$ given by 
 \begin{align} 
 B=
\begin{pmatrix}
 0 & 0 & 0 & 0 & 1 & 0 \\
 0 & 0 & 0 & 0 & 0 & 1 \\
 1 & 0 & 0 & 0 & 0 & 0 \\
 0 & 1 & 0 & 0 & 0 & 0 \\
 0 & 0 & 1 & 0 & 0 & 0 \\
 0 & 0 & 0 & 1 & 0 & 0
\end{pmatrix}
\,, 
\end{align}
which determines a non-trivial braid invariant sublattice of the junction lattice generated by
\begin{center}
\begin{tabular}{c|c|c} 
 Junction $J$ & $(J,J)$ & $a(J)$ \\ \hline
$(1, 0, 1, 0, 1, 0)$ & $-4$ & $(0, 0)$ \\
$(0, 1, 0, 1, 0, 1)$ & $-4$ & $(0, 0)$
\end{tabular}.
\end{center}
We see that the invariant junctions $\beta_i$ are a linear combination 
\begin{align}
 \beta_1=\alpha_1+2 \alpha_2+\alpha_4\,,\qquad\beta_2=\alpha_1+\alpha_3\,.
\end{align}
of the simple roots of $G_f=SO(8)$, which generate an algebra $G=SU(3)$.
The asymptotic charges transform as $(a_1,a_2)\mapsto(a_2-a_1,-a_1)$, maintaining the asymptotic
charge of $a(J)=0$ junctions, and thus giving an automorphism of the roots. The action on the simple roots is
\begin{align}
\alpha_1 \mapsto \alpha_1+\alpha_2+\alpha_3+\alpha_4\,,\quad
\alpha_2 \mapsto -\alpha_3\,,\quad
\alpha_3 \mapsto -\alpha_2-\alpha_4\,,\quad
\alpha_4 \mapsto \alpha_2+\alpha_3\,,
\end{align}
which leaves $\beta_1$ and $\beta_2$ invariant.
Note that even though the action on the simple roots differs from the $b=2$ case (in fact, one is the inverse of the other), the invariant combination of simple roots $\beta_i$ are the same. 

\textbf{$\boldsymbol{b=5}$:\quad}
From (\ref{eqn:braidactiononbasisftube}) the braid induces an action on the junction basis $\Gamma_i=e_i\in \bZ^{6}$ given by 
 \begin{align} 
 B=
\begin{pmatrix}
 0 & 0 & -1 & 0 & 0 & 0 \\
 0 & 0 & 0 & -1 & 0 & 0 \\
 0 & 0 & 0 & 0 & -1 & 0 \\
 0 & 0 & 0 & 0 & 0 & -1 \\
 -1 & 0 & 0 & 0 & 0 & 0 \\
 0 & -1 & 0 & 0 & 0 & 0
\end{pmatrix}
\,, 
\end{align}
which leaves no junctions invariant; the invariant sublattice has dimension $0$; thus $G_h=\emptyset$.
The asymptotic charges map as $(a_1,a_2)\mapsto(a_2,a_2-a_1)$, which leave fixed the asymptotic charge of
junctions with $a(J)=0$, and the braid therefore induces an automorphism of the roots. The simple roots map as
\begin{align}
\alpha_1 \mapsto -\alpha_1-\alpha_2-\alpha_3\,,\quad
\alpha_2 \mapsto -\alpha_2-\alpha_4\,,\quad
\alpha_3 \mapsto \alpha_2\,,\quad
\alpha_4 \mapsto \alpha_2+\alpha_3+\alpha_4\,.
\end{align}
This map does not have an eigenspace with eigenvalue 1, which of course matches the same fact about $B$.

\subsubsection*{The \texorpdfstring{$g$}{g}-tube analysis}
An $I_0^*$ fiber is obtained in the $g$-tube when $b=3$, in which case the $I$-matrix is given by
\begin{align}
 I=\left(
\begin{array}{cccccc}
 -1 & \frac{1}{2} & 0 & \frac{1}{2} & 0 & \frac{1}{2} \\
 \frac{1}{2} & -1 & -\frac{1}{2} & 0 & -\frac{1}{2} & 0 \\
 0 & -\frac{1}{2} & -1 & \frac{1}{2} & 0 & \frac{1}{2} \\
 \frac{1}{2} & 0 & \frac{1}{2} & -1 & -\frac{1}{2} & 0 \\
 0 & -\frac{1}{2} & 0 & -\frac{1}{2} & -1 & \frac{1}{2} \\
 \frac{1}{2} & 0 & \frac{1}{2} & 0 & \frac{1}{2} & -1 \\
\end{array}
\right)\,.
\end{align}
The resulting root junctions are
\begin{align}
\begin{split}
 \alpha_1&=\{0, 0, 0, 1, 0, -1\}\,,\quad~~
 \alpha_2=\{0, 0, 1, 0, -1, 0\}\,,\\
 \alpha_3&=\{0, 1, -1, -1, 1, 0\}\,,\quad
 \alpha_4=\{1, 0, -1, 0, 0, 0\}\,. 
\end{split}
\end{align}

\textbf{$\boldsymbol{a=1}$:\quad}
From (\ref{eqn:braidactiononbasisgtube}) the braid induces an action on the junction basis $\Gamma_i=e_i\in \bZ^{6}$ given by 
 \begin{align} 
 B=
\begin{pmatrix}
 0 & 0 & 0 & -1 & 0 & 0 \\
 0 & 0 & 0 & 0 & 1 & 0 \\
 0 & 0 & 0 & 0 & 0 & -1 \\
 1 & 0 & 0 & 0 & 0 & 0 \\
 0 & -1 & 0 & 0 & 0 & 0 \\
 0 & 0 & 1 & 0 & 0 & 0
\end{pmatrix}
\,, 
\end{align}
which leaves no junctions invariant; the invariant sublattice has dimension $0$; thus $G_h=\emptyset$.

The maps induced by the braid acts as $(a_1,a_2)\mapsto(-a_2,a_1)$ which leaves the asymptotic charge of junctions $J$
with $a(J)=0$ fixed, thereby giving an automorphism on roots. The map on simple roots is
\begin{align}
\alpha_1 \mapsto -\alpha_4\,,\quad
\alpha_2 \mapsto -\alpha_1-\alpha_2-\alpha_3\,,\quad
\alpha_3 \mapsto \alpha_1+2 \alpha_2+\alpha_3+\alpha_4\,,\quad
\alpha_4 \mapsto \alpha_1\,,
\end{align}
and there is no invariant subspace of the $\alpha_i$ under this map, as expected from $B$.

\textbf{$\boldsymbol{a=2}$:\quad}
From (\ref{eqn:braidactiononbasisgtube}) the braid induces an action on the junction basis $\Gamma_i=e_i\in \bZ^{6}$ given by 
 \begin{align} 
 B=
\begin{pmatrix}
 -1 & 0 & 0 & 0 & 0 & 0 \\
 0 & -1 & 0 & 0 & 0 & 0 \\
 0 & 0 & -1 & 0 & 0 & 0 \\
 0 & 0 & 0 & -1 & 0 & 0 \\
 0 & 0 & 0 & 0 & -1 & 0 \\
 0 & 0 & 0 & 0 & 0 & -1
\end{pmatrix}
\,, 
\end{align}
which leaves no junctions invariant; the invariant sublattice has dimension $0$; thus $G_h=\emptyset$.

Note that this case gives rise to a $(6,6)$ torus link and is thus completely identical to the previously discussed case in the $f$-tube with $b=3$. It is a nice cross-check that we find the braid action to be minus the identity as well. Hence the asymptotic charges and simple roots are mapped to their negatives and the invariant subspace is trivial.

\textbf{$\boldsymbol{a=3}$:\quad}
From (\ref{eqn:braidactiononbasisgtube}) the braid induces an action on the junction basis $\Gamma_i=e_i\in \bZ^{6}$ given by 
 \begin{align} 
 B=
\begin{pmatrix}
 0 & 0 & 0 & 1 & 0 & 0 \\
 0 & 0 & 0 & 0 & -1 & 0 \\
 0 & 0 & 0 & 0 & 0 & 1 \\
 -1 & 0 & 0 & 0 & 0 & 0 \\
 0 & 1 & 0 & 0 & 0 & 0 \\
 0 & 0 & -1 & 0 & 0 & 0
\end{pmatrix}
\,, 
\end{align}
which leaves no junctions invariant; the invariant sublattice has dimension $0$; thus $G_h=\emptyset$.

The asymptotic charges are mapped as $(a_1,a_2)\mapsto(a_2,-a_1)$. The simple roots are mapped as
\begin{align}
\alpha_1 \mapsto \alpha_4\,,\quad
\alpha_2 \mapsto \alpha_1+\alpha_2+\alpha_3\,,\quad
\alpha_3 \mapsto -\alpha_1-2 \alpha_2-\alpha_3-\alpha_4\,,\quad
\alpha_4 \mapsto -\alpha_1\,,
\end{align}
which is the same as in the $a=1$ case up to a total minus sign. There is no invariant subspace of this map,
as expected from the more general map $B$ on general junctions.

\subsection{Type \texorpdfstring{$\IV^*$}{IV*} fibers}
The $\IV^*$ fibers occur in the $g$-tube for $b=4$. The corresponding $I$-matrix is
\begin{align}
I=\left(
\begin{array}{cccccccc}
 -1 & \frac{1}{2} & 0 & \frac{1}{2} & 0 & \frac{1}{2} & 0 & \frac{1}{2} \\
 \frac{1}{2} & -1 & -\frac{1}{2} & 0 & -\frac{1}{2} & 0 & -\frac{1}{2} & 0 \\
 0 & -\frac{1}{2} & -1 & \frac{1}{2} & 0 & \frac{1}{2} & 0 & \frac{1}{2} \\
 \frac{1}{2} & 0 & \frac{1}{2} & -1 & -\frac{1}{2} & 0 & -\frac{1}{2} & 0 \\
 0 & -\frac{1}{2} & 0 & -\frac{1}{2} & -1 & \frac{1}{2} & 0 & \frac{1}{2} \\
 \frac{1}{2} & 0 & \frac{1}{2} & 0 & \frac{1}{2} & -1 & -\frac{1}{2} & 0 \\
 0 & -\frac{1}{2} & 0 & -\frac{1}{2} & 0 & -\frac{1}{2} & -1 & \frac{1}{2} \\
 \frac{1}{2} & 0 & \frac{1}{2} & 0 & \frac{1}{2} & 0 & \frac{1}{2} & -1 \\
\end{array}
\right)
\end{align}
For the simple root basis we find
\begin{align}
\begin{split}
 \alpha_1&=\{0, 1, -1, -1, 0, -1, 1, 1\}\,,\quad\,
 \alpha_2=\{0, 0, 0, 0, 0, 1, 0, -1\}\,,\\
 \alpha_3&=\{0, 0, 0, 0, 1, 0, -1, 0\}\,,\qquad~~
 \alpha_4=\{0, 0, 1, 0, -1, 0, 0, 0\}\,,\\
 \alpha_5&=\{1, 0, -1, 0, 0, 0, 0, 0\}\,,\qquad~~
 \alpha_6=\{0, 0, 0, 1, -1, -1, 1, 0\}\,.  
\end{split}
\end{align}
There is an outer automorphism which folds $E_6$ to $F_4$, which is,
however, not realized by any of the braid action in these
examples. Thus as in the previous cases (as well as in the cases to be
discussed subsequently), the braid induces inner automorphisms only.

\textbf{$\boldsymbol{a=1}$:\quad}
From (\ref{eqn:braidactiononbasisgtube}) the braid induces an action on the junction basis $\Gamma_i=e_i\in \bZ^{8}$ given by 
 \begin{align} 
 B=
\begin{pmatrix}
 0 & 0 & 0 & 0 & 0 & -1 & 0 & 0 \\
 0 & 0 & 0 & 0 & 0 & 0 & 1 & 0 \\
 0 & 0 & 0 & 0 & 0 & 0 & 0 & -1 \\
 1 & 0 & 0 & 0 & 0 & 0 & 0 & 0 \\
 0 & -1 & 0 & 0 & 0 & 0 & 0 & 0 \\
 0 & 0 & 1 & 0 & 0 & 0 & 0 & 0 \\
 0 & 0 & 0 & -1 & 0 & 0 & 0 & 0 \\
 0 & 0 & 0 & 0 & 1 & 0 & 0 & 0
\end{pmatrix}
\,, 
\end{align}
which determines a non-trivial braid invariant sublattice of the junction lattice generated by
\begin{center}
\begin{tabular}{c|c|c} 
 Junction $J$ & $(J,J)$ & $a(J)$ \\ \hline
$(1, -1, -1, 1, 1, -1, -1, 1)$ & $-12$ & $(0, 0)$
\end{tabular}.
\end{center}
We see that the invariant simple root junction $\beta_i$ is the linear combinations 
\begin{align}
 \beta_1=-\alpha_1-2 \alpha_2-\alpha_4+\alpha_5\,
\end{align}
of the simple roots of $G_g=E_6$, which themselves generate an algebra $G_h=SU(2)$.

The asymptotic charges are mapped by the braid as $(a_1,a_2)\mapsto(-a_2,a_1)$, fixing the asymptotic
charge of junctions $J$ with $a(J)=0$, and thereby giving an automorphism on the roots. For the simple roots, we find
\begin{align}
\begin{split}
\alpha_1 &\mapsto \alpha_1+\alpha_2+\alpha_3+\alpha_4+\alpha_5+\alpha_6\,,\quad 
\alpha_2 \mapsto -\alpha_5\,,\quad 
\alpha_3 \mapsto -\alpha_1-2 \alpha_2-2 \alpha_3-\alpha_4-\alpha_6\,,\\
\alpha_4 &\mapsto \alpha_2\,,\quad 
\alpha_5 \mapsto \alpha_3+\alpha_6\,,\quad 
\alpha_6 \mapsto \alpha_1+2 \alpha_2+3 \alpha_3+2 \alpha_4+\alpha_5+\alpha_6\,.
\end{split}
\end{align}
The invariant subspace has dimension one and is precisely that subspace spanned by $\beta_1$.

\textbf{$\boldsymbol{a=2}$:\quad}
From (\ref{eqn:braidactiononbasisgtube}) the braid induces an action on the junction basis $\Gamma_i=e_i\in \bZ^{8}$ given by 
 \begin{align} 
 B=
\begin{pmatrix}
 0 & 0 & -1 & 0 & 0 & 0 & 0 & 0 \\
 0 & 0 & 0 & -1 & 0 & 0 & 0 & 0 \\
 0 & 0 & 0 & 0 & -1 & 0 & 0 & 0 \\
 0 & 0 & 0 & 0 & 0 & -1 & 0 & 0 \\
 0 & 0 & 0 & 0 & 0 & 0 & -1 & 0 \\
 0 & 0 & 0 & 0 & 0 & 0 & 0 & -1 \\
 -1 & 0 & 0 & 0 & 0 & 0 & 0 & 0 \\
 0 & -1 & 0 & 0 & 0 & 0 & 0 & 0
\end{pmatrix}
\,, 
\end{align}
which determines a non-trivial braid invariant sublattice of the junction lattice generated by
\begin{center}
\begin{tabular}{c|c|c} 
 Junction $J$ & $(J,J)$ & $a(J)$ \\ \hline
$(1, 0, -1, 0, 1, 0, -1, 0)$ & $-4$ & $(0, 0)$ \\
$(0, 1, 0, -1, 0, 1, 0, -1)$ & $-4$ & $(0, 0)$
\end{tabular}.
\end{center}
We see that the invariant junctions $\beta_i$ are the linear combinations 
\begin{align}
 \beta_1=\alpha_3+\alpha_5\,,\qquad \beta_2=\alpha_1+2 \alpha_2+\alpha_3+\alpha_4\,.
\end{align}
of the simple roots of $G_g=E_6$, which themselves generate an algebra $G_h=SU(3)$.

The asymptotic charge map $(a_1,a_2)\mapsto(-a_1,-a_2)$ acts trivially on junctions with $a(J)=0$,
so that the braid gives an automorphism of roots. The simple roots map as
\begin{align}
\begin{split}
\alpha_1 &\mapsto \alpha_1 +2 \alpha_2+3 \alpha_3+2 \alpha_4+\alpha_5+2 \alpha_6\,,\quad 
\alpha_2 \mapsto -\alpha_3-\alpha_6\,,\quad 
\alpha_3 \mapsto -\alpha_4\,,\\ 
\alpha_4 &\mapsto -\alpha_5\,,\quad 
\alpha_5 \mapsto \alpha_3+\alpha_4+\alpha_5\,,\quad 
\alpha_6 \mapsto -\alpha_1-\alpha_2-\alpha_3\,.
\end{split}
\end{align}
The invariant subalgebra is spanned by the two simple roots.
Inspection of the Cartan matrix \eqref{eq:CartanMatrix} associated to $\beta_i$ reveals that the invariant flavor algebra of the original $E_6$ is $SU(3)$. Note that this is in accordance with the result from the \ab23 case in the $f$-tube. There, the $\IN_{0}^*$ singularity was reduced to an invariant $SU(3)$ subgroup.

\textbf{$\boldsymbol{a=3}$:\quad}
From (\ref{eqn:braidactiononbasisgtube}) the braid induces an action on the junction basis $\Gamma_i=e_i\in \bZ^{8}$ given by 
 \begin{align} 
 B=
\begin{pmatrix}
 0 & 0 & 0 & 0 & 0 & 0 & 0 & 1 \\
 -1 & 0 & 0 & 0 & 0 & 0 & 0 & 0 \\
 0 & 1 & 0 & 0 & 0 & 0 & 0 & 0 \\
 0 & 0 & -1 & 0 & 0 & 0 & 0 & 0 \\
 0 & 0 & 0 & 1 & 0 & 0 & 0 & 0 \\
 0 & 0 & 0 & 0 & -1 & 0 & 0 & 0 \\
 0 & 0 & 0 & 0 & 0 & 1 & 0 & 0 \\
 0 & 0 & 0 & 0 & 0 & 0 & -1 & 0
\end{pmatrix}
\,, 
\end{align}
which determines a non-trivial braid invariant sublattice of the junction lattice generated by
\begin{center}
\begin{tabular}{c|c|c} 
 Junction $J$ & $(J,J)$ & $a(J)$ \\ \hline
$(1, -1, -1, 1, 1, -1, -1, 1)$ & $-12$ & $(0, 0)$
\end{tabular}.
\end{center}
We see that the invariant simple root junction $\beta_1$ is the linear combination 
\begin{align}
 \beta_1=-\alpha_1-2 \alpha_2-\alpha_4+\alpha_5\,,
\end{align}
of the simple roots of $G_g=E_6$, which generates an algebra $G_h=SU(2)$.

This case yields $(a_1,a_2)\mapsto(a_2,-a_1)$ for the asymptotic charges, so the braid
generates an automorphism on roots, and map the simple roots 
\begin{align}
\begin{split}
\alpha_1 &\mapsto \alpha_2+2 \alpha_3+2 \alpha_4+\alpha_5+\alpha_6\,,\quad 
\alpha_2 \mapsto -\alpha_3-\alpha_4-\alpha_5\,,\quad 
\alpha_3 \mapsto -\alpha_2\,,\\
\alpha_4 &\mapsto -\alpha_3-\alpha_6\,,\quad 
\alpha_5 \mapsto -\alpha_1-\alpha_2-\alpha_3-\alpha_4\,,\quad 
\alpha_6 \mapsto \alpha_2+\alpha_3\,.
\end{split}
\end{align}
Note that the invariant combination of root $\beta_1$ is the same as the invariant simple root found in the $a=1$ case.

\subsection{Type \texorpdfstring{$\III^*$}{III*} fibers}
Type $\III^*$ occur in the $f$-tube for $a=3$. The $I$-matrix reads
\begin{align}
 I=\left(
\begin{array}{ccccccccc}
 -1 & \frac{1}{2} & -\frac{1}{2} & 0 & \frac{1}{2} & -\frac{1}{2} & 0 & \frac{1}{2} & -\frac{1}{2} \\
 \frac{1}{2} & -1 & \frac{1}{2} & -\frac{1}{2} & 0 & \frac{1}{2} & -\frac{1}{2} & 0 & \frac{1}{2} \\
 -\frac{1}{2} & \frac{1}{2} & -1 & \frac{1}{2} & -\frac{1}{2} & 0 & \frac{1}{2} & -\frac{1}{2} & 0 \\
 0 & -\frac{1}{2} & \frac{1}{2} & -1 & \frac{1}{2} & -\frac{1}{2} & 0 & \frac{1}{2} & -\frac{1}{2} \\
 \frac{1}{2} & 0 & -\frac{1}{2} & \frac{1}{2} & -1 & \frac{1}{2} & -\frac{1}{2} & 0 & \frac{1}{2} \\
 -\frac{1}{2} & \frac{1}{2} & 0 & -\frac{1}{2} & \frac{1}{2} & -1 & \frac{1}{2} & -\frac{1}{2} & 0 \\
 0 & -\frac{1}{2} & \frac{1}{2} & 0 & -\frac{1}{2} & \frac{1}{2} & -1 & \frac{1}{2} & -\frac{1}{2} \\
 \frac{1}{2} & 0 & -\frac{1}{2} & \frac{1}{2} & 0 & -\frac{1}{2} & \frac{1}{2} & -1 & \frac{1}{2} \\
 -\frac{1}{2} & \frac{1}{2} & 0 & -\frac{1}{2} & \frac{1}{2} & 0 & -\frac{1}{2} & \frac{1}{2} & -1 \\
\end{array}
\right)\,.
\end{align}
We can choose the set of simple root junctions
\begin{align}
\begin{split}
  \alpha_1&=\{1, -1, -1, 0, 1, 0, -1, 0, 1\}\,,\quad
  \alpha_2=\{0, 1, 0, -1, -1, 0, 0, -1, -1\}\,,\\
  \alpha_3&=\{0, 0, 0, 0, 0, 0, 1, 1, 1\}\,,\qquad\quad~
  \alpha_4=\{0, 0, 0, 0, 0, 1, 0, 0, -1\}\,,\\
  \alpha_5&=\{0, 0, 0, 0, 1, 0, 0, -1, 0\}\,,\qquad~\,
  \alpha_6=\{0, 0, 1, 0, -1, -1, 0, 1, 0\}\,,\\
  \alpha_7&=\{0, 0, 0, 1, 0, -1, -1, 0, 1\}\,. 
\end{split}
 \end{align}
All automorphisms of $E_7$ are inner.

\textbf{$\boldsymbol{b=1}$:\quad}
From (\ref{eqn:braidactiononbasisftube}) the braid induces an action on the junction basis $\Gamma_i=e_i\in \bZ^{9}$ given by 
 \begin{align} 
 B=
\begin{pmatrix}
 0 & 0 & 0 & 0 & 0 & 0 & 0 & -1 & 0 \\
 0 & 0 & 0 & 0 & 0 & 0 & 0 & 0 & -1 \\
 -1 & 0 & 0 & 0 & 0 & 0 & 0 & 0 & 0 \\
 0 & -1 & 0 & 0 & 0 & 0 & 0 & 0 & 0 \\
 0 & 0 & -1 & 0 & 0 & 0 & 0 & 0 & 0 \\
 0 & 0 & 0 & -1 & 0 & 0 & 0 & 0 & 0 \\
 0 & 0 & 0 & 0 & -1 & 0 & 0 & 0 & 0 \\
 0 & 0 & 0 & 0 & 0 & -1 & 0 & 0 & 0 \\
 0 & 0 & 0 & 0 & 0 & 0 & -1 & 0 & 0
\end{pmatrix}
\,, 
\end{align}
which leaves no junctions invariant; the invariant sublattice has dimension $0$; thus $G_h=\emptyset$.

The asymptotic charges are mapped according to $(a_1,a_2)\mapsto(a_1-a_2,a_1)$ and the simple roots according to
\begin{align}
 \begin{split}
  \alpha_1 &\mapsto -\alpha_2-\alpha_3-\alpha_4-\alpha_5-\alpha_6\,,\\
  \alpha_2 &\mapsto \alpha_1+2 \alpha_2+3 \alpha_3+3 \alpha_4+2 \alpha_5+\alpha_6+\alpha_7\,,\\
  \alpha_3 &\mapsto -\alpha_1-2 \alpha_2-3 \alpha_3-3 \alpha_4-2 \alpha_5-\alpha_6-2 \alpha_7\,,\\
  \alpha_4 &\mapsto \alpha_2+\alpha_3+\alpha_4+\alpha_5+\alpha_7\,,\\
  \alpha_5 &\mapsto \alpha_1+\alpha_2+\alpha_3+2 \alpha_4+\alpha_5+\alpha_6+\alpha_7\,,\\
  \alpha_6 &\mapsto -\alpha_1-\alpha_2-\alpha_3-2 \alpha_4-2 \alpha_5-\alpha_6-\alpha_7\,,\\
  \alpha_7 &\mapsto -\alpha_2-\alpha_3-2 \alpha_4-\alpha_5-\alpha_7\,.
 \end{split}
\end{align}
There is no non-trivial invariant subalgebra.

\textbf{$\boldsymbol{b=2}$:\quad}
From (\ref{eqn:braidactiononbasisftube}) the braid induces an action on the junction basis $\Gamma_i=e_i\in \bZ^{9}$ given by 
 \begin{align} 
 B=
\begin{pmatrix}
 0 & 0 & 0 & 0 & 0 & 1 & 0 & 0 & 0 \\
 0 & 0 & 0 & 0 & 0 & 0 & 1 & 0 & 0 \\
 0 & 0 & 0 & 0 & 0 & 0 & 0 & 1 & 0 \\
 0 & 0 & 0 & 0 & 0 & 0 & 0 & 0 & 1 \\
 1 & 0 & 0 & 0 & 0 & 0 & 0 & 0 & 0 \\
 0 & 1 & 0 & 0 & 0 & 0 & 0 & 0 & 0 \\
 0 & 0 & 1 & 0 & 0 & 0 & 0 & 0 & 0 \\
 0 & 0 & 0 & 1 & 0 & 0 & 0 & 0 & 0 \\
 0 & 0 & 0 & 0 & 1 & 0 & 0 & 0 & 0
\end{pmatrix}
\,, 
\end{align}
which determines a non-trivial braid invariant sublattice of the junction lattice generated by
\begin{center}
\begin{tabular}{c|c|c} 
 Junction $J$ & $(J,J)$ & $a(J)$ \\ \hline
$(1, 1, 1, 1, 1, 1, 1, 1, 1)$ & $-6$ & $(0, 0)$
\end{tabular}.
\end{center}
We see that the invariant simple root junction $\beta_i$ is a linear combination 
\begin{align}
 \beta_1=\alpha_1+2 \alpha_2+5 \alpha_3+6 \alpha_4+4 \alpha_5+2 \alpha_6+3 \alpha_7\,.
\end{align}
of the simple roots of $G_f=E_7$, which generates $G_h=SU(2)$.
This matches $g$-tube case of the same $(a,b)$, which has a \IV fiber in the $g$-tube and was also reduced to $SU(2)$.

The maps are in this case $(a_1,a_2)\mapsto(-a_2,a_1-a_2)$ and 
\begin{align}
 \begin{split}
  \alpha_1 &\mapsto -\alpha_2-\alpha_3-\alpha_4\,,\qquad\hspace{33mm}
  \alpha_2 \mapsto -\alpha_3-\alpha_4-\alpha_5-\alpha_6-\alpha_7\,,\\
  \alpha_3 &\mapsto \alpha_2+2 \alpha_3+3 \alpha_4+2 \alpha_5+\alpha_6+2 \alpha_7\,,\qquad~
  \alpha_4 \mapsto \alpha_1+\alpha_2+\alpha_3+\alpha_4+\alpha_5+\alpha_6\,,\\
  \alpha_5 &\mapsto -\alpha_4-\alpha_5-\alpha_6\,,\quad\hspace{37mm}
  \alpha_6 \mapsto -\alpha_1-\alpha_2-\alpha_3-\alpha_4-\alpha_7\,,\\
  \alpha_7 &\mapsto -\alpha_1-2 \alpha_2-2 \alpha_3-2 \alpha_4-2 \alpha_5-\alpha_6-\alpha_7\,.
 \end{split}
\end{align}
for the asymptotic charges and the simple roots, respectively.

\textbf{$\boldsymbol{b=3}$:\quad}
From (\ref{eqn:braidactiononbasisftube}) the braid induces an action on the junction basis $\Gamma_i=e_i\in \bZ^{9}$ given by 
 \begin{align} 
 B=
\begin{pmatrix}
 0 & 0 & 0 & -1 & 0 & 0 & 0 & 0 & 0 \\
 0 & 0 & 0 & 0 & -1 & 0 & 0 & 0 & 0 \\
 0 & 0 & 0 & 0 & 0 & -1 & 0 & 0 & 0 \\
 0 & 0 & 0 & 0 & 0 & 0 & -1 & 0 & 0 \\
 0 & 0 & 0 & 0 & 0 & 0 & 0 & -1 & 0 \\
 0 & 0 & 0 & 0 & 0 & 0 & 0 & 0 & -1 \\
 -1 & 0 & 0 & 0 & 0 & 0 & 0 & 0 & 0 \\
 0 & -1 & 0 & 0 & 0 & 0 & 0 & 0 & 0 \\
 0 & 0 & -1 & 0 & 0 & 0 & 0 & 0 & 0
\end{pmatrix}
\,, 
\end{align}
which leaves no junctions invariant; the invariant sublattice has dimension $0$; thus $G_h=\emptyset$.

The asymptotic charges are mapped as $(a_1,a_2)\mapsto(-a_1,-a_2)$ and the simple root string junctions as
\begin{align}
 \begin{split}
  \alpha_1 &\mapsto -\alpha_2-\alpha_3-\alpha_4-\alpha_5\,,\qquad
  \alpha_2 \mapsto \alpha_1+2 \alpha_2+3 \alpha_3+4 \alpha_4+3 \alpha_5+\alpha_6+2 \alpha_7\,,\\
  \alpha_3 &\mapsto -\alpha_3-2 \alpha_4-\alpha_5-\alpha_7\,,\quad~\,
  \alpha_4 \mapsto -\alpha_5-\alpha_6\,,\\
  \alpha_5 &\mapsto -\alpha_2-\alpha_3-\alpha_4-\alpha_7\,,\qquad
  \alpha_6 \mapsto \alpha_2+\alpha_3+2 \alpha_4+\alpha_5+\alpha_6+\alpha_7\,,\\
  \alpha_7 &\mapsto -\alpha_1-\alpha_2-\alpha_3-\alpha_4\,.
 \end{split}
\end{align}
The flavor algebra after the quotient by the braid action is trivial, matching the corresponding $\IN_0^*$ case.

\textbf{$\boldsymbol{b=4}$:\quad}
From (\ref{eqn:braidactiononbasisftube}) the braid induces an action on the junction basis $\Gamma_i=e_i\in \bZ^{9}$ given by 
 \begin{align} 
 B=
\begin{pmatrix}
 0 & 1 & 0 & 0 & 0 & 0 & 0 & 0 & 0 \\
 0 & 0 & 1 & 0 & 0 & 0 & 0 & 0 & 0 \\
 0 & 0 & 0 & 1 & 0 & 0 & 0 & 0 & 0 \\
 0 & 0 & 0 & 0 & 1 & 0 & 0 & 0 & 0 \\
 0 & 0 & 0 & 0 & 0 & 1 & 0 & 0 & 0 \\
 0 & 0 & 0 & 0 & 0 & 0 & 1 & 0 & 0 \\
 0 & 0 & 0 & 0 & 0 & 0 & 0 & 1 & 0 \\
 0 & 0 & 0 & 0 & 0 & 0 & 0 & 0 & 1 \\
 1 & 0 & 0 & 0 & 0 & 0 & 0 & 0 & 0
\end{pmatrix}
\,, 
\end{align}
which determines a non-trivial braid invariant sublattice of the junction lattice generated by
\begin{center}
\begin{tabular}{c|c|c} 
 Junction $J$ & $(J,J)$ & $a(J)$ \\ \hline
$(1, 1, 1, 1, 1, 1, 1, 1, 1)$ & $-6$ & $(0, 0)$
\end{tabular}.
\end{center}
We see that the invariant junction $\beta_1$ is the linear combination 
\begin{align}
 \beta_1=\alpha_1+2 \alpha_2+5 \alpha_3+6 \alpha_4+4 \alpha_5+2 \alpha_6+3 \alpha_7\,,
\end{align}
of the simple roots of $G_f=E_7$, which generates an algebra $G_h=SU(2)$.
This invariant $SU(2)$ algebra matches the reduced flavor group of the $\IV^*$ case of the $g$-tube.

We find $(a_1,a_2)\mapsto(a_2-a_1,-a_1)$ and 
\begin{align}
 \begin{split}
  \alpha_1 &\mapsto -\alpha_1-2 \alpha_2-2 \alpha_3-3 \alpha_4-2 \alpha_5-\alpha_6-\alpha_7\,,\qquad
  \alpha_2 \mapsto \alpha_1+\alpha_2\,,\\
  \alpha_3 &\mapsto \alpha_3+\alpha_4\,,\hspace{62.5mm}
  \alpha_4 \mapsto \alpha_5\,,\\
  \alpha_5 &\mapsto \alpha_4+\alpha_7\,,\hspace{62.5mm}
  \alpha_6 \mapsto \alpha_2+\alpha_3\,,\\
  \alpha_7 &\mapsto \alpha_6\,.
 \end{split}
\end{align}

\textbf{$\boldsymbol{b=5}$:\quad}
From (\ref{eqn:braidactiononbasisftube}) the braid induces an action on the junction basis $\Gamma_i=e_i\in \bZ^{9}$ given by 
 \begin{align} 
 B=
\begin{pmatrix}
 0 & 0 & 0 & 0 & 0 & 0 & 0 & 0 & -1 \\
 -1 & 0 & 0 & 0 & 0 & 0 & 0 & 0 & 0 \\
 0 & -1 & 0 & 0 & 0 & 0 & 0 & 0 & 0 \\
 0 & 0 & -1 & 0 & 0 & 0 & 0 & 0 & 0 \\
 0 & 0 & 0 & -1 & 0 & 0 & 0 & 0 & 0 \\
 0 & 0 & 0 & 0 & -1 & 0 & 0 & 0 & 0 \\
 0 & 0 & 0 & 0 & 0 & -1 & 0 & 0 & 0 \\
 0 & 0 & 0 & 0 & 0 & 0 & -1 & 0 & 0 \\
 0 & 0 & 0 & 0 & 0 & 0 & 0 & -1 & 0
\end{pmatrix}
\,, 
\end{align}
which leaves no junctions invariant; the invariant sublattice has dimension $0$; thus $G_h=\emptyset$.

For this case the maps are given by $(a_1,a_2)\mapsto(a_2,a_2-a_1)$ for the asymptotic charges and by
\begin{align}
  \alpha_1 &\mapsto -\alpha_1-2 \alpha_2-2 \alpha_3-2 \alpha_4-\alpha_5-\alpha_7\,,\qquad~~
  \alpha_2 \mapsto \alpha_1+\alpha_2+2 \alpha_3+2 \alpha_4+\alpha_5+\alpha_7\,,\nonumber\\
  \alpha_3 &\mapsto -\alpha_1-\alpha_2-2 \alpha_3-2 \alpha_4-\alpha_5-\alpha_6-\alpha_7\,,\quad
  \alpha_4 \mapsto \alpha_1+\alpha_2+\alpha_3+2 \alpha_4+\alpha_5+\alpha_6+\alpha_7\,,\nonumber\\
  \alpha_5 &\mapsto -\alpha_4\,,\hspace{61mm}
  \alpha_6 \mapsto -\alpha_7\,,\nonumber\\
  \alpha_7 &\mapsto -\alpha_1-\alpha_2-\alpha_3-2 \alpha_4-2 \alpha_5-\alpha_6-\alpha_7\,.
\end{align}
for the simple roots. The invariant flavor algebra is trivial.

\subsection{Type \texorpdfstring{$\II^*$}{II*} fibers}
Finally, we obtain type $II^*$ fibers in the $g$-tube for $b=5$. We find the $I$-matrix
\begin{align}
 I=\left(
\begin{array}{cccccccccc}
 -1 & \frac{1}{2} & 0 & \frac{1}{2} & 0 & \frac{1}{2} & 0 & \frac{1}{2} & 0 & \frac{1}{2} \\
 \frac{1}{2} & -1 & -\frac{1}{2} & 0 & -\frac{1}{2} & 0 & -\frac{1}{2} & 0 & -\frac{1}{2} & 0 \\
 0 & -\frac{1}{2} & -1 & \frac{1}{2} & 0 & \frac{1}{2} & 0 & \frac{1}{2} & 0 & \frac{1}{2} \\
 \frac{1}{2} & 0 & \frac{1}{2} & -1 & -\frac{1}{2} & 0 & -\frac{1}{2} & 0 & -\frac{1}{2} & 0 \\
 0 & -\frac{1}{2} & 0 & -\frac{1}{2} & -1 & \frac{1}{2} & 0 & \frac{1}{2} & 0 & \frac{1}{2} \\
 \frac{1}{2} & 0 & \frac{1}{2} & 0 & \frac{1}{2} & -1 & -\frac{1}{2} & 0 & -\frac{1}{2} & 0 \\
 0 & -\frac{1}{2} & 0 & -\frac{1}{2} & 0 & -\frac{1}{2} & -1 & \frac{1}{2} & 0 & \frac{1}{2} \\
 \frac{1}{2} & 0 & \frac{1}{2} & 0 & \frac{1}{2} & 0 & \frac{1}{2} & -1 & -\frac{1}{2} & 0 \\
 0 & -\frac{1}{2} & 0 & -\frac{1}{2} & 0 & -\frac{1}{2} & 0 & -\frac{1}{2} & -1 & \frac{1}{2} \\
 \frac{1}{2} & 0 & \frac{1}{2} & 0 & \frac{1}{2} & 0 & \frac{1}{2} & 0 & \frac{1}{2} & -1 \\
\end{array}
\right)
\end{align}
From the root junctions of $E_8$ we choose the following set of simple roots:
\begin{align}
\begin{split}
 \alpha_1&=\{1,-1,1,1,0,1,-1,0,-1,-1\}\,,\qquad
 \alpha_2=\{0,1,-2,-1,0,-1,1,0,1,1\}\,,\\
 \alpha_3&=\{0,0,1,0,-1,0,0,0,0,0\}\,,\quad\hspace{12.8mm}
 \alpha_4=\{0,0,0,0,1,0,-1,0,0,0\}\,,\\
 \alpha_5&=\{0,0,0,0,0,0,1,0,-1,0\}\,,\quad\hspace{12.8mm}
 \alpha_6=\{0,0,0,0,0,0,0,1,0,-1\}\,,\\
 \alpha_7&=\{0,0,0,1,-1,-1,0,-1,1,1\}\,,\quad\hspace{7mm}
 \alpha_8=\{0,0,0,0,0,1,-1,-1,1,0\}\,.
\end{split}
\end{align}
Note that all automorphisms of $E_8$ are inner.

\textbf{$\boldsymbol{a=1}$:\quad}
From (\ref{eqn:braidactiononbasisgtube}) the braid induces an action on the junction basis $\Gamma_i=e_i\in \bZ^{10}$ given by 
 \begin{align} 
 B=
\begin{pmatrix}
 0 & 0 & 0 & 0 & 0 & 0 & 0 & -1 & 0 & 0 \\
 0 & 0 & 0 & 0 & 0 & 0 & 0 & 0 & 1 & 0 \\
 0 & 0 & 0 & 0 & 0 & 0 & 0 & 0 & 0 & -1 \\
 1 & 0 & 0 & 0 & 0 & 0 & 0 & 0 & 0 & 0 \\
 0 & -1 & 0 & 0 & 0 & 0 & 0 & 0 & 0 & 0 \\
 0 & 0 & 1 & 0 & 0 & 0 & 0 & 0 & 0 & 0 \\
 0 & 0 & 0 & -1 & 0 & 0 & 0 & 0 & 0 & 0 \\
 0 & 0 & 0 & 0 & 1 & 0 & 0 & 0 & 0 & 0 \\
 0 & 0 & 0 & 0 & 0 & -1 & 0 & 0 & 0 & 0 \\
 0 & 0 & 0 & 0 & 0 & 0 & 1 & 0 & 0 & 0
\end{pmatrix}
\,, 
\end{align}
which leaves no junctions invariant; the invariant sublattice has dimension $0$; thus $G_h=\emptyset$.
 This matches the reduction from the corresponding $f$-tube, where the $G_f=SU(2)$ flavor algebra was reduced to a trivial one as well.

We find that the asymptotic charges are mapped via $(a_1,a_2)\mapsto(-a_2,a_1)$. The map for the simple roots is
\begin{align}
 \begin{split}
  \alpha_1 &\mapsto -\alpha_2-\alpha_3\,,\qquad
  \alpha_2 \mapsto \alpha_2+\alpha_3+\alpha_4+\alpha_5+\alpha_6+\alpha_7\,,\qquad~\,
  \alpha_3 \mapsto \alpha_5+\alpha_8\,,\\
  \alpha_4 &\mapsto \alpha_6\,,\quad
  \alpha_5 \mapsto -\alpha_2-2 \alpha_3-3 \alpha_4-4 \alpha_5-3 \alpha_6-\alpha_7-2 \alpha_8\,,\quad
  \alpha_6 \mapsto -\alpha_1-\alpha_2\\
  \alpha_7 &\mapsto \alpha_1+2 \alpha_2+2 \alpha_3+3 \alpha_4+3 \alpha_5+2 \alpha_6+\alpha_7+2 \alpha_8\,,\\
  \alpha_8 &\mapsto \alpha_1+2 \alpha_2+3 \alpha_3+4 \alpha_4+5 \alpha_5+3 \alpha_6+\alpha_7+2 \alpha_8\,.
 \end{split}
\end{align}
This action does not have an invariant subspace, as already deduced from $B$.

\textbf{$\boldsymbol{a=2}$:\quad}
From (\ref{eqn:braidactiononbasisgtube}) the braid induces an action on the junction basis $\Gamma_i=e_i\in \bZ^{10}$ given by 
 \begin{align} 
 B=
\begin{pmatrix}
 0 & 0 & 0 & 0 & -1 & 0 & 0 & 0 & 0 & 0 \\
 0 & 0 & 0 & 0 & 0 & -1 & 0 & 0 & 0 & 0 \\
 0 & 0 & 0 & 0 & 0 & 0 & -1 & 0 & 0 & 0 \\
 0 & 0 & 0 & 0 & 0 & 0 & 0 & -1 & 0 & 0 \\
 0 & 0 & 0 & 0 & 0 & 0 & 0 & 0 & -1 & 0 \\
 0 & 0 & 0 & 0 & 0 & 0 & 0 & 0 & 0 & -1 \\
 -1 & 0 & 0 & 0 & 0 & 0 & 0 & 0 & 0 & 0 \\
 0 & -1 & 0 & 0 & 0 & 0 & 0 & 0 & 0 & 0 \\
 0 & 0 & -1 & 0 & 0 & 0 & 0 & 0 & 0 & 0 \\
 0 & 0 & 0 & -1 & 0 & 0 & 0 & 0 & 0 & 0
\end{pmatrix}
\,, 
\end{align}
which leaves no junctions invariant; the invariant sublattice has dimension $0$; thus $G_h=\emptyset$.
The $G_g=E_8$ flavor algebra is again reduced to a trivial algebra, as is the case for the corresponding $SO(8)$ flavor algebra in the $f$-tube analysis.

The asymptotic charges and the simple root maps are $(a_1,a_2)\mapsto(-a_1,-a_2)$ and
\begin{align}
  \alpha_1 &\mapsto -\alpha_2-\alpha_3-\alpha_4-2 \alpha_5-\alpha_6-\alpha_7-\alpha_8\,,\quad
  \alpha_2 \mapsto \alpha_2+\alpha_3+\alpha_4+\alpha_5+\alpha_6+\alpha_7+\alpha_8\,,\nonumber\\
  \alpha_3 &\mapsto \alpha_1+\alpha_2+\alpha_3+\alpha_4+\alpha_5\,,\hspace{26.5mm}
  \alpha_4 \mapsto -\alpha_1-\alpha_2\,,\nonumber\\
  \alpha_5 &\mapsto -\alpha_3\,,\hspace{58.8mm}
  \alpha_6 \mapsto -\alpha_4-\alpha_5-\alpha_6-\alpha_7\,,\\
  \alpha_7 &\mapsto \alpha_1+2 \alpha_2+3 \alpha_3+4 \alpha_4+5 \alpha_5+4 \alpha_6+2 \alpha_7+2 \alpha_8\,,\nonumber\\
  \alpha_8 &\mapsto -\alpha_2-\alpha_3-2 \alpha_4-2 \alpha_5-\alpha_6-\alpha_8\,,\nonumber
\end{align}
respectively. There is no invariant subspace for this map, as already deduced from $B$.

\textbf{$\boldsymbol{a=3}$:\quad}
From (\ref{eqn:braidactiononbasisgtube}) the braid induces an action on the junction basis $\Gamma_i=e_i\in \bZ^{10}$ given by 
 \begin{align} 
 B=
\begin{pmatrix}
 0 & 1 & 0 & 0 & 0 & 0 & 0 & 0 & 0 & 0 \\
 0 & 0 & -1 & 0 & 0 & 0 & 0 & 0 & 0 & 0 \\
 0 & 0 & 0 & 1 & 0 & 0 & 0 & 0 & 0 & 0 \\
 0 & 0 & 0 & 0 & -1 & 0 & 0 & 0 & 0 & 0 \\
 0 & 0 & 0 & 0 & 0 & 1 & 0 & 0 & 0 & 0 \\
 0 & 0 & 0 & 0 & 0 & 0 & -1 & 0 & 0 & 0 \\
 0 & 0 & 0 & 0 & 0 & 0 & 0 & 1 & 0 & 0 \\
 0 & 0 & 0 & 0 & 0 & 0 & 0 & 0 & -1 & 0 \\
 0 & 0 & 0 & 0 & 0 & 0 & 0 & 0 & 0 & 1 \\
 -1 & 0 & 0 & 0 & 0 & 0 & 0 & 0 & 0 & 0
\end{pmatrix}
\,, 
\end{align}
which leaves no junctions invariant; the invariant sublattice has dimension $0$; thus $G_h=\emptyset$.
This matches with the reduction of $E_7$ to a trivial algebra in the $f$-tube.

We find for the asymptotic charges that $(a_1,a_2)\mapsto(a_2,-a_1)$ and for the simple roots
\begin{align}
  \begin{split}
    \alpha_1 &\mapsto -\alpha_1-2 \alpha_2-2 \alpha_3-2 \alpha_4-2 \alpha_5-\alpha_6-\alpha_7-\alpha_8\,,\\
    \alpha_2 &\mapsto \alpha_1+3 \alpha_2+4 \alpha_3+5 \alpha_4+6 \alpha_5+4 \alpha_6+2 \alpha_7+3 \alpha_8\,,\\
    \alpha_3 &\mapsto -\alpha_2-2 \alpha_3-2 \alpha_4-2 \alpha_5-\alpha_6-\alpha_8\,,\qquad
    \alpha_4 \mapsto -\alpha_4-\alpha_5-\alpha_6-\alpha_7\,,\\
    \alpha_5 &\mapsto -\alpha_5-\alpha_8\,,\hspace{49mm}
    \alpha_6 \mapsto \alpha_5\,,\\
    \alpha_7 &\mapsto \alpha_3+\alpha_4+\alpha_5+\alpha_6+\alpha_7+\alpha_8\,,\hspace{16.3mm}
    \alpha_8 \mapsto \alpha_4+\alpha_5+\alpha_8\,. 
  \end{split}
\end{align}
There is no invariant sublattice, as already deduced from $B$.

\end{appendices}

\bibliographystyle{chetref}
\bibliography{refs}
\end{document}